\documentclass[11pt, a4paper, twoside]{scrartcl}

\usepackage[left=4cm, right=2cm, top=2cm, bottom=3cm]{geometry}

\usepackage{pgffor} 

\usepackage[british]{babel}
\usepackage[utf8]{inputenc}
\usepackage[T1]{fontenc}
\usepackage{lmodern}
\usepackage{xcolor}
\usepackage{amsmath} 
\usepackage{amsfonts} 
\usepackage{multicol} 

\usepackage{pdfpages} 

\usepackage{booktabs}
\usepackage{multirow} 
\usepackage{graphicx}

\usepackage{float}

\usepackage{rotating}

\usepackage{csquotes}

\usepackage{url}
\usepackage{ulem} 

\usepackage{ragged2e}
\usepackage{nameref}
\makeatletter
\newcommand*{\currentname}{\@currentlabelname}
\makeatother

\usepackage{soul} 
\usepackage{xspace} 

\usepackage[locale=UK, round-mode=places, exponent-product=\cdot, separate-uncertainty=true]{siunitx} 
\usepackage{subcaption}

\usepackage{listings}

\usepackage{csquotes}
\usepackage[final]{microtype}
\usepackage[onehalfspacing]{setspace}

\usepackage{pgfplots} 
\usepackage{pgfplotstable} 
\usepackage{colortbl} 
\usepackage{array}
\pgfplotsset{width=10cm,compat=newest} 

\usepackage{hyperref}
\hypersetup{
    colorlinks,
    linkcolor={red!70!black},
    citecolor={green!60!black},
    urlcolor={blue!90!black}
}

\usepackage{wasysym} 

\usepackage[headsepline]{scrlayer-scrpage}

\newcommand{\inputimg}[4]{ 
	\begin{figure}
		\centering
		\includegraphics[#2]{img/#1}
        \caption{#4}
		\label{dia:#3}
	\end{figure}
}

\newcommand{\inputsidewaysimg}[4]{ 
\begin{sidewaysfigure}
		\centering
		\includegraphics[#2]{img/#1}
        \caption{#4}
		\label{dia:#3}
\end{sidewaysfigure}
}

\newcommand{\inputimgh}[4]{ 
	\begin{figure}[h!]
		\centering
		\includegraphics[#2]{img/#1}
        \caption{#4}
		\label{dia:#3}
	\end{figure}
}

\newcommand{\inputsubfigureimg}[5]{ 
	\begin{subfigure}[t]{#3}
		\centering
		\includegraphics[#2]{img/#1}
        \caption{#5}
		\label{dia:#4}
	\end{subfigure}
}

\renewcommand{\thesection}{\arabic{section}} 
\renewcommand{\thesubsection}{\thesection.\arabic{subsection}} 

\usepackage{titlesec} 

\titleformat{\section}{\normalfont\Large\bfseries}{\thesection}{1em}{}
\titleformat{\subsection}{\normalfont\large\bfseries}{\thesubsection}{1em}{}

\newcommand{\nb}{$\mathrm{Nb}$\ }          
\newcommand{\oxygen}{$\mathrm{O}$\ }       
\newcommand{\pentoxide}{$\mathrm{Nb}_2\mathrm{O}_5$\ } 
\newcommand{\dioxide}{$\mathrm{Nb}\mathrm{O}_2$\ }     
\newcommand{\monooxide}{$\mathrm{Nb}\mathrm{O}$\ }     

\newcommand{\facid}{$\mathrm{HF}$}        
\newcommand{\sacid}{$\mathrm{H_2 SO_4}$}  
\newcommand{\nacid}{$\mathrm{HNO_3}$}     
\newcommand{\pacid}{$\mathrm{H_3PO_4}$}   

\subject{
\includegraphics[height=2cm]{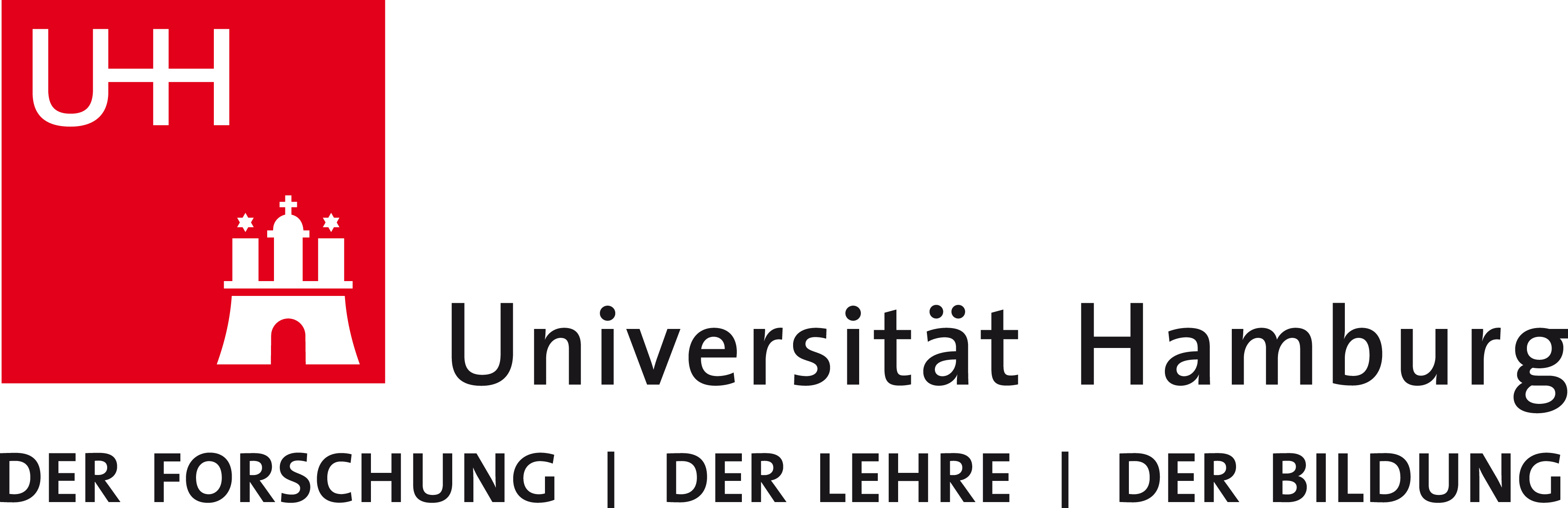}\\[2em]
bachelor thesis}

\title{\color{green!50!black} Investigations and Considerations\\of Oxygen Diffusion Profiles in\\Superconducting Mid-T RF Cavities\\Utilizing EXAFS Spectroscopy}

\author{\textsc{Niels Eckert}}

\date{}

\usepackage[backend=biber, style=alphabetic]{biblatex} 
\begin{document}
\sffamily
\pagenumbering{gobble}

\maketitle

\hfill
\vspace{-4em}
\begin{figure}[H]
\begin{center}
  \includegraphics[height=6cm]{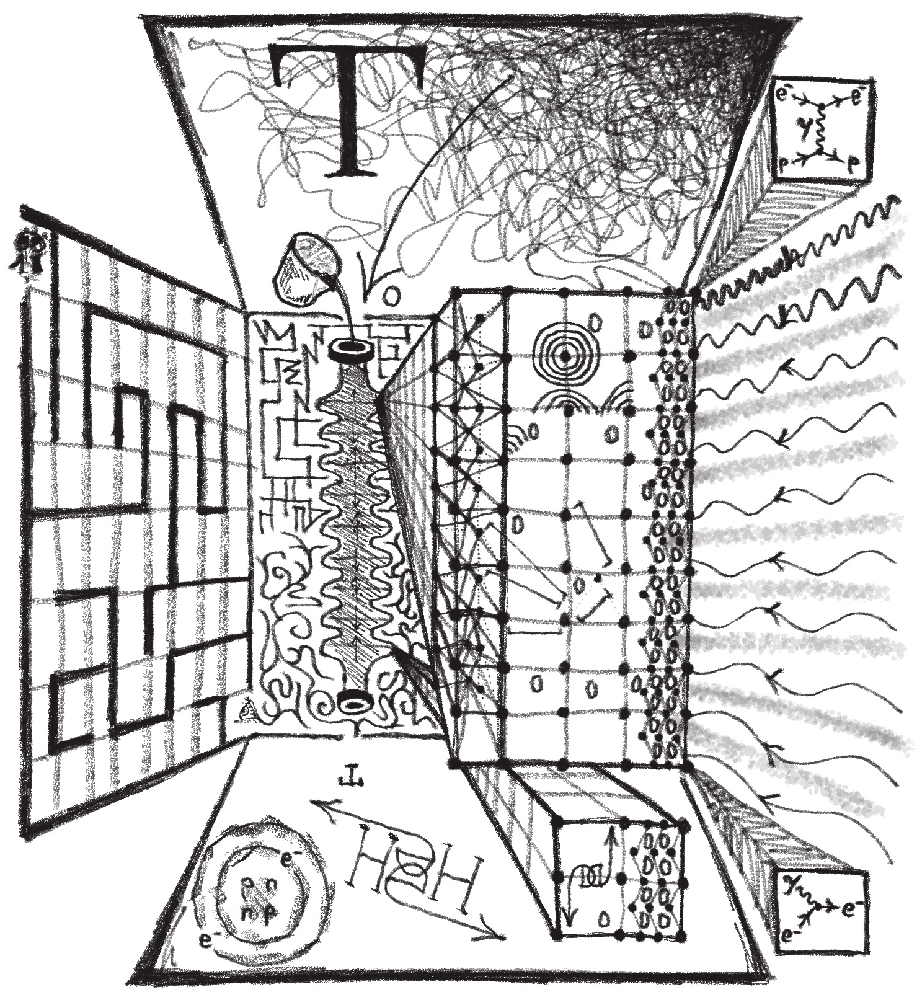}\\[3em] 
\end{center}
\end{figure}

\begin{multicols}{2}
  Submitted on \textrm{20. October 2025} \\[1em]

  Supervisors: \\[-0.7cm]

  \begin{itemize}
    \item[$1^\textrm{st}$] \textsc{Prof. Dr. Wolfgang Hillert} \\[-0.9cm]
    \item[$2^\textrm{nd}$] \textsc{Dr. Marc Wenskat}
  \end{itemize} 
  \columnbreak 
  \begin{minipage}[t]{0.42\textwidth} 
    \raggedright 
    \textrm{Physics B.Sc.} \\
    Matriculation number: $7615659$ \\
    Mail address: \textrm{niels.eckert}$@$\textrm{proton.me} \\
  \end{minipage}
\end{multicols}

\begin{center}
\small
\rmfamily
\end{center}

\clearpage 
\thispagestyle{empty}
\hfill
\clearpage

\setul{}{1.5pt} 
\begin{center}
  \vspace*{\fill}
    \LARGE
    \textrm{\textit{Für meine Oma}\\\&\\\textit{jene, die mit meinem Opa für sie da sind.}}
  \vspace*{\fill}
\end{center}

\newpage

\clearpage 
\thispagestyle{empty}
\hfill
\clearpage

\newpage

\begin{center}
  \vspace*{\fill}
  \includegraphics[width=\textwidth]{img/thesis-decoration.png}  
  \vspace*{\fill}
\end{center}

\newpage

\clearpage 
\thispagestyle{empty}
\hfill
\clearpage

\begin{center}
    \textbf{Zusammenfassung}\\[1ex]
    Die Untersuchung von Sauerstofftiefenprofilen in mid-T behandelten SRF Kavitäten
    ist zentral, um physikalische Zusammenhänge zwischen dem mikroskopischen Kavitätengitter
    und der entsprechenden Leistungsfähigkeit im Betrieb entschlüsseln zu können -- eine Fragestellung, die
    Beschleunigungsphysiker/innen seit Jahren beschäftigt.
    Diese Arbeit bietet die Analyse dieser Profile
    dreier unterschiedlich behandelter Proben: Zwei des mid-T Typs und die Dritte
    nach bewährtem EuXFEL Rezept. Die Messmethode greift auf die EXAFS-Spektroskopie zurück
    und wurde am DELTA Speicherring in Dortmund durchgeführt.
    Das Ergebnis ist stark verrauscht und erlaubt kaum quantitative Rückschlüsse.
    Qualitativ konnte keine Abweichung der Profile von gängigen Modellen
    nachgewiesen werden, wie auch die Messungen nicht im Widerspruch zu vorigen Untersuchungen stehen.
    Nichtsdestotrotz wird das Vorgehen im Experiment bis ins Detail dokumentiert,
    Interpretationen, soweit möglich, vorgenommen und theoretische Überlegungen,
    bezüglich der Fehlerabschätzung und potentieller EXAFS-Simulationen, für künftige Forschungen angestellt.\\
    Ein erneuter Anlauf an der PETRA III Anlage wird vorgeschlagen.

\end{center}

\vfill

\begin{center}    
    \textbf{Abstract}\\[1ex]
    A screening of oxygen profiles in mid-T treated SRF cavities is crucial, in order to infer physical
    correlations between the microscopic cavity lattice and cavity performance -- a problem concerning
    acceleration physicists for years. This thesis provides an analysis
    of oxygen diffusion profiles for three differently treated samples: Two mid-T baked
    and the third with the standard EuXFEL recipe. The measurement method
    utilizes EXAFS spectroscopy and was carried out at the DELTA facility in Dortmund.
    The result suffers heavily under noise, making the quantity of the result barely
    useable. Qualitatively, no deviations of current models regarding the profiles, could be proven,
    and no results of previous studies were contradicted.
    The experimental analysis is described in precision, interpretations of the possible are undertaken
    and theoretical considerations regarding error estimation and possible EXAFS simulations
    for future attempts are provided.
    A repitition at the PETRA III facility is indicated.

\end{center}

\newpage

\clearpage 
\thispagestyle{empty}
\hfill
\clearpage

{
\renewcommand{\baselinestretch}{1.3}\normalsize 
\tableofcontents
}

\newpage

\clearpage 
\thispagestyle{empty}
\hfill
\clearpage

\pagenumbering{arabic} 
\chead{\textsc{}}                                
\ihead{\footnotesize \textsc{\nb Lattice Analysis via EXAFS}} 
\chead{\footnotesize \textsc{$2025$}} 
\ohead{\footnotesize \textsc{Bachelor Thesis}}  
\renewcommand*{\footfont}{\itshape} 
\ofoot{%
  \vspace{-8mm}
  \hfill\rule{\textwidth}{0.4pt}\hfill \\ 
  \vspace{1.5mm}
  \pagemark 
}

\section{Introduction}\label{sec:introduction}
With the historical implementation of superconductivity in the field of radio frequency (RF) cavities
in the $1990\mathrm{s}$, major progress was made in terms of the operation and development of
modern particle accelerators,
utilized especially in particle physics, material science, biology and medicine.
This progress can be expressed by the $\sim 6$ orders of magnitude improved quality factor of SRF cavities,
compared to normally conducting ones.\cite{cas-linacs}
This is causing an enormous reduction in needed RF energy accordingly and outweighing additional cooling costs
below critical temperature.
Accelerators like the European XFEL in Hamburg and the LCLS-II at SLAC in Stanford
are already based on this SRF technology and even the LHC at CERN near Geneva uses partially
superconducting cavities, in order to benefit from their advantages.
Damages and impurities of the surface during the cavity fabrication process
led to the standard procedure of performing chemical and heat treatments
of SRF cavities before usage in acceleration operation, in order to polish and purify their surfaces.
The unexpected observation of quality factor improvements due to intentional impurities
of certain elements on the other hand, initiated the
effort of analyzing those improvements extent and understanding it's theoretical
background, up until today. A promising approach in recent years hereby,
are heat treatments in the medium temperature range between roughly \SI[round-precision=0]{200}{\celsius}
and \SI[round-precision=0]{400}{\celsius}, thus called mid-T baking, where
oxygen atoms on the surface due to the natural oxidation process of the cavity metal exposed to air,
are expected to diffuse further into the surface.\cite{posen-2019}
The conventional model of Fickian diffusion suggests in this case an oxygen diffusion profile
which is based on the cavity metal beeing a perfect crystal lattice.
Crystal defects, especially grains, suggest this profile applies over long distances,
but that inhomogeneous diffusion in the near-surface range cannot average out and thus causes deviations.
This thesis addresses this question by analyzing mid-T treated samples
in terms of their oxygen concentration profile
in the near-surface range and furthermore carries out a comparison with
the treatment recipe performed for the (already in operation) EuXFEL cavities.
First, in section \ref{sec:theory} major concepts in the context of
this analysis are introduced, in section \ref{sec:treatments} the samples are reviewed,
followed by the description and evaluation of an experiment
at the DELTA accelerator at TU Dortmund in section \ref{sec:inductive},
with which the samples have been analyzed. The underlying method of investigation is
utilizing EXAFS spectroscopy in fluoresence mode, which is determining the samples crystal structure
through absorption measurements of X-rays. This approach was already pursued
last year by Luth \cite{luth-2024} and turned out to suffer heavily on
measurement noise. The lessons we have learned in terms of the experimental
execution, motivated a second attempt, which is presented here.
Let it be said in advance, that data quality didn't improve significantly,
thus a detailed analysis of what is possible and
general considerations regarding error analysis (section \ref{sec:error-analysis})
are undertaken. Moreover, the aim of simulating an EXAFS measurement, in order to
judge whether the EXAFS measurements agree with presumptions of the sample lattices, is pursued
(see section \ref{sec:deductive}). For this, two software programs are in consideration.
In section \ref{sec:discussion}, errors and method are discussed, alternative techniques of analysis pointed out,
and an outlook for future research is provided.
Additionally, section \ref{sec:seref} proposes an unconventional analysis for a thesis, focusing on self referencing structures that arise in the context of this study. 

\section{\color{violet}Self Referenciality}\label{sec:seref}

This section carries out the proposed program of the last paragraph
in appendix \ref{app:self-referencialism} for the context of the present thesis, using the terms
introduced in it's previous paragraphs. References to the relevant sections in this thesis
are provided.\\

\textit{seref} $\cdot$ \textit{launism} $\cdot$ \textit{calcism} $\cdot$ \textit{serlau pleasure} $\cdot$ \textit{serlau doubt} $\cdot$ \textit{flarameter}\\

Seref contexts can be recognized in terms of
the description of XAFS spectroscopy as a whole,           
it's error analysis,                                       
the beamline experiment at the DELTA facility,             
the mysterious effect of impurities on cavity performance 
and benefits of the SRF technology overall.\\              

First of all, XAFS depends on scattering processes,
the implications of which were of great debate
during the last decades and is up until today
(see sec. \ref{sec:theory-exafs}, \cite{lytle-1999-xas-exafs-history}).
Because the behaviour of the photo electron depends on
scattering, whose stimulation depends on the photo electron again,
XAFS shows seref properties.
The launistic attempts of describing XAFS
in a calculistic manner in the community can be taken as opportunity for
the serlau pleasure induced proposition,
that upcoming thoughts will result into
a deeper description of XAFS, eliminating the need for
comparing XAFS spectra with literature
in order to gain information about the single measurement.\\

Connected to that is the current widespread handling of not calculating
uncertainties for XAFS spectra in the community
(see sec. \ref{sec:error-analysis}, \cite{booth-2009}, \cite{morrison-1981}).
Error estimation serves the purpose of comparing different measurements
with one another, to consult on a rating in the seref context of society
(and rather not to state a \textit{true error}, because it has no clear definition).
Thus the proposition of no need for error estimation, can be doubted serlauily by
stating that the seref scientific discussion relies on transparent uncertainties.\\

Moreover, regarding the specific EXAFS experiment in this thesis (see sec. \ref{sec:inductive}),
the situation unfolds that samples got analyzed via radiation of an acceleration facility.
The outcome of which is dedicated to improve those acceleration facilities themselves.
This obvious seref relationship refers to the reliability of particle dynamics and
production of synchrotron radiation. Thus, the performed EXAFS analysis in this thesis
is potentially vulnerable to serlau doubts, stating that the SRF-expertise through
this kind of analysis is contradicting itself and the acceleration process is indeed
incorrectly described. But this could only apply in the very bold case that
main groundwork on particle acceleration and synchrotron radiation was false all along.\\

Furthermore, the empirical observation that cavity impurities cause
a lowering of the surface resistance and thus improvements in the $Q$ curve (see sec. \ref{sec:theory-mid-t-bake}),
raises questions about the BCS theory and the interaction of particles in electrical current
with the cavity lattice. The whole BCS theory in the postulate of Cooper pairs
and how they emerge (see sec. \ref{sec:theory-superconductors}), utilizes
principles and methods of quantum physics which are deeply seref.\footnote{As expressed in the materialistic illustration by Werner Heisenberg in 1965, that the uncertainty principle can be understood as lack of knowledge of both position and momentum of an electron, because each position measurement via a photon influences it's momentum and vice versa.}\\
This would allow the serlau induced proposition, that oxygen interstitials intermediate
the emergence of other Cooper pairs (other than electrons) in order to cause an effect
not present in pure cavities.\\

And finally, the main benefits of SRF cavities, which lie in the lower energy consumption
and thus, not least, in the saving of research costs, connects to the
seref system of society. Therefore serlau induced propositions about the behaviour of society,
can lead to inspiration regarding strategic research in order to ongoingly satisfy curiosity.
Such seref propositions may say something about the political community
deciding about research funding or the economic system as a whole.
Time as the flarameter will continuously raise new situations.
\section{Foundations of SRF Cavities and Crystallographic Dynamics}\label{sec:theory}

To fully understand the following analysis of SRF cavity lattices,
it is necessary to be clear about some underlying knowledge and concepts in
accelerator physics, low temperature physics, thermodynamics and atomic spectroscopy.
Thus I summarize the significance of cavity structures for the field of
accelerator physics, aspecially in the realization of the superconducting accelerator technology. 
So called mid-T heat treatments of SRF cavities are beeing explained,
followed by an introduction into the diffusion principles which
are crucial for describing the heating processes appropriately.
Lastly, the atomic physics theory of EXAFS is discussed, in order to
disclose the analyzation method for the oxygen concentrations in the niobium lattice.

\subsection{Importance of SRF Cavities for Accelerator Physics}

The history of acceleration facilities has gone through many steps throughout the $20^\mathrm{th}$ century,
steadily increasing their performance and applicability. Currently, the technological gold standard
in accelerating relativistic particles are cavity resonators, in which an alternating electromagnetic field
is induced at radio frequency\footnote{Radio frequency ranges per definition between \SI[round-precision=0]{20}{\kilo\hertz} and \SI[round-precision=0]{300}{\giga\hertz}, while in accelerators frequencies in the range of \unit{\giga\hertz} are commonly used.}
and the particles are accelerated by the Coulomb force due to that field mode.
The conceivable cavity shapes
are diverse 
and fulfill various optimization requirements (for the EM fields, those modes
decisively depend on the resonator shape), like field magnitudes on the beam axis
or transverse beam focussing. Here a currently important standard is
given by the \textit{TESLA} shape introduced in
2000\footnote{Cavities in the TESLA shape are already installed e.g. in the EuXFEL and furthermore intended to be installed in the yet to be planned and build international linear collider (ILC).}.\cite{tesla-manifest}
The TESLA standard is designed for superconducting operation, meaning those cavities consist of niobium and are cooled
in order to enter the superconducting phase, which led to the need of a cumbersome cryogenic infrastructure
around the cavities while accelerating. 
The benefit on the other hand is a massive improvement of
the (oscillation related) quality factor of the field modes; while copper cavities show $Q$ factors of about $10^4$,
superconducting niobium cavities reach values in the order of $10^{10}$.\cite{cas-linacs}
It is calculated in general by
\begin{align}
    Q = \frac{\omega U}{P_\mathrm{diss}}
\end{align}

with the cavity modes eigenfrequency $\omega$, the stored energy in the cavity $U$ and the dissipated power
into the cavity wall $P_\mathrm{diss}$. These quantities are useful for measuring $Q$, but
it is of interest as well to express it in terms of the surface resistance $R_\mathrm{s}$ of the cavity,
which is possible via the expression
\begin{align}\label{equ:q-factor}
    Q = \frac{G}{R_\mathrm{s}}
\end{align}

because $P_\mathrm{diss}$ is proportional to $R_\mathrm{s}$\footnote{It is assumed $R_\mathrm{s}$ is constant over the cavity surface.}.
For a given field mode $G$ represents a purely geometric factor, independent of material properties.\\

When analyzing the performance of an SRF cavity, the $Q$ value gets important.
In test facilities like the AMTF at DESY, such performance tests are run systematically
by successively increasing the EM field inside the cavity.
The result is characterized by the \textit{quality curve},
where the $Q$ factor is plotted against the acceleration gradient $E_\mathrm{acc}$ (average electric field, which the accelerating particle is exposed to, during it's flight through the cavity).
In (BCS) theory, the $Q$ curve would be assumed constant until the cavity quenches.\cite{webster-2008-q-curve}
However, we see multiple phenomena in experiments that prove differently.
First there is an increase in $Q$ for small values of $E_\mathrm{acc}$ (low field),
a slight drop for medium values (medium field) and a very steep drop for high values (high field)
(see fig. \ref{dia:schematic-q-curve}). In some cases of heat treatments (see sec. \ref{sec:theory-mid-t-bake})
we instead see an increase in the medium field (named \textit{anti-$Q$-slope}, see fig. \ref{dia:antiq-comparison}).
Aspecially the high-field $Q$ slope (often referred to as HFQS or just \textit{$Q$-drop}) is of great interest
and limits the possibilities of cavities, it is historically usually called
the \enquote{European headache}.\cite{bauer-2006-q-slope-paper-collection}\\

\begin{figure}[h!] 
    \centering
    \inputsubfigureimg{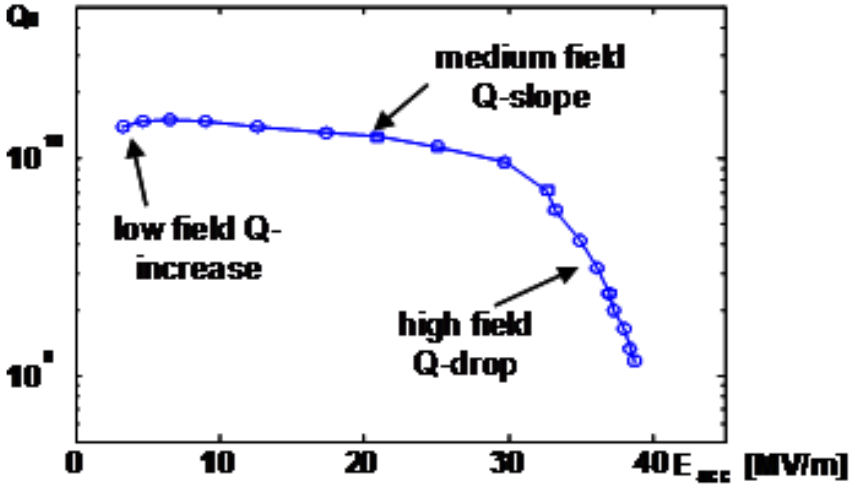}{width=\textwidth}{0.485\textwidth}{schematic-q-curve}{Schematic $Q$ curve of cavities in performance tests. All three regimes (low-, medium- and high-field) are part of current research.\cite{bauer-2006-q-slope-paper-collection}}
    \hfill
    \inputsubfigureimg{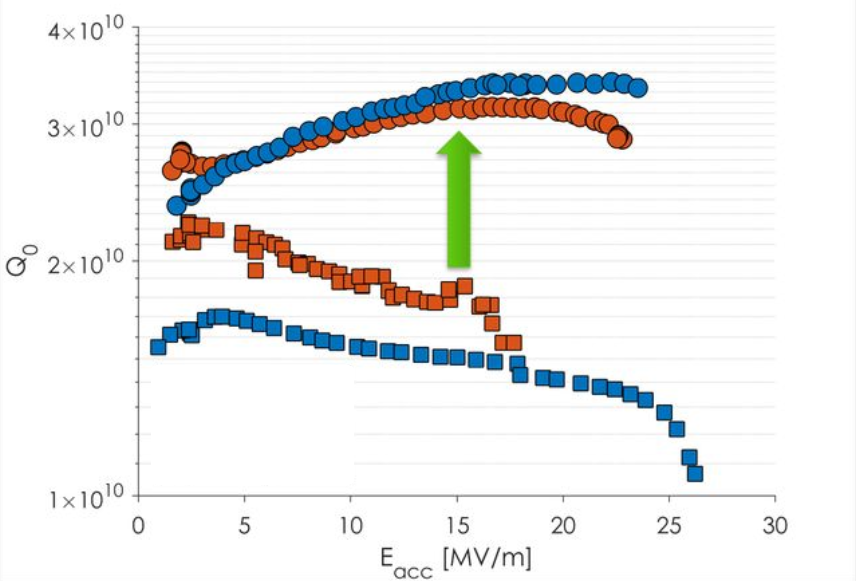}{width=\textwidth}{0.485\textwidth}{antiq-comparison}{Comparison of two cavity $Q$ curves both before (MF $Q$-slope, squares) and after (anti-$Q$-slope, circles) a specific treatment demonstrates improvements in the $Q$ factor. Data taken at SLAC.}
    \caption{}
    \label{fig:q-curves}
\end{figure}

The explanation is subject of current research and, not least, this thesis.
But assumed are loss mechanisms which depend on the accelerating field
and make it at least comprehensible, to see those phenomena.\cite{webster-2008-q-curve}

\subsection{Groundwork of Superconductors}\label{sec:theory-superconductors}

Superconductors as they are utilized for particle acceleration as
SRF cavities, originate in the phenomenon of a completely vanishing electrical resistance
for some conductors under specific circumstances.
Those circumstances are a temperature below some critical value $T_{c}$, a possibly existing external magnetic field
underneath a threshold $H_{c1}$ (both material-dependent) and an applied DC. This results into the conductor entering the
thermodynamical Meißner phase.
Even further, those conductors show not only the property of an ideal conductor but an ideal diamagnet as well, which is known as
the Meißner-Ochsenfeld effect. After these discoveries in the first half of the $20^{\textrm{th}}$ century, in $1957$
the so called BCS theory (named after John Bardeen, Leon Neil Cooper and John Robert Schrieffer) gave a microscopic explanation for superconductivity by postulating a correlated quantum state
between two electrons in the conduction band of the material. Those so called Cooper pairs would therefore only form
for (very) low temperatures and due to
their nature as bosons eventually lead to a vanishing electrical resistance in the material.
Important hereby is that this full disappearance only applies to DC, an alternating current is only
associated with a greatly reduced resistance.
This is due to the property that, simply stated, Cooper pairs prefer to form in the \enquote{slipstream} of other electrons and
only the formation of this slipstream causes resistance, but not it's flow.
Since the field reversal in AC repeatedly interrupts this slipstream, electrical loss occurs.\\

Even though the Meißner-Ochsenfeld effect claims an ideal diamagnet, an external magnetic field actually
decreases exponentially into the superconductors surface

\begin{align}
    B(z) = B_0 e^{-z/\lambda_L}
\end{align}

This is described by the London equations
and results into a London penetration depth

\begin{align}
    \lambda_L = \sqrt{\frac{m}{\mu_0 q^2 n_s}} 
\end{align}

where $n_s$ is associated to the number density of charge carriers, $m$ the mass and $q$ the charge of a carrier.
In case of the BCS theory, not electrons are assumed as charge carriers but Cooper pairs, so
$m = 2m_e$, $q = 2e$ and $n_s = n_e/2$.\cite{cas-2004}
If $\lambda_L$ is depending on the depth into the superconductor
(as is the case for an oxygen diffusion profile in a niobium cavity,
see sec. \ref{sec:theory-mid-t-bake}),
the London penetration depth and magnetic field are connected by a partial differential equation\cite{checchin}

\begin{align}
    \lambda_L^2 B''(z) + 2\lambda_L \lambda_L' B'(z) - B(z) = 0
\end{align}

The discussed properties alone are defining for the class of type I superconductors. Furthermore, elements like
the metal of interest, niobium, show as superconductors of type II an additional phase called the Shubnikov phase in direct succession to the Meißner phase.
It prevails until a second critical magnetic field $H_{c2}$ (while the temperature still has to be below $T_c$).
This phase is characterized by magnetic flux vertices entering the superconductor
(in case an external magnetic field is applied) which cause the electrical resistance to rise, but cannot be explained by BCS theory.\\

Now SRF cavities use niobium, which makes them type II superconductors and they are operated at $\num[round-precision=1]{1.8} - \SI[round-precision=0]{2}{\kelvin}$,
which is below the critical temperature for the Meißner phase of $T_{c} = \SI[round-precision=1]{9.2}{\kelvin}$.
The surface resistance $R_S$ consists of two parts 

\begin{align}
    R_S = R_\mathrm{BCS} + R_\mathrm{res}
\end{align}

Here $R_\mathrm{BCS}$ is the surface resistance due to the alternating
RF field as explained by BCS theory
and shows a decrease for lower temperatures with convergence to zero.
On the other hand $R_\mathrm{res}$ is the so called residual resistance. It is caused by lattice distortions,
grain boundaries and in case of an external magnetic field, normal conducting
flux vertices in the Shubnikov phase, as discussed above.
The latter is especially for cavities the case, where the magnetic field of the
accelerating mode penetrates into the cavity wall.
Furthermore, $R_\mathrm{res}$ is temperature independent
and it's decrease is target of cavity treatments.\cite{cas-2004}

\subsection{Mid-T Bake of SRF Cavities}\label{sec:theory-mid-t-bake}
Taking SRF cavities and carrying out so called \textit{heat treatments} under ultra high vacuum (UHV) with them can significantly 
improve their performance (recognizable by their quality curve).
The proximate cause thereby are modifications in the crystallographic structure
of the cavity material and the precise recipe of the treatment can be assumed to be crucial
for these modifications. Here it should be kept in mind that minor amounts of impurity atoms are
present in every real crystal either by production or due to environmental
influence afterwards. Aspecially the niobium crystals will oxidize on their surface simply
due to air exposure, resulting into three natural layers of oxides:
\pentoxide being the uppermost and thickest, \monooxide and \dioxide laying underneath,
both being collectively referred to as \textit{suboxides}.\\ 

\inputimgh{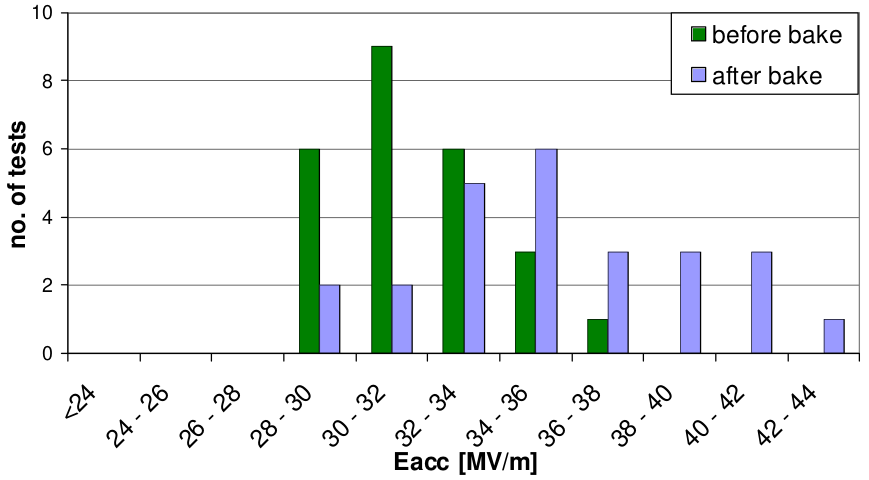}{width=0.75\textwidth}{baking-statistics}{Analysis of maximum $E_\mathrm{acc}$ for $25$ EP treated cavities before and after baking at $100$--\SI[round-precision=0]{140}{\celsius}, study from 2013 at DESY.\cite{desy-baking-statistics}}
\newpage
An instructive plot for the improvements of performing heat treatments for cavities
can be seen in fig. \ref{dia:baking-statistics}.
In 2019, it was found that by using a new procedure for heat treatments in the temperature range
between $\SI[round-precision=0]{200}{\celsius}$ and $\SI[round-precision=0]{400}{\celsius}$
can indeed cause performance improvements.\cite{posen-2019} Because this range lies between lower
temperature treatments (like $\SI[round-precision=0]{120}{\celsius}$ for the EuXFEL recipe) and higher temperature treatments 
above $\SI[round-precision=0]{600}{\celsius}$,
those treatments were introduced as the \textit{mid-T heat treatments}.
Remarkable about those treatments is, that the temperature is
high enough to dissolve the discussed natural oxidation layer and thus
oxygen atoms diffuse into the material. Therefore an improvement in performance
is observed, not by purifying the cavity, but by impurifying it.\\

To precisely execute a mid-T treatment you need an oven and four steps consisting of
chemical and baking stages. First an electropoliture (EP) is carried out where the
cavity is exposed to an acid mixture consisting of hydrofluoric (\facid) and sulphuric (\sacid) acid. 
Followed by a default annealing at $\SI[round-precision=0]{800}{\celsius}$ and afterwards
another chemical treatment in form of either another EP or a
buffer chemical polishing (BCP) with an acid mixture consisting of \facid,
nitric (\nacid) and phosphoric (\pacid) acid. In the latter case, the \facid\ is utilized for the destruction of
the \pentoxide layer as in the EP, the \nacid\ serves for regrowing that layer thereafter,
meanwhile the \pacid\ acts like a buffer due to impurities in the oven. 
Finally, in the fourth stage the cavity is baked again while temperature and duration are free parameters for the treatment.
In case other treatments would be desired, the last stage is customized.\\

\subsection{Principles of Atomic Diffusion}\label{sec:theory-diffusion} 
In the considered mid-T treatments, oxygen atoms diffuse into the \nb lattice, which is
described primarily by Fickian diffusion. For the substance concentration $c$ (particles per volume, $\left[c\right] = \unit{\m^{-3}}$) and diffusion flux $\vec{\jmath}$ (passing particles per time and area, $\left[\vec{\jmath}\,\right] = \unit{\per\s^{-1}\m^{-2}}$),
the following equations apply
\begin{multicols}{2}
\begin{align}
    \textrm{First law:}\quad\vec{\jmath} = -D\nabla c 
\end{align}
        \vfill
\columnbreak
\begin{align}
    \textrm{Second law:}\quad\partial_t c = \nabla (D \nabla c)
\end{align}
\end{multicols}

where $D$ is a diffusion coefficient ($\left[D\right] = \unit{\m^2 s^{-1}}$)
and the second law follows directly from the first using the continuity equation. 
In case of a fixed surface concentration ($c(z=0,t) \equiv c_0 = \mathrm{const.}$),
as it applies for oxygen diffusion during mid-T treatments, Fick's second law can be
solved \textit{sufficiently} by\\
\begin{align}
    c(z) = c_0\,\mathrm{erfc}\left(\frac{z}{\delta}\right)
\end{align}
With $\mathrm{erfc}$ the complementary error function\footnote{$\mathrm{erfc}(x) \equiv 1-\frac{2}{\sqrt{\pi}}\int_{0}^{x}e^{-t^2}\mathrm{d}t$}, $c_0$ the surface concentration and the depth $z$.
The diffusion length $\delta$ is time-dependent and in case of a time-independent diffusion coefficient
$D$ given simply by
\begin{align}
    \delta_\mathrm{hom} \equiv 2\sqrt{Dt} 
\end{align}
However in the general case of a treatment $T(t)$ the diffusion coefficient is given by
an Arrhenius equation
\begin{align}\label{equ:arrhenius-equation}
    D(T) = D_0\,e^{-\frac{E_a}{k_B T(t)}}
\end{align}
and thus time-dependent\cite{bate-2024-sims}, which results into the integral
\begin{align}\label{equ:penetration-depth-Tt}
    \delta_\mathrm{inhom} = 2\sqrt{\int_{t_0}^{t_1} D(t)\,\mathrm{d}t}
\end{align}
For eq. \ref{equ:arrhenius-equation}, $D_0$ is the initial
diffusion coefficient and $E_a$ the activation energy, both of which depend on the medium and diffusing particles.

\subsection{Theory of EXAFS}\label{sec:theory-exafs} 
Taking a thin foil of some material and radiating photons onto it, results into
a transmitted radiation with lower intensity on the other side because of absorption.
Plotting the absorption coefficient $\mu$ with respect to the photon energy $E$,
we see distinct absorption edges for specific energies (see fig. \ref{dia:absorption-edges-example}). This is due to
the ionization energy of the various bound electrons (meaning the atomic orbitals) to their nucleus in the material --
a photon energetic enough to ionize an electron upon interaction will
most likely not pass through the foil. Furthermore, a general downwards trend in the absorption can be
observed. This can be explained by quantum mechanical reasons, where multiple interaction processes
occur and vary in their dominance (see fig. \ref{dia:total-photon-attenuation}). \\

\inputimgh{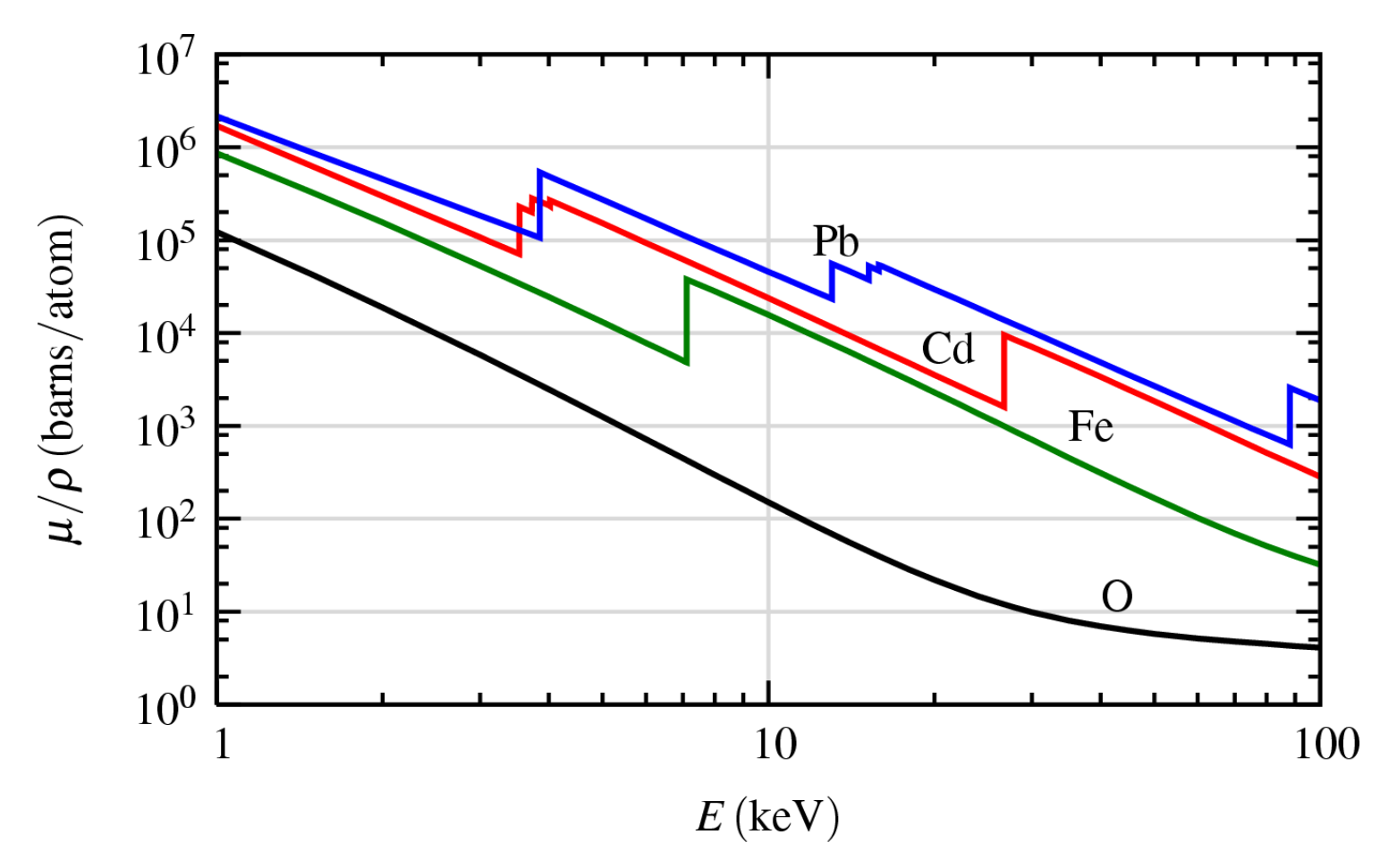}{width=0.75\textwidth}{absorption-edges-example}{Examples for absorption edges of some chemical elements in a log-log plot for the absorption cross-section over the photon energy.\cite{newville-2004}}

\inputimgh{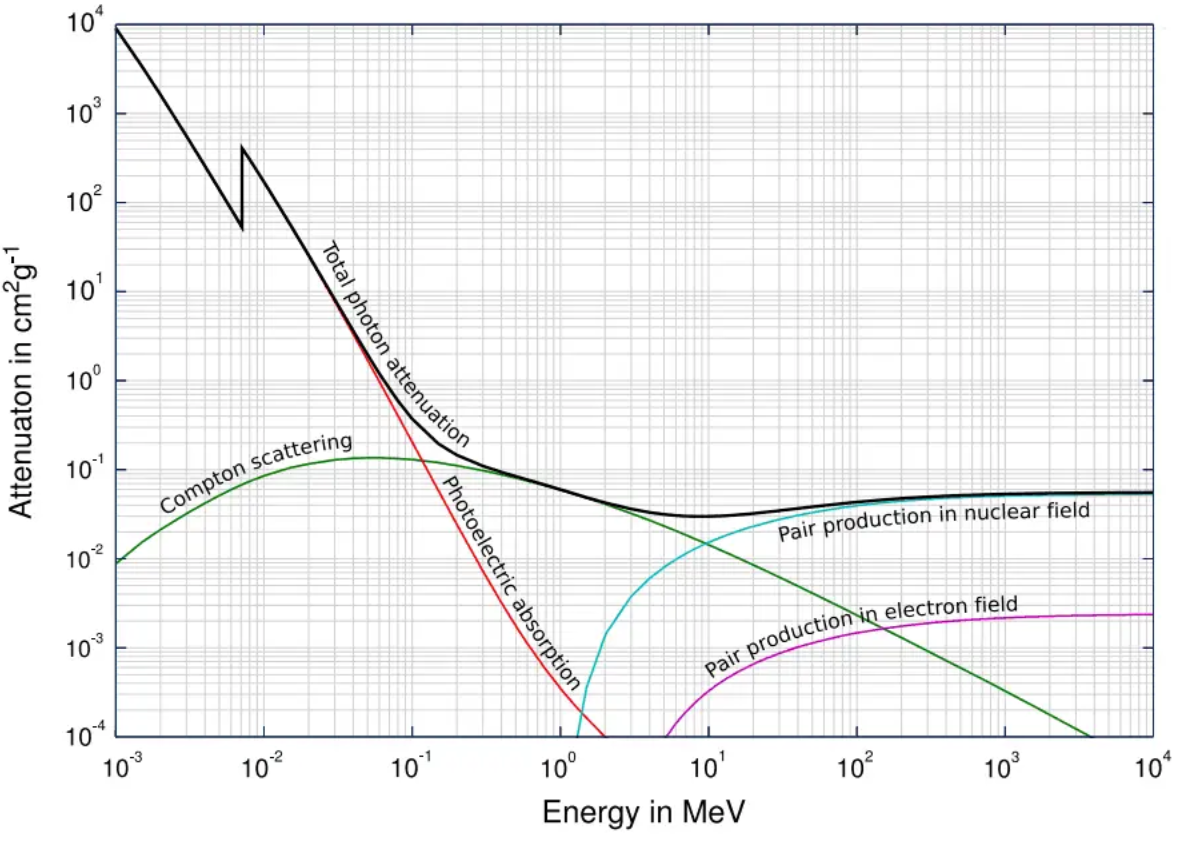}{width=0.75\textwidth}{total-photon-attenuation}{General term of attenuation in an exemplary material as a function of the photon energy. Multiple interaction processes alternate, in EXAFS spectroscopy the photoelectric absorption process is dominating (order of $10^{-2}\,$\unit{\mega\electronvolt}).}

In the $1920\mathrm{s}$ it was first discovered, that above an absorption edge, the absorption curve $\mu(E)$ 
of a crystallographic solid shows a fine structure with an attenuated trend for
increasing energies (see fig. \ref{dia:schematic-xafs-curve}).
This so called \textit{X-ray absorption fine structure} (XAFS) depends not only on the
absorption edge and chemical element, but profoundly on the crystal structure of the foil.
It lies in the back scattering processes of the ionized (photo-)electron with bound electrons in the crystal.
Expressed in modern (quantum mechanical) terms, the wave function of the electron encountered
by a photon, has a pertubated transition between \textit{bound} and \textit{ionized} due to the presence of neighboring
(scattering) atoms. This effect can be described quantitatively by a pertubating hamiltonian and
application of Fermi's golden rule (utilizing time-dependent pertubation theory).
The qualitative explanation hereby can be given by argumenting with resonance conditions
inside the absorbing atoms vicinity\footnote{\textit{Nachbarschaft} in german.}:
When a scattering atom reflects the photo-electron wave function, the initial wave is interfering
with the reflecting one and either constructive or destructive interference occurs.
In the former case, the transition of \textit{bound to ionized} is favored, while in the latter
case it is suppressed. Due to the fixed distance between atoms, the interference changes
with increasing energy periodically, which implies a lattice-related oscillation above the absorption edge, as
shown in fig. \ref{dia:schematic-xafs-curve}. This makes XAFS a non-destructive method
for structural analysis of materials.\cite{newville-2004}\\

\inputimgh{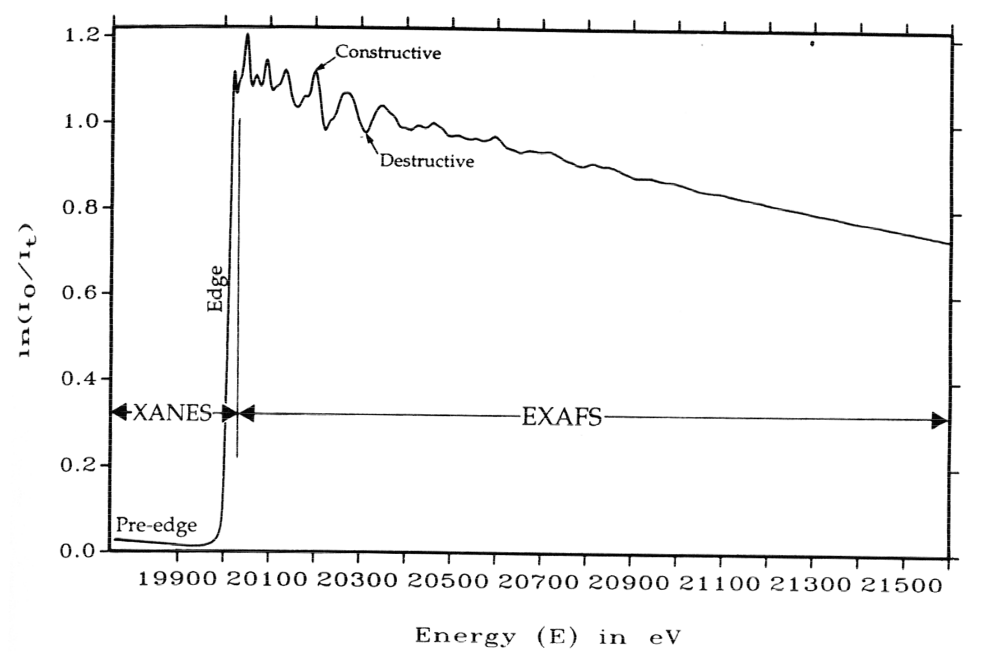}{width=0.85\textwidth}{schematic-xafs-curve}{Exemplary XAFS curve (transmittive absorption $\mu(E)$) of the $\mathrm{K}$-edge of molybdenum. The NEXAFS ($\equiv$ XANES) and EXAFS regimes can be distinguished.\cite{antonio-exafs-history}}

The to be analyzed structures in this thesis consist of a bcc lattice of $\mathrm{Nb}$ atoms.
A major term to investigate lattices is the \textit{$j$-th coordination sphere},
which describes all atoms in the $j$-th nearest vicinity of a reference atom (see fig. \ref{dia:coordination-radii} for the associated \textit{coordination radii}).
Additionally an unknown
amount of \oxygen interstitials in the octahedral void of the lattice is assumed.\\

\inputimg{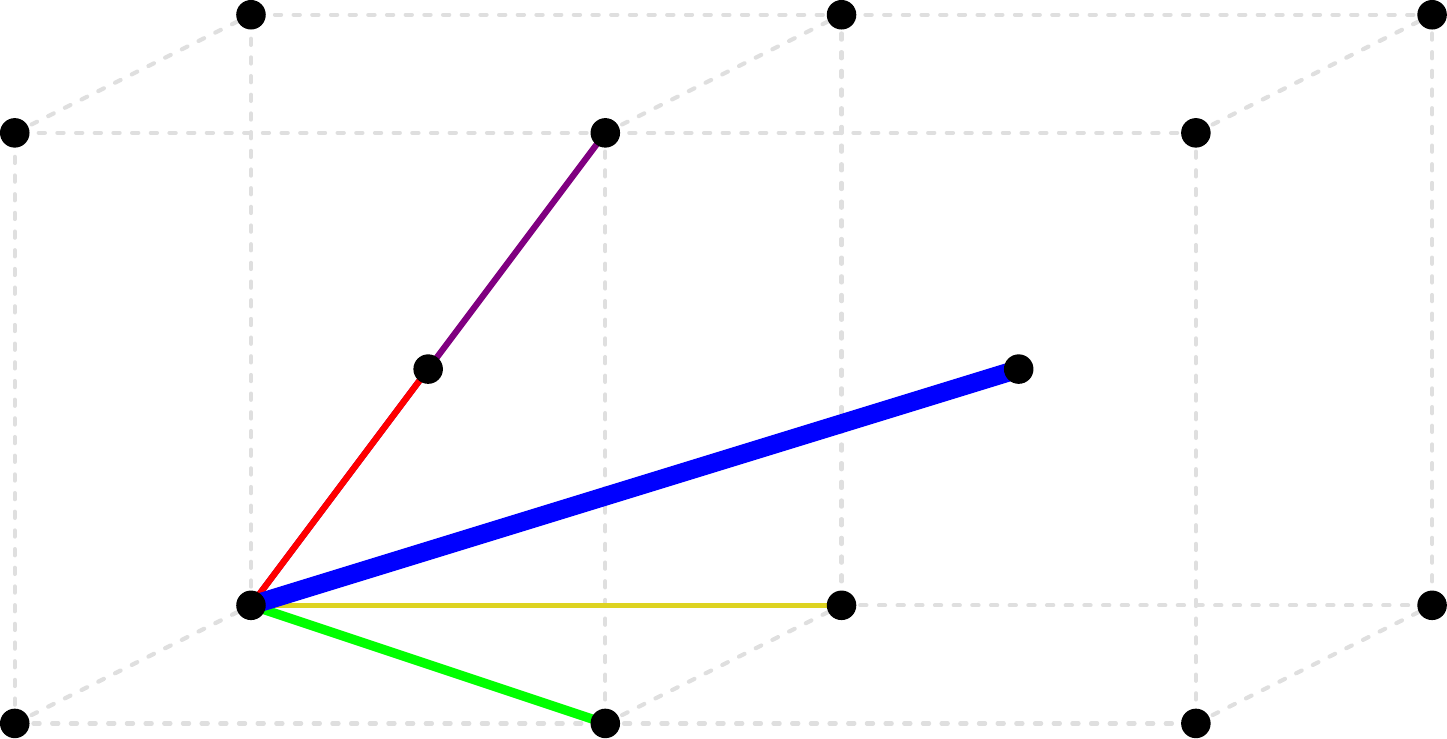}{width=0.5\textwidth}{coordination-radii}{A bcc lattice with the first five coordination radii, the drawn line thickness is a measure for the corresponding coordination number ($8$, $6$, $12$, $24$ and $8$, in ascending order). The positions of the octahedral void are drawn in light blue.}

Furthermore XAFS is divided into the
\textit{near edge XAFS}
(NEXAFS\footnote{Synonymous with the historically older term \textit{XANES} (X-ray absorption near edge structure) which is also used in literature.\cite{bianconi-1980-inventing-xanes-term}})
which applies until about \SI[round-precision=0]{50}{\electronvolt} 
and the \textit{extended XAFS} (EXAFS) which follows afterwards. 
But their boundary is not sharply defined, it consists of the fact that
EXAFS is mostly due to single-scattering processes and carries information about
the vicinity of an absorbing atom (inter-atomic distances/ lattice constants, coordination numbers/ lattice system, defects, element of neighboring atom)
while NEXAFS relies on multiple-scattering processes and contains information about the inner-atomic properties (oxidation state, coordination chemistry, band structure, etc.).\cite{behrens-2009} \\

The quantitative approach of EXAFS starts with the following equation:

\begin{align}
    \chi(E) = \frac{\mu(E) - \mu_0(E)}{\Delta\mu_0}
\end{align}

$\chi$ is called fine structure of the absorption curve $\mu$ and furthermore depends
on a background function $\mu_0$ and the absorption difference of the absorption edge itself.
While there is no full theoretical and quantitative description of the rather complex NEXAFS phenomenon up until today,
EXAFS can be understood as a short-range-order phenomenon.
This led $1971$ to the so called EXAFS equation
\begin{align}\label{equ:exafs-equation}
    \chi(k) = \sum_j \frac{N_j\, e^{-2k^2 \sigma_j^2} \, e^{-2R_j/\lambda(k)}\, F_j(k)}{kR_j^2}\sin{(2kR_j + \Phi_j(k))}
\end{align}
where the fine structure $\chi$ is depending on the mean-free-path of the photo-electron
$\lambda(k)$, for each coordination sphere $j$ there is a radius $R_j$, number of neighbors $N_j$,
variance of thermal displacement $\sigma_j^2$ and two optical scattering properties
$F_j(k)$ (backscattering amplitude) and $\Phi_j(k)$ (backscattering phase).\cite{lytle-1999-xas-exafs-history}
The wave number of the photo-electron $k$ hereby arises from the non-relativistic quantum mechanical relationship
\begin{align}\label{equ:e2k-relation}
    k = \sqrt{\frac{2m_e}{\hbar^2}(E-E_0)}
\end{align}
with the photon energy $E$ and edge position $E_0$.\\

It can be seen directly from the equation that besides some complicated expression for the amplitude,
the EXAFS consists mainly of a linear combination of harmonic oscillations
related to the coordination radii.
For this reason, performing a fourier transform and considering it's absolute value is standard to obtain $R_j$ as peaks
and thus crystallographic information. But caution is advised: The optical phase $\Phi_j(k)$ can
be monotonous in $k$ and cause the peaks to be shifted and the optical amplitude $F_j(k)$
can have a local extremum, which leads to a beat\footnote{\textit{Schwebung} in german.} in the fine structure and a splitting of
two peaks in the fourier space even though they represent only one radius. The latter phenomenon
is called Ramsauer-Townsend effect and can be tricky to identify.\cite{ravel-2012-ramsauer-townsend}
And in general, a peak that is no perfect gaussian shows side lobes in the fourier space
which can be misinterpreted as independent lattice distances.\\

It is crucial here that the EXAFS analysis in this thesis does not use the transmission
of an X-ray beam through a sample, but (due to the thickness of the samples)
the fluorencence emitted in it's attenuation.
Ionized electrons leave holes which are filled by electrons in higher shells and
cause the emission of fluorecence photons.
This is a major method for practicing EXAFS spectroscopy, besides measuring the transmission
or the direct analysis of Auger electrons. But it goes along with the fact, that
in fluorescence only a quantity proportional to the absorption coefficient can be
measured.
Additionally even the proportionality isn't valid strictly because of
the so called \textit{self-absorption}.\cite{ravel-2011-self-absorption} When a fluorencence photon is emitted
(whose probability is actually proportional to the absorptance)
it has to travel out through the material again. This process includes
secondary absorption of the fluorescence photons and increases for higher
primary absorption. Furthermore, the X-ray beams penetration
depth into the material (mostly weakly) depends on the photon energy,
which results in addition into a varying extent of the self-absorption effect.
Thus self-absorption dimms the EXAFS oscillations and
leads to a non-linear relationship between transmission and fluoresence EXAFS. 

\section{Characterization of Sample Treatments}\label{sec:treatments}

\begin{table}[htbp]
    \centering
    \caption{Final treatment stage of all three analyzed samples.}

\begin{tabular}{crrcl}
    \toprule
    Sample N° & Kategory & Treatment & Temperature & Duration \\
    \midrule
    243 & mid-T & regular mid-T & $300$\,\unit{\celsius} & $3$\,\unit{\hour} \\
    245 & mid-T & long duration mid-T & $250$\,\unit{\celsius} & $20$\,\unit{\hour} \\
    247 & EuXFEL recipe & low-T baseline & $120$\,\unit{\celsius} & $48$\,\unit{\hour} \\
    \bottomrule
\end{tabular}
    \label{tab:np-samples}
\end{table}

The investigation included the ex situ analysis of three \nb samples with a thickness of \SI[round-precision=1]{2.8}{\mm}.
Each sample has undergone two chemical and two thermal treatments alternately.
First a coarse chemistry, followed by a heating up to \SI[round-precision=0]{800}{\celsius}
for \SI[round-precision=0]{3}{\hour} and a fine chemistry. Lastly each sample got again heat treated
but with individual parameters (see sec. \ref{sec:theory-mid-t-bake}). Samples $243$ and $245$ thereby got a mid-T heating and sample $247$ was treated with the
standard EuXFEL recipe (see tab. \ref{tab:np-samples}).\\

\inputimgh{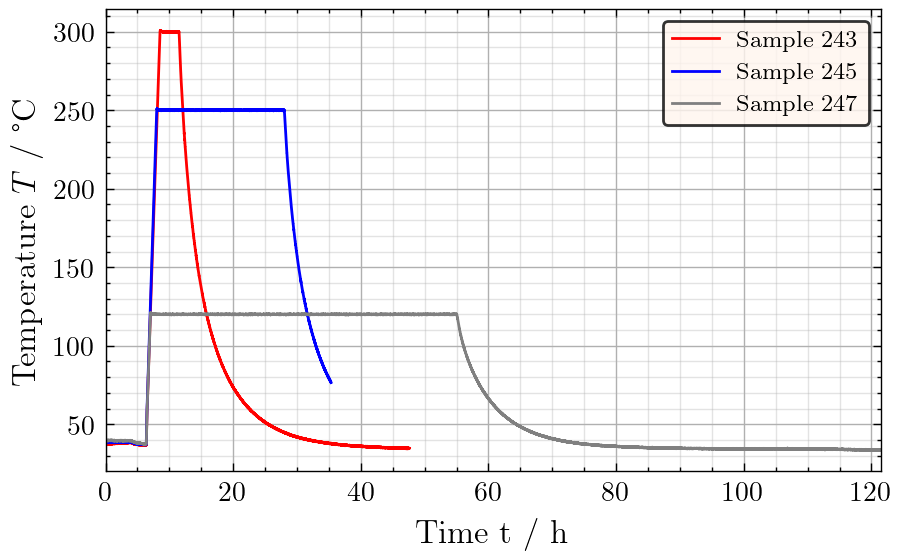}{width=0.75\textwidth}{treatments}{Temperature curve of the final treatment stage of the investigated samples $243$ (regular mid-T), $245$ (long duration mid-T) and $247$ (low-T baseline).}

All final stages had a time-dependent temperature $T(t)$,
the curves of which are shown in fig. \ref{dia:treatments}.
Thus the Fickian diffusion is described by a
time-dependent diffusion coefficient $D(t)$ and
an integration has to be performed in order to
obtain the theoretical diffusion length $\delta$ (see eq. \ref{equ:penetration-depth-Tt}).
Two quantities of the Arrhenius equation (see eq. \ref{equ:arrhenius-equation})
have to be assumed here, the first is the initial diffusion
length $D_0 = \SI[round-precision=1]{1.5e-2}{\centi\m\squared\per\s}$
and the second is the activation energy $E_a = \SI[round-precision=1]{1.2}{\electronvolt}$.
Both values apply for diffusing oxygen in a \nb lattice with some
inhomogeneous influence of grains.\cite{bate-2024-sims}
The calculation yields $\delta_{243} = \SI{1442.56}{\nm}$,
$\delta_{245} = \SI{1102.31}{\nm}$ and $\delta_{247} = \SI{20.87}{\nm}$ for each
sample respectively, whereby it should be noted that $\delta_{245}$ is slightly
bigger due to the artificial cutoff in the $T(t)$ curve.
In reality, the lattice shows distortions like point defects and grain boundaries
whose influence is only estimated by the literature values of $D_0$ and $E_a$
and the material (cavity wall) is relatively thick compared to the diffusion length
but not infinite (precisely the diffusion process runs on both sides of the cavity wall, even though one process can be assumed to not affect the other).
Thus it is necessary for the diffusion lengths to be checked by the following analysis.\\

Regarding the purity of the \nb samples in the production process,
aspecially tantalum and hydrogen succeeded by some others
occur as trace elements. Looking at the atomic percentage
of these elements in such a produced niobium ingot yields
$257\,\mathrm{ppm\,at.}\%$ for tantalum and $184\,\mathrm{ppm\,at.}\%$ for hydrogen
as upper limits, following purity requirements for SRF
cavities (other trace elements lie below these concentrations).\cite{wenskat-2015} Comparing this with the
atomic percentage of oxygen interstitials due to mid-T treatments,
which lies in the order of $\mathrm{at.}$\permil\,makes one confident that EXAFS will especially not be disturbed by trace elements. 

\section{Inductive EXAFS Examination of Niobium Samples}\label{sec:inductive}

The niobium examinations discussed here were carried out at the Dortmund Electron Accelerator (DELTA)
which is operated by the Technische Universität Dortmund as a storage ring for electrons.
It delivers the needed X-ray radiation as synchrotron radiation of a superconducting wiggler (SCW, see fig. \ref{dia:scw}).
The beamline 8 (BL8) was used, during the period from the $12^\mathrm{th}$ to the $16^\mathrm{th}$ of May $2025$.
This beamline is operated by the Bergische Universität Wuppertal and
the \nb samples are owned by and were prepared at the MSL group at the Deutsches Elektronensynchrotron (DESY).
This way, our experiment represents a tripartite cooperation.

\inputimgh{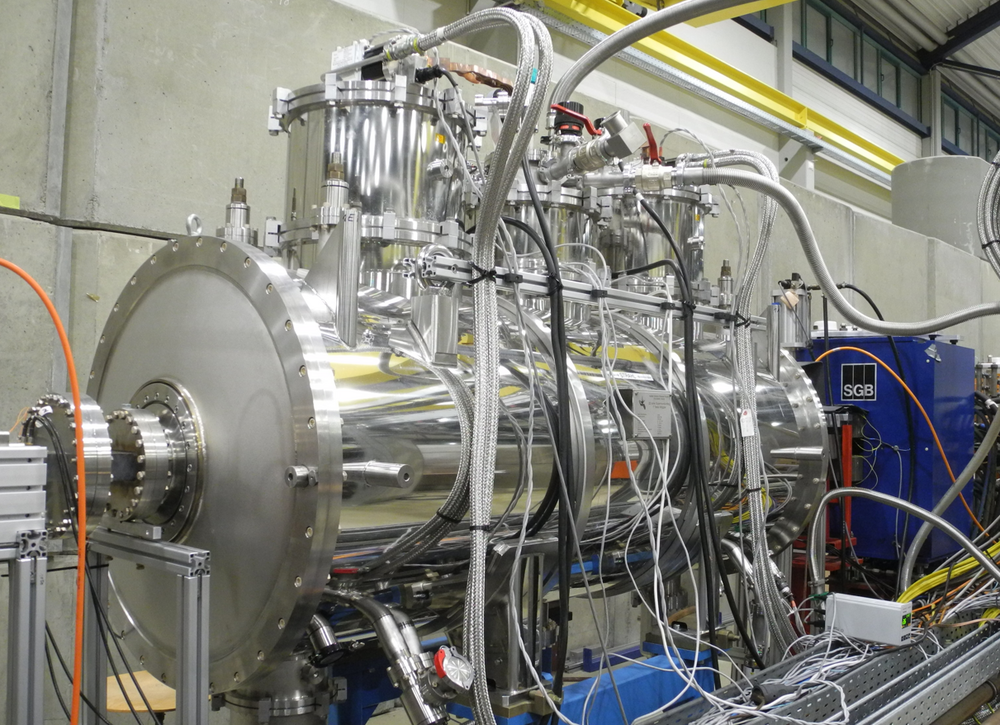}{width=0.75\textwidth}{scw}{The utilized superconducting wiggler (SCW) provided the used synchrotron radiation through manipulation of \SI[round-precision=1]{1.5}{\giga\electronvolt} electrons in a magnetic field of \SI[round-precision=0]{7}{\tesla} and with a period length of \SI[round-precision=0]{122}{\milli\m} (Source: \cite{wiggler}).}

\subsection{Setup of the Beamline Experiment}\label{sec:setup}
The setup of the EXAFS experiment (see fig. \ref{dia:setup-ziesche} respectively \ref{dia:setup-photograph})
is given first by the incoming synchrotron radiation traversing multiple instruments before
radiating into the \nb sample, a fraction passing it and traveling through a reference setup
and finally to be absorbed by a lead block.  
Initially there is a crystal pair monochromator consisting of silicon with a piezo element
connected to one of the crystals for calibration reasons.
This is succeeded by an element for transfering the radiation from the vacuum in the synchrotron
into the air. It consists of a $\mathrm{Be}$-window and a Kapton foil, filled with $\mathrm{He}$
in between to both ensure the beam quality and keep the vacuum inside the storage ring.
Otherwise the $\mathrm{Be}$-window alone would break under the pressure difference.\\

Subsequently the radiation is passing an aperture $\mathrm{S}3$ and an ionization chamber
to measure the initial X-ray intensity $I_0$. 
Directly afterwards the \nb sample is attached on a sample mount which in turn is placed on a table.
The sample has the shape of a truncated cone and is rotated by an angle $\theta$ at it's geometric
center (perpendicular to the plane of projection in fig. \ref{dia:setup-ziesche}). 
Above the sample, a detector consisting of silicon called \textit{PIPS}
(passivated implanted planar silicon) is capturing the fluorescent photons emitted
in the sample
in a widespread solid angle. Thus the \nb sample is measured in fluorescence.\\

Due to the beam diameter, part of the radiation passes by the sample and travels through
another aperture $\mathrm{S}4$. After that again an ionization chamber $I_1$ is installed followed
by a reference sample and a third ionization chamber $I_2$ measuring the reference sample
in transmission. At the end, a lead block absorbs the remaining radiation.\\ 

\inputsidewaysimg{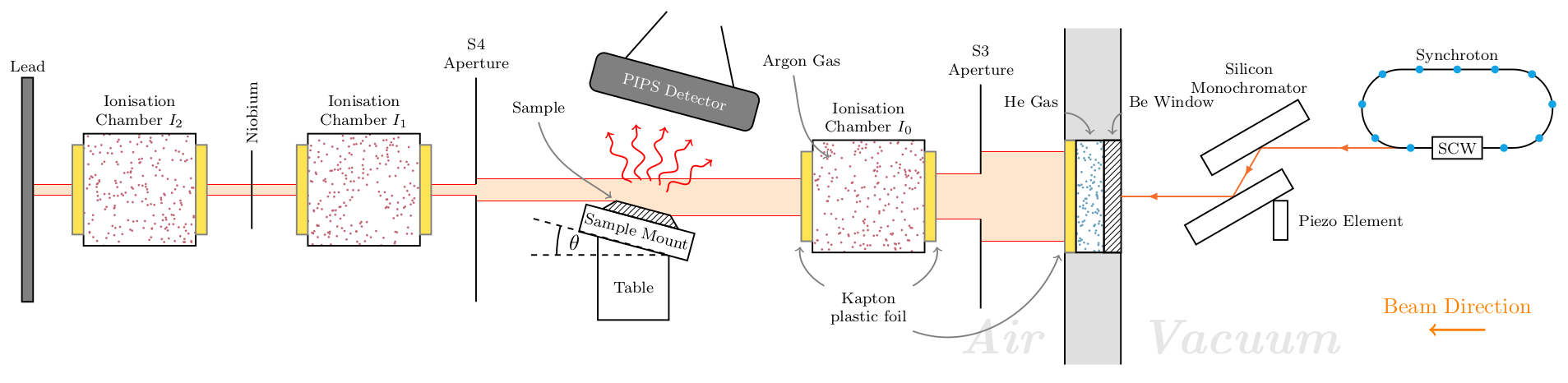}{width=\textwidth}{setup-ziesche}{Experimental setup as described in the text. The synchrotron beam is drawn two dimensional after the vacuum-air transition, beam width and distances are not to scale. The author of the graphic is Ivo Ziesche, Friedrich-Schiller-Universit\"at Jena (see Acknowledgements, last page).}    
\inputimg{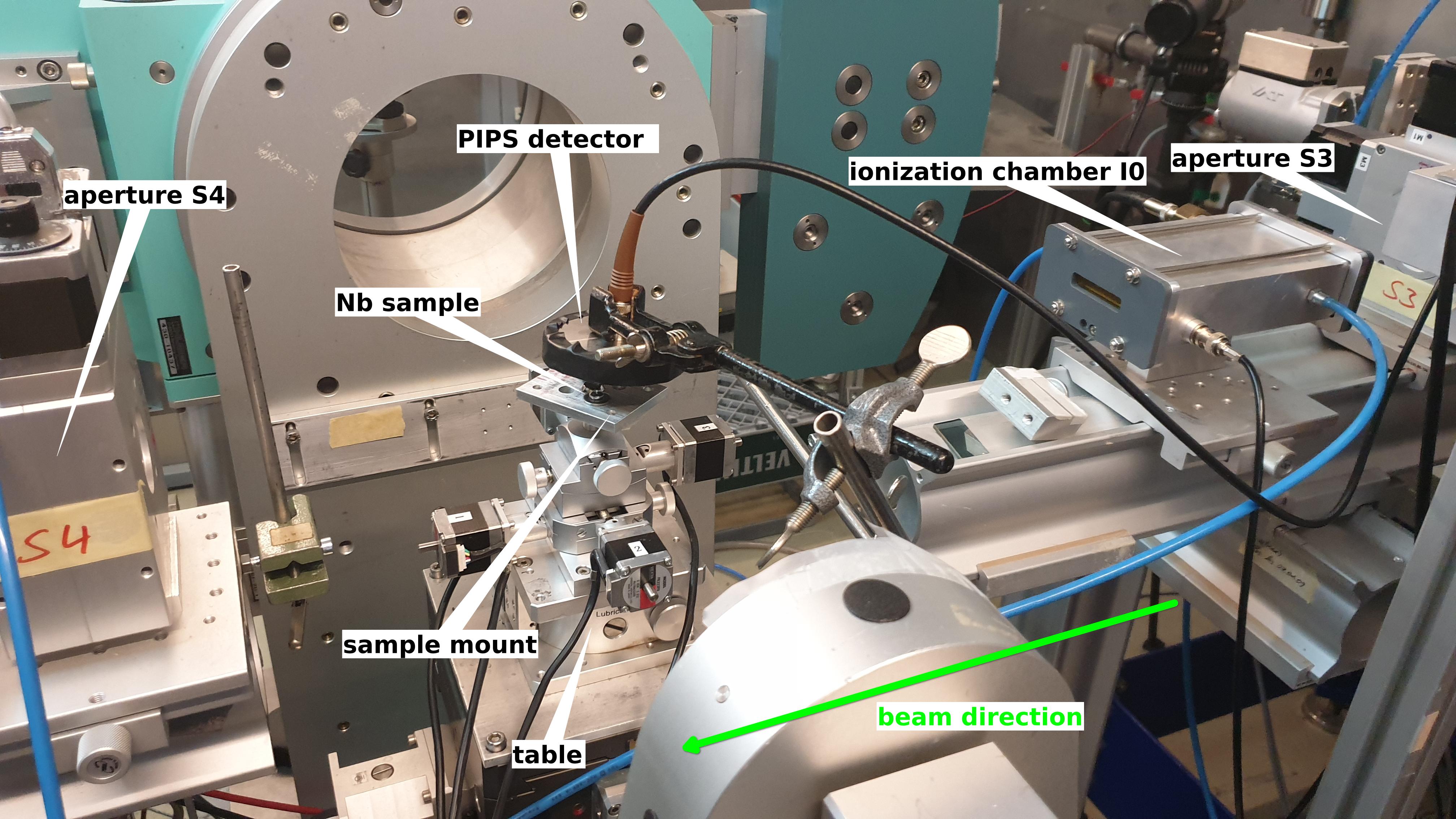}{width=\textwidth}{setup-photograph}{Photograph of the experimental setup.}

\subsection{Acquisition of Measurement Spectra}
The sample of interest was first fixed to the mount and placed on the table in the setup.
Then a calibration process was undertaken, involving the position of the table height and the angle $\theta$ of the mount.
The beam intensity behind the sample $I_1$ was plotted against the table height, which results into a logistic-like curve,
between the adjustments where the beam passes completely past the sample and where it is completely blocked by it.
The turning point of this plot was then set as the calibrated table height. Afterwards, the angle $\theta$ was varied
and $I_1$ plotted against $\theta$. Here we got a triangular function, where $I_1$ is at maximum when the sample is completely
parallel to the beam. The maximum point was chosen for $\theta$ and defined as zero.
After this procedure, the data aquisition was run. For eight different angles
\begin{align*}
    \theta \in \{\num[round-precision=1]{0.1},\,\,\,\num[round-precision=2]{0.15},\,\,\,\num{0.17},\,\,\,\num[round-precision=1]{0.2},\,\,\,\num[round-precision=1]{0.3},\,\,\,\num[round-precision=1]{0.5},\,\,\,\num[round-precision=1]{0.7},\,\,\,\num[round-precision=0]{1}\}\unit{\degree}
\end{align*}
the beam energy was varied between \SI[round-precision=3]{18.837}{\kilo\electronvolt} and \SI[round-precision=3]{19.407}{\kilo\electronvolt}.
The energy step between consecutive datapoints and the integration time were chosen separately on four intervals (see tab. \ref{tab:spectrum-intervals}). The beginning and ending of each scan
was of rather little interest, while the intermediate energies carried the more decisive information.
The intensities $I_0$ and $\mathrm{PIPS}$ were measured, together with $I_1$ and $I_2$.
Such a single \textit{exafsscan} on an energy range of \SI[round-precision=0]{570}{\electronvolt} took about \SI[round-precision=0]{23}{\minute}.\\

Between \textit{exafsscans} of this kind, the piezo element of the monochromator was usually used for recalibration by
varying the piezo voltage and comparing this with the intensity $I_0$. The piezo voltage
was set to where $I_0$ is at it's maximum. This ensured, that the monochromator crystals remained aligned parallel to each other.\\ 

After the scan of all angles $\theta$ the samples were switched and the setup again calibrated.
Each setting for sample and angle was measured redundantly multiple times and at different times of day.
The procedure was supervised by living beings during the day (between about $8$\,\textsc{am} and $11$\,\textsc{pm})
and executed automatically by terminal commands during the night.\\

Each ionization chamber was filled with argon at a pressure
of \SI[round-precision=0]{1}{\bar}, which resulted into about $18\,\%$ of
the transmitting radiation to be absorbed. As for all ionization chambers, the electrons and
ions of ionized argon atoms travel to the electrodes and cause an
electrical current (in the order of \unit{\nano\ampere}).
Next, a current amplifier converts this current into a voltage,
normalized to \SI[round-precision=0]{10}{\volt} at full scale\footnote{\textit{Vollausschlag} in german.} of the amplifier 
and a voltage-to-frequency converter transcribes the signal into
a rectangular pulse. Electronically, a software counts the pulses, which finally represent the photon counts.
The PIPS detector on the other side, detects radiation by them producting electron-hole pairs
in the detector, which results into a current that is measured. 

\begin{table}[htbp]
    \centering
    \caption{Details of the aquisition of an \textit{exafsscan}. The relative information refers to the $K$ edge of niobium at \SI[round-precision=4]{18.9855}{\kilo\electronvolt}.}

\begin{tabular}{crrlc}
    \toprule
    Interval                & Relative interval & \# Datapoints & Step & Integration time \\
    \midrule
    $[18.837, 18.957]\,\unit{\kilo\electronvolt}$ & $[-150, -30]\,\unit{\eV}$ & $40$ & $3$\,\unit{\eV} & $1$\,\unit{\s} \\
    $[18.957, 19.047]\,\unit{\kilo\electronvolt}$ & $[-30, 60]\,\unit{\eV}$ & $180$ & $0.5$\,\unit{\eV} & $2$\,\unit{\s} \\
    $[19.047, 19.137]\,\unit{\kilo\electronvolt}$ & $[60, 150]\,\unit{\eV}$ & $120$ & $0.75$\,\unit{\eV} & $3$\,\unit{\s} \\
    $[19.137, 19.407]\,\unit{\kilo\electronvolt}$ & $[150, 420]\,\unit{\eV}$ & $150$ & $1.8$\,\unit{\eV} & $4$\,\unit{\s} \\    
    \bottomrule
\end{tabular}
    \label{tab:spectrum-intervals}
\end{table}

\subsection{Extracting the Fine Structure}\label{sec:extracting-the-fine-structure}
The raw spectrum of a scan is given by the quotient $\mu(E) = \mathrm{PIPS}(E)/I_0(E)$.
It is then further processed using the software \textsc{Athena}, which is part of the
\textsc{Demeter} package.\cite{demeter}\footnote{\textsc{Demeter} is a software package
for the data processing and analysis in the field of x-ray absorption spectroscopy (XAS),
where XAFS is a subfield of XAS. It uses the library IFEFFIT for computations.}
It's purpose is to extract the fine structures $\chi(k)$ out of the raw measurements $\mu(E)$
and perform a specific fourier transform $|\tilde{\chi}(R)|$ for gaining spacial information.
This extraction method ($\mu(E) \rightarrow \chi(k), |\tilde{\chi}(R)|$) is standard in the
XAFS field and well documented, e.g. in the \textsc{Athena}
documentation.\footnote{See \url{https://bruceravel.github.io/demeter/documents/Athena/index.html}}\\

First, poor spectra (high noise level, systematic errors, etc.) are sorted out
(see tab. \ref{tab:measurements-sorted-out} in appendix)
and apparent glitches in the remaining spectra are removed (see sec. \ref{sec:error-discussion}).
Afterwards, both the edge position $E_0$ and pre-edge and post-edge lines are fitted into each spectrum.
The former is defined by the the first maximum of the first derivative (difference quotient of the
discrete data), while the latter are determined by polynomial regression in manually chosen ranges
before and after $E_0$. The pre-edge lines are linear regression curves while the post-edge lines
are mostly cubic polynomials (otherwise quadratic, choice was made manually depending on best
suitability\footnote{The differences between quadratic and cubic order weren't significant for the needed regressions whatsoever.}).
The ranges were chosen, so that in the pre-edge region no significant rise in $\mu(E)$ occured
due to the absorption edge and no distorting effect of noise appeared. And similarly the
post-edge region was selected up to the point where significant noise occured.\\

Subtracting those regression curves from the pre-edge and post-edge parts of each spectra and
normalizing it, provides us a standardized form for each absorption curve, whatever the
individual conditions were in the experiment. Then the redundant spectra for the same sample
and angle of incidence are averaged and a background curve is calculated, whose difference
to the averaged spectrum after the edge corresponds to the fine structure $\chi(E)$ (and thus $\chi(k)$).
This background curve is the result of the \textsc{Athena}-intern \texttt{AUTOBK} algorithm that
uses multiple parameters to compute a spline function, mathematically spoken it shall consist of the
lowest frequencies of the absorption spectrum.\footnote{The \texttt{rbkg} parameter was set
uniformly to $1$ and the \texttt{spline clamps} were set to \textit{None} for \texttt{low}
and \textit{Strong} for \texttt{high}. They influence how strictly the background function
follows the absorption curve.}
Also a $k$-weighting is being done. The exponent chosen depends in general on the shape of
the data, here a quadratic weight is used so $k^2\chi(k)$ is considered. 
The purpose of $k$-weighting is to mathematically compensate the attenuation of the
fine structure amplitude, which can be observed in every empirically EXAFS (seen later in fig. \ref{fig:measured-spectra})
and motivated with the $\frac{1}{k}$ factor in the
EXAFS equation (see eq. \ref{equ:exafs-equation}).
The chosen exponent however doesn't necessarily need to be $1$, because main goal
is an overall constant amplitude of the EXAFS, so that
a linear combination of non-attenuating oscillations
(as expected for distinct atoms in real space) can be analyzed.
Instead, even a slightly higher weighting can be beneficial
in order to damp the influence of low $k$ fine structure
which is part of NEXAFS and not EXAFS.
Besides that, because of the measuring in fluorescence, a correction of the self-absorption
(see sec. \ref{sec:theory-exafs}) is performed,
because it cannot be eliminated by measurement devices, only corrected artificially.\\
Therefore another \textsc{Athena}-intern algorithm\footnote{The algorithm \texttt{Fluo} is used,
the documentation can be accessed under \url{https://www3.aps.anl.gov/haskel/FLUO/Fluo-manual.pdf}.},
which uses tabulated data for an approximation, is applied.\\

Finally the fourier transform provides information used for structural deductions.
For this the already weighted fine structure $k^2\chi(k)$ is additionally weighted by a \textit{Hanning} window
in the range $[1, \num[round-precision=1]{10.5}]\,Å^{-1}$ (so called \textit{fourier range}).
This helps to further reduce the fine structure of the NEXAFS area we're not interested in.
After applying a discrete fourier transformation, we get $|\tilde{\chi}(R)|$.

\subsection{Results}

\begin{figure} 
    \centering
    
    \inputsubfigureimg{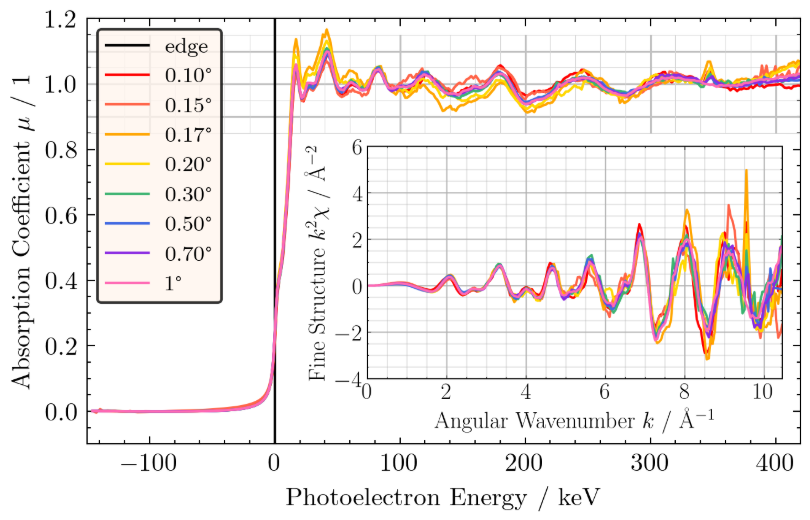}{width=\textwidth}{0.75\textwidth}{}{Sample $243$}
    \vfill
    \inputsubfigureimg{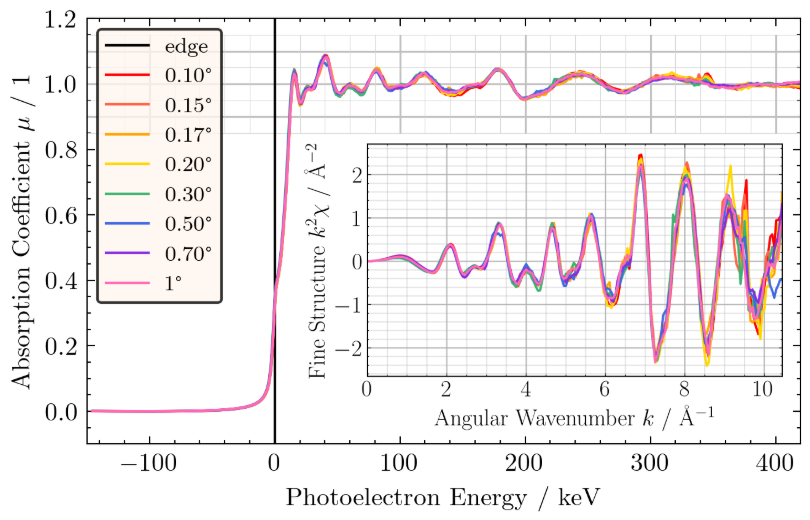}{width=\textwidth}{0.75\textwidth}{}{Sample $245$}
    \vfill
    \inputsubfigureimg{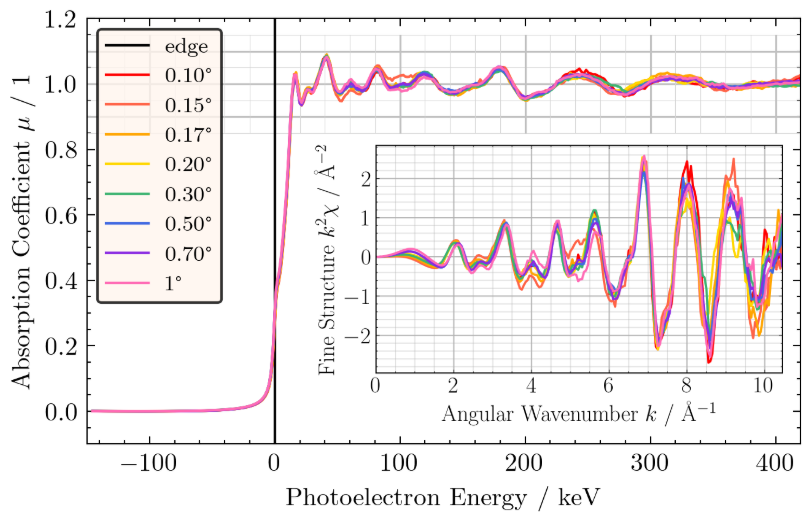}{width=\textwidth}{0.75\textwidth}{}{Sample $247$}
    
    \caption{Normalized absorption curves for each analyzed sample and angle of incidence $\theta$. The abscissa was calibrated to the absorption edge, so it indicates the energy of the ionized photoelectron. In addition, the fine structure (EXAFS) of each spectrum weighted by $k^2$, is plotted.}
    \label{fig:measured-spectra}
\end{figure}

\begin{figure}
    \centering
    
    \inputsubfigureimg{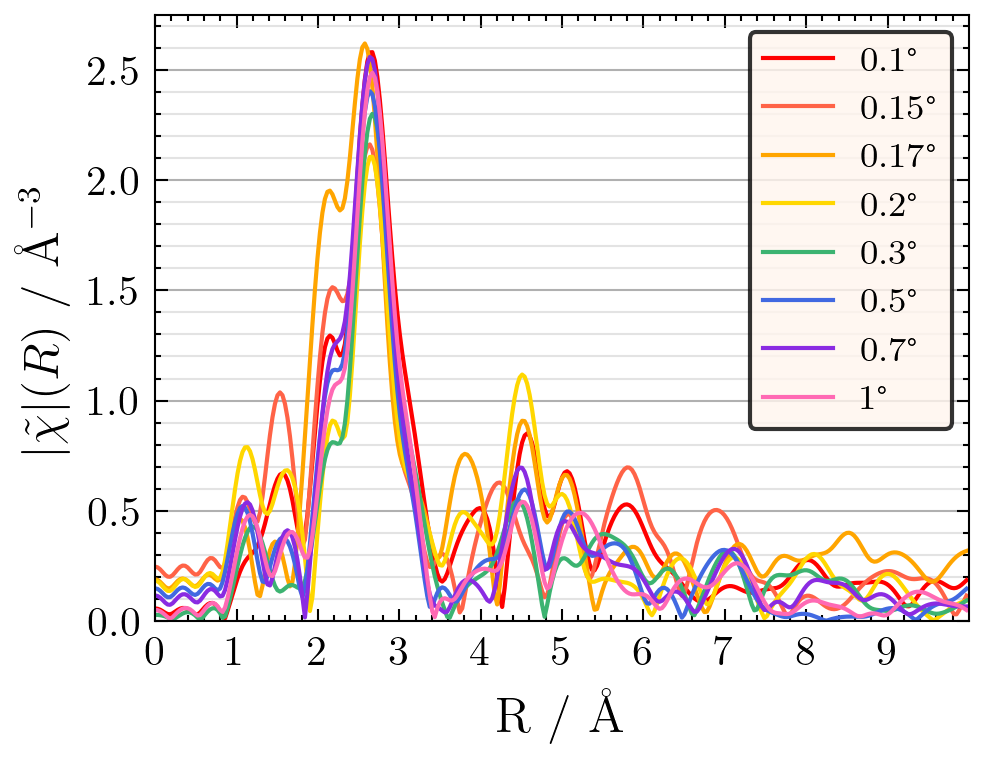}{width=\textwidth}{0.6\textwidth}{}{Sample $243$}
    \inputsubfigureimg{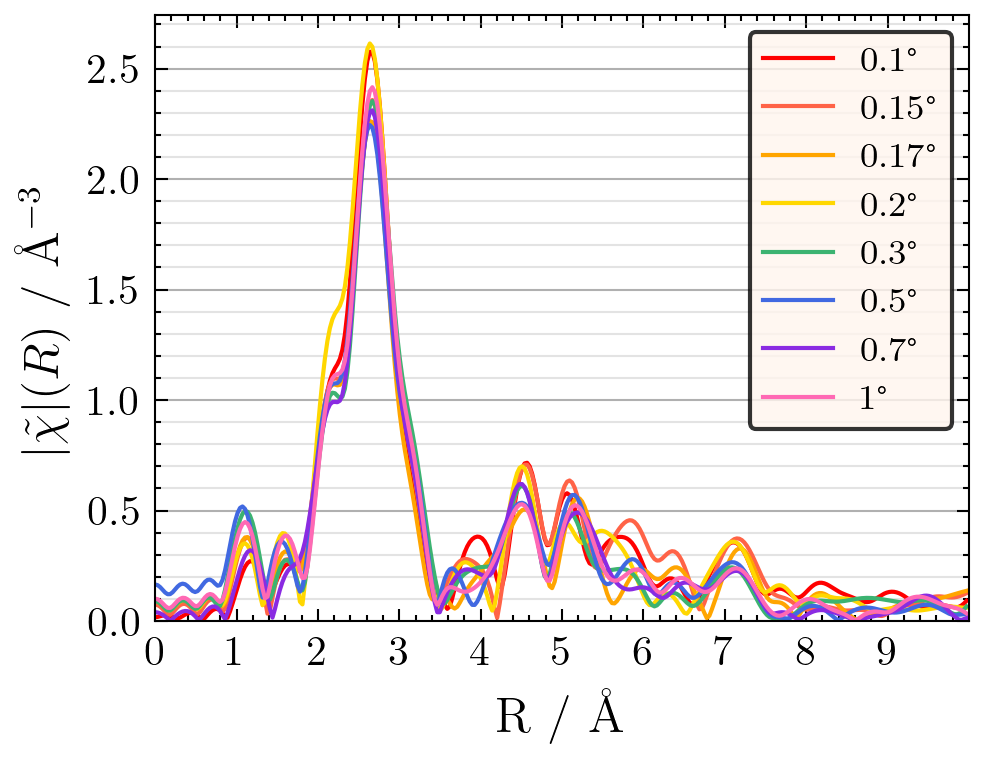}{width=\textwidth}{0.6\textwidth}{}{Sample $245$}
    \inputsubfigureimg{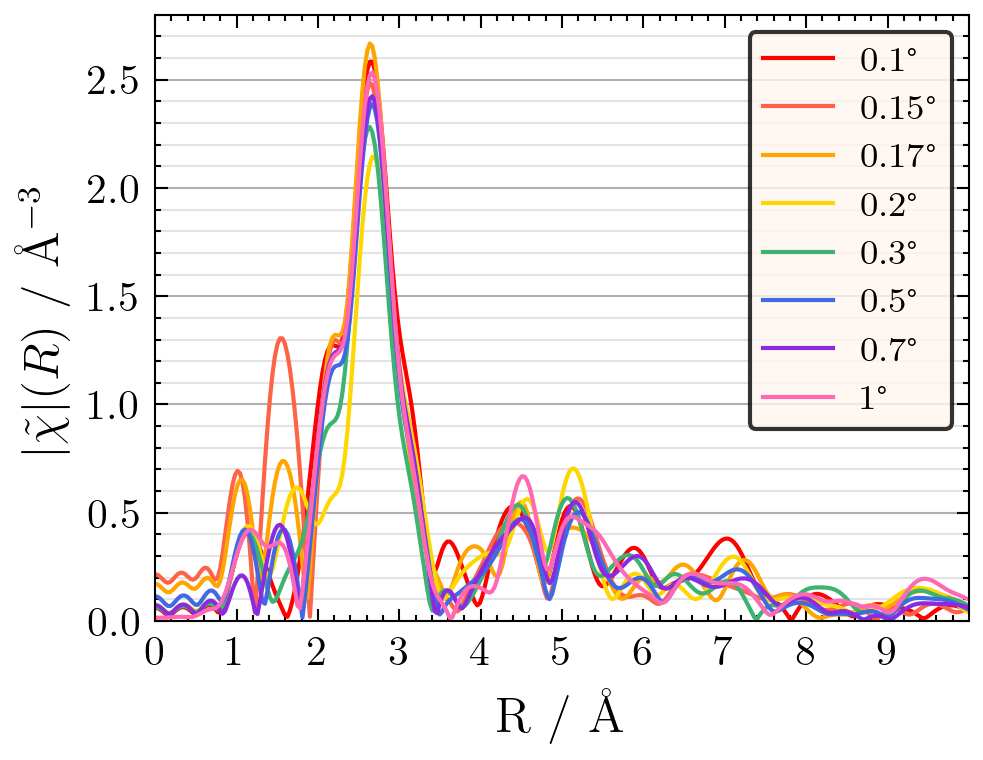}{width=\textwidth}{0.6\textwidth}{}{Sample $247$}

    \caption{Sample measurements in $R$ space of the spectra shown in fig. \ref{fig:measured-spectra}. At about \SI[round-precision=1]{2.6}{\angstrom}, every plot shows a strong and broad peak, together with less distinct but additional peaks.}
    \label{fig:measured-fourier-angle}
\end{figure}

The measured spectra show similar behaviour for different angles of the same sample
(see fig. \ref{fig:measured-spectra}), aspecially sample $245$ shows little changes, as does
sample $247$, whereas sample $243$ exhibits the strongest fluctuations with the angle $\theta$.
At about \SI[round-precision=0]{345}{\electronvolt}, we see a feature for both mid-T treated samples which decreases for higher $\theta$.
It doesn't show up for the EuXFEL sample.
There is no shift in the edge position identifiable and we see, as expected, a stronger
noise level for higher energies/ wavenumbers. In $R$ space the measurements show a
strong and broad peak at about \SI[round-precision=1]{2.6}{\angstrom}
(see fig. \ref{fig:measured-fourier-angle}). Also some peaks at higher and lower radii seem
characteristic. A look onto the measured reference spectra shows no distortions and
gives confidence that no fundamental malfunctions occured while beam time.\\

For further evaluation, a conversion of angles of incidence $\theta$ into corresponding
penetration depths $z_{1/e}$ (depth up to an attenuation of $1/e$) is helpful, so a more solid state related analysis can be done.
The underlying formula\cite{parratt-1954}
is a result of electromagnetic optics applied particularly to the X-ray range and given by

\begin{align}\label{equ:penetration-depth}
    z_{1/e} = \frac{\lambda}{\sqrt{2}\pi}\left[\sqrt{(\theta^2-\theta_c^2)^2+4\beta^2}-(\theta^2-\theta_c^2)\right]^{-1/2}
\end{align}

where the critical angle of incidence $\theta_c$ is given approximately by
$\theta_c \approx \sqrt{2\delta}$ and the photon wavelength $\lambda$
only depends on the photon energy $E$. The optical parameters $\delta$ and $\beta$
are connected to the real and imaginary part of the complex refractive index $n$ of
a given medium (defined as $n = 1-\delta+i\beta$). They
depend on the lattice type and energy and
were taken from the database of the Center for X-ray Optics (CXRO).\footnote{See \url{https://henke.lbl.gov/optical_constants/getdb2.html}}
We obtain a relationship as shown in fig. \ref{dia:penetration-depths-cxro}.\\

\inputimg{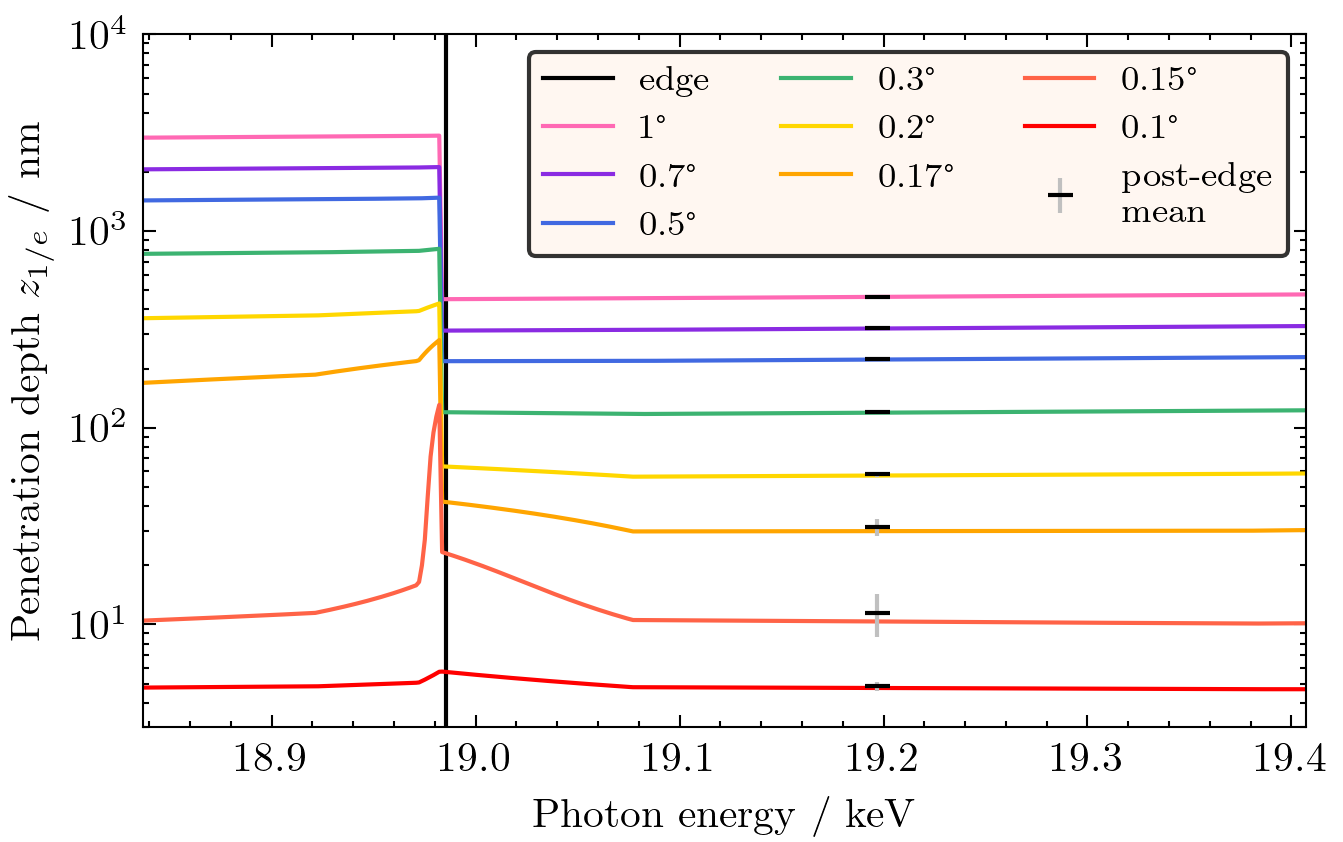}{width=0.9\textwidth}{penetration-depths-cxro}{Penetration depth into a pure \nb lattice as a function of the used X-ray radiation energy, calculated after eq. \ref{equ:penetration-depth}. The behaviour of $z_{1/e}$ is approximately constant beyond the $K$ edge, it's mean value and standard deviation is calculated for each angle of incidence. Optical parameters $\delta$, $\beta$ for calculation were taken from the CXRO database, because the database only stores discrete values, the plot looks edgy.}

As to be assumed, $z_{1/e}$ rises for steeper $\theta$.
The penetration depth is actually a function of the photon energy but due to
it's low dependence after the absorption edge,
a mapping between $\theta$ and $z_{1/e}$ can be done.
For this, the mean value of $z_{1/e}$ beyond the $K$ edge is chosen.\\

\begin{figure}
    \centering
    
    \inputsubfigureimg{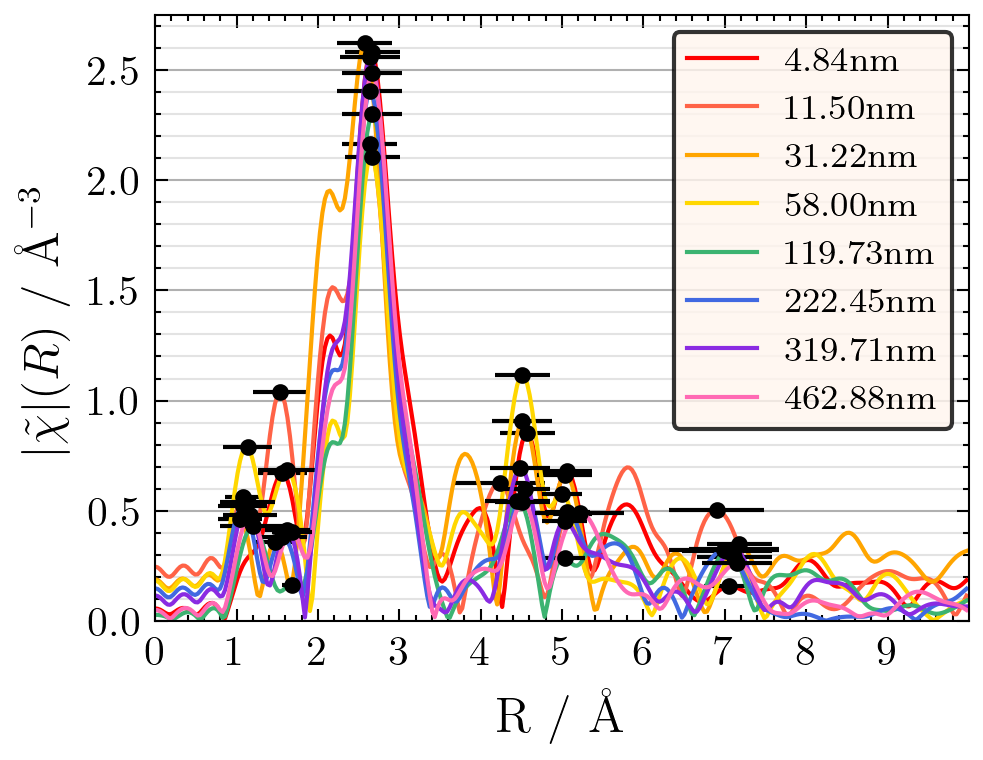}{width=\textwidth}{0.6\textwidth}{}{Samples $243$}
    \inputsubfigureimg{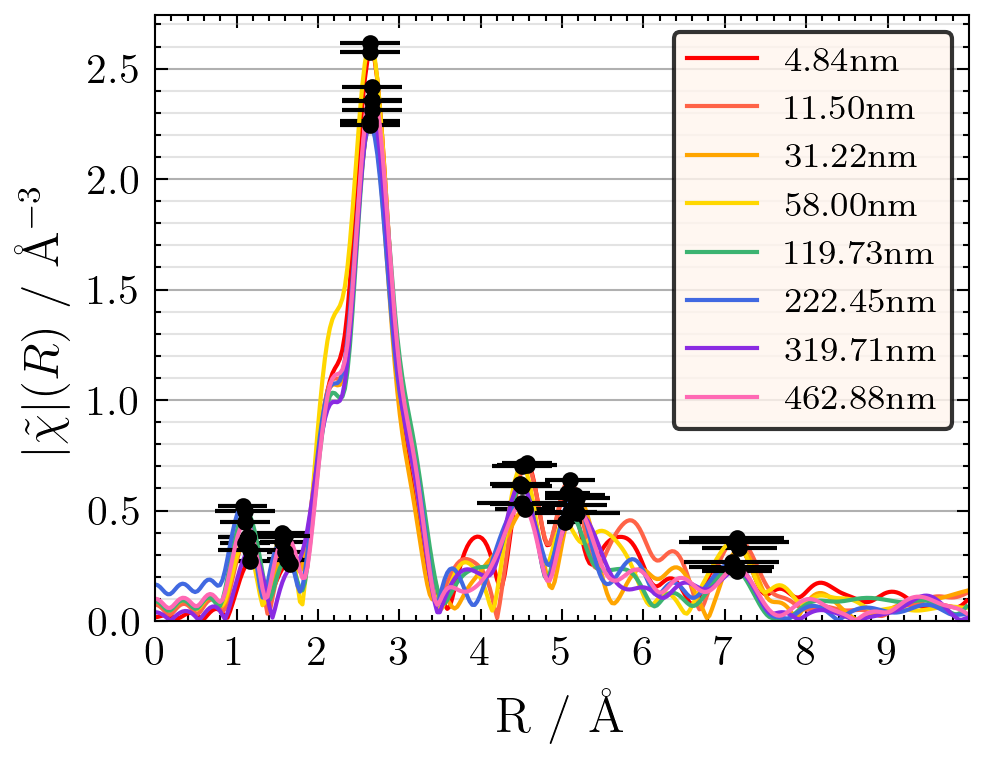}{width=\textwidth}{0.6\textwidth}{}{Samples $245$}
    \inputsubfigureimg{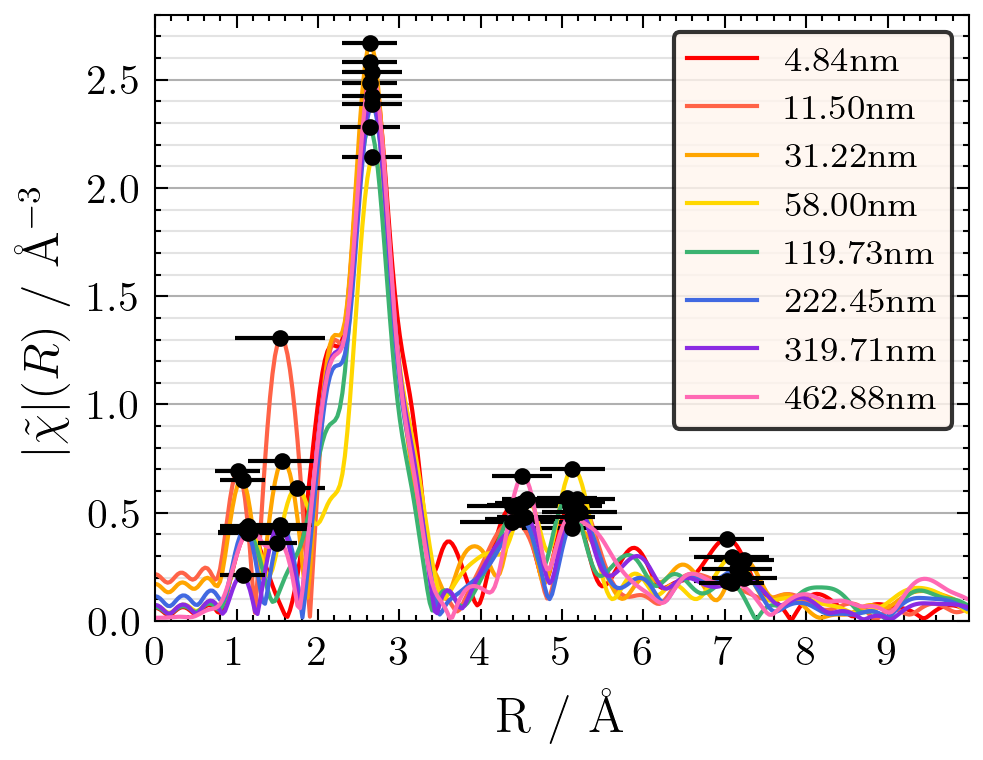}{width=\textwidth}{0.6\textwidth}{}{Samples $247$}

    \caption{Identification of $6$ characteristic peaks in the $R$ space of the measurements, defined as in tab. \ref{tab:peak-identification}. Their positions, amplitudes and widths are marked. The width of a peak is defined mathematically as the distance between the turning points next to that peak. The angles of incidence have been mapped to penetrations depths following eq. \ref{equ:penetration-depth}.}
    \label{fig:measured-fourier-pendepth}
\end{figure}

This allows to further classify the peaks in fig. \ref{fig:measured-fourier-angle}.
First $6$ different peaks are defined by a local maximum in $6$ specific ranges
(see tab. \ref{tab:peak-identification}).
Additionally, the peaks width is calculated by the distance of the turning points before and
after that peak.\footnote{Points where the second difference quotient of the discrete data has a change of sign.}
In fig. \ref{fig:measured-fourier-pendepth} the identified peaks are plotted together with the
assigned penetration depths as discussed.\\

\begin{table}[htbp]
    \centering
    \caption{Definition ranges and identification of peaks in the $R$ space measurements.}

\begin{tabular}{ccc}
    \toprule
    Peak no.                & definition range & identification \\
    \midrule
    1&$[\num[round-precision=1]{0.9}, \num[round-precision=1]{1.4}]\,\unit{\angstrom}$ & 1. $\mathrm{Nb-O}$ distance \\
    2&$[\num[round-precision=1]{1.4}, \num[round-precision=1]{2.0}]\,\unit{\angstrom}$ & 2. $\mathrm{Nb-O}$ distance \\
    3&$[\num[round-precision=1]{2.5}, \num[round-precision=1]{3.0}]\,\unit{\angstrom}$ & 1. coordination radius ($\mathrm{Nb-Nb}$) \\
    4&$[\num[round-precision=1]{4.2}, \num[round-precision=1]{4.8}]\,\unit{\angstrom}$ & 3. coordination radius ($\mathrm{Nb-Nb}$) \\
    5&$[\num[round-precision=1]{4.8}, \num[round-precision=1]{5.4}]\,\unit{\angstrom}$ & 4. coordination radius ($\mathrm{Nb-Nb}$) \\
    6&$[\num[round-precision=1]{6.8}, \num[round-precision=1]{7.4}]\,\unit{\angstrom}$ & 5. coordination radius ($\mathrm{Nb-Nb}$) \\
    \bottomrule
\end{tabular}

    \label{tab:peak-identification}
\end{table}

It has to be kept in mind that the $R$ space of a fine structure does not show
the radial distribution function inside the sample lattice.\cite{ravel-2012-ramsauer-townsend}
Instead the backscattering phase $\Phi_j$ of a reflecting atom can shift a peaks position and
the backscattering amplitude $F_j$ can cause the Ramsauer-Townsend effect where additional
peaks occur without representing real lattice distances (see sec. \ref{sec:theory-exafs}).
Also, measurement uncertainties and thermal fluctuations inside the lattice can lead to broader
peaks, thus a bigger peak may swallow a smaller one.
And in general, the fourier transform of a peak will produce multiple peaks as side lobes
of that main peak, not representing independent lattice distances themselves.\\

But an assignment can be done and a comparison with previous analyses
allows to identify the coordination shells and characteristic lattice distances.\cite{klaes-2020}
That way, the marked peaks in the measurements show two $\mathrm{Nb-O}$ distances and the
first, third, fourth and fifth coordination radii.
The second coordination radius is not visible, because it is merged into the main peak of
the first coordination radius, but the suspicious looking feature left to the main peak
on the other hand is uncharacteristic;
it is a side lobe of the first coordination radius.\\


Furthermore the peak migration was analyzed, meaning the changes in position, amplitude and width
for varying penetration depths. This is of interest, because changes in the crystal structure
imply changes in the position (expanding and contracting lattice) and amplitude (amount of atoms in that vicinity) of peaks in $|\tilde{\chi}|$.
The results are displayed in app. \ref{app:peak-migration}
and show mainly insignificant and chaotic behaviour. However the amplitude related to
the second $\mathrm{Nb-O}$ distance shows a distinct attenuation for increasing depth
in agreement with the assumption of a decreasing concentration of
\oxygen interstitials (see fig. \ref{dia:afterprepeak}).
Due to the original normalization of the spectra, only a fitting of a concentration distribution
can be accomplished, without any statement regarding quantity.\\
In fig. \ref{dia:concentration-fits}, those fits are shown. The underlying model is a
complementary error function (see sec. \ref{sec:theory-diffusion}) with
the surface concentration $c_0$, diffusion length $\delta$ and an offset as parameters.
The data for sample $245$ shows no significant trend for a concentration curve to fit to,
while sample $243$ and $247$ do, the former with a greater diffusion length ($\delta \approx \SI[round-precision=2]{73.69}{\nm}$) than the latter ($\delta \approx \SI[round-precision=2]{44.56}{\nm}$).

\inputimgh{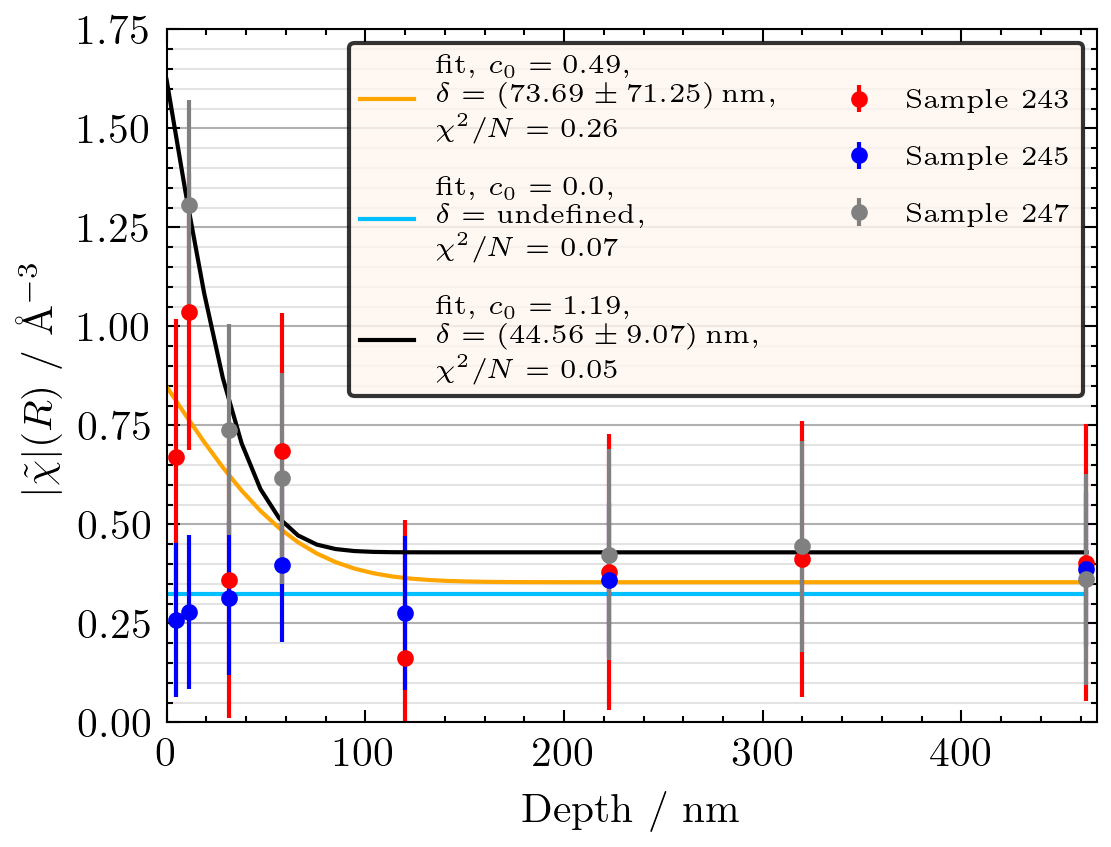}{width=0.75\textwidth}{concentration-fits}{Least squares fit of an interstitial concentration profile to the amplitude behaviour of the 2. $\mathrm{Nb-O}$ distance (see. fig. \ref{dia:afterprepeak}), the measurements for sample $245$ show no significance for a concentration profile. The fitting function is a complementary error function as described in sec. \ref{sec:theory-diffusion}. The error bars have a uniform value for each sample respectively (calculated following sec. \ref{sec:error-analysis}). All $\chi^2/N$ values are way below $1$.}

\section{Error Analysis in XAFS}\label{sec:error-analysis}

Researchers in the field of XAFS spectroscopy have a special relationship to
the estimation of measurement uncertainties. Often uncertainties aren't
extimated at all, and when they are, many researchers are showing skepticism whether
uncertainties got calculated properly and with the correct method.\cite{booth-2009}
This seems to be due to the highly context-dependent evaluation of XAFS data (see sec. \ref{sec:theory-exafs}).
Taking the EXAFS equation following eq. \ref{equ:exafs-equation}, it has in case of one considered
coordination sphere alone $6$ different parameters to be measured/ fitted. This amount of degrees of freedom
typically has to be reduced by context-related assumptions, like a crystal lattice and atomic properties
(e.g. in the case of this thesis, a \underline{bcc} lattice of \underline{\textrm{Nb}} and \underline{\textrm{O}} as an unknown amount
of \underline{interstitials}). In contrast, other analyses make different assumptions, e.g. for the thermal displacements $\sigma_j$
or the mean-free-path of the photo-electron $\lambda(k)$, thus the type of information an
EXAFS measurement delivers, varies from analysis to analysis.
With that the starting point for an error analysis is fluent and thus an universal convention is missing (yet).\\

Nevertheless, methods for estimating uncertainties exist and vary in their level of detail.
First of all, it can be assumed that the energy resolution $\delta E$\footnote{It should be noted that $\delta E$ here is not the energy step, but rather the measurement uncertainty.}
fulfills the proportionality $\delta E = WE$ where $W = \num[round-precision=0]{7e-4}$
is the fractional energy resolution in a rough estimate,
following \cite{morrison-1981}\footnote{The paper by Morrison et al. names the double value for $W$. This is because they discuss the FWHM as $\delta E$.}.
With regard to eq. \ref{equ:e2k-relation} and by applying the definition of the total differential,
the resolution $\delta k$ can be obtained by\\

\begin{align}\label{equ:delta-k}
    \delta k = \frac{kEW}{2(E-E_0)}
\end{align}

Transfering into the $R$ space, in the scope of this thesis, the uncertainty of the radius $\delta R$ is approximated by the
simple relation

\begin{align}\label{equ:delta-r}
    \delta R = \frac{\pi}{2 k_\mathrm{max}}
\end{align}

with $k_\mathrm{max} = \SI[round-precision=1]{10.5}{\per\angstrom}$ for this analysis.\\

For the fluctuations of $\chi$ and $|\tilde{\chi}|$ on the other hand this thesis relies on a paper discussing high-$R$ components
of an XAFS spectrum to measure the RMS\footnote{root mean square} as an approximation for the uncertainty $\varepsilon_R$ of $|\tilde{\chi}|$.\cite{newville-1999}
The reason is that those components are equivalent to frequencies that are beyond
structural information and only consist of noise.
In general it would be appropriate to take components far greater than the farthest identified
coordination radius (to avoid structural influence), but due to the small $R$-range measured, components between
\SI[round-precision=0]{8}{\angstrom} and \SI[round-precision=0]{10}{\angstrom}
are going to be used, just above the peak of the fifth coordination radius. 
Furthermore, $\varepsilon_R$ can be converted into the uncertainty $\varepsilon_k$
of $\chi$ by considering Parseval's theorem and assume a $k$- and $R$-independence.
This leads to the relationship

\begin{align}\label{equ:epsilon-k}
    \varepsilon_k = \varepsilon_R\,\sqrt{\frac{5\pi}{\delta k\,(k^5_{\mathrm{max}} - k^5_{\mathrm{min}})}}
\end{align}

for a $k^2$-weighted fine structure, where $\delta k$ is the resolution of $k$ as introduced and
$k_\mathrm{min} =$ \SI[round-precision=0]{1}{\per\angstrom},
$k_\mathrm{max} =$ \SI[round-precision=1]{10.5}{\per\angstrom} is
the fourier range (see sec. \ref{sec:extracting-the-fine-structure}).
The resulting uncertainties $\delta k$, $\varepsilon_k$, $\delta R$ and $\varepsilon_R$
are exemplarily plotted in a single curve in appendix \ref{app:error-plots} to give a sense of the dimensions of
the uncertainties\footnote{It is necessary to notice that it seems not clear from Newville et al. (\cite{newville-1999}) that a $k$-weighting of the fluctuation $\varepsilon_k$ is needed. I decided to do it anyways in fig. \ref{dia:fs-error}, after careful consideration that the errors would otherwise be implausibly small.}.

\section{Considerations for EXAFS Simulations}\label{sec:deductive}

Performing a simulation is helpful
in order to rate the possibilities of an EXAFS analysis,
but also to verify the consistency of a model on the basis of a
well EXAFS measurement.
Hereby it is crucial to distinguish between EXAFS measurements
in transmission and fluorescence mode.
Transmission measurements can be described through
the EXAFS equation of Lytle et al. (see eq. \ref{equ:exafs-equation}),
while the measurable fluorescence photons intensity suffers from the self-absorption effect
and additionally from it's dependence on the X-ray photon energy
(penetration depth into the sample depends on the photon energy,
see last paragraph in sec. \ref{sec:theory-exafs}).
This non-linear relationship makes it unsuitable to model fluorescence measurements
with the EXAFS equation, even though the spectra look similar.
Instead a feasible approach would be to utilize Fresnel theory
in order to composite multiple known lattices
and calculate their fluoresence EXAFS theoretically.
In other words, reference EXAFS spectra of already measured
lattices are used to extract optical parameters
and simulate a multilayer system whose fluorescence including
self-absorption can be calculated. Hereby the strong
assumption is made, that a \nb lattice with \oxygen interstitials
behaves kind of similar to a lattice of niobium suboxides (see sec. \ref{sec:theory-mid-t-bake}).
Because interstitials do not form a lattice but rather lattice defects
they cannot be simulated directly and a suboxide lattice has to be approximated
for this purpose.
Thus such a simulation of fluorescence EXAFS is not suitable for
verifying the performed measurement, but rather to investigate
similarities in the behaviour of suboxides and \oxygen interstitials.
The software programs under previous consideration serve this purpose and are therefore
unsuitable for simulating an oxygen diffusion profile
(see back in introduction, sec. \ref{sec:introduction}).\\

Simulations of transmission EXAFS (following \cite{newville-2004}) on the other hand, work quite straightforward
by first making assumptions for properties of the lattice structure ($N$, $R$, $\sigma$ and edge $E_0$)
and scattering process ($F_j(k)$, $\Phi_j(k)$ and $\lambda(k)$),
the latter can be calculated e.g. using the \textsc{FEFF} programm\footnote{See \url{https://monalisa.phys.washington.edu/}.}.
These can then be inserted directly into the EXAFS equation, where each coordination sphere is
represented by an independent summand and a comparison of calculated and measured EXAFS
and provides a measure for the similarity of the real lattice to the model
(e.g. consisting only of the first two coordination spheres).
Multiple of such comparisons can help getting a feeling for the dominant
structual parts in the real lattice.
Thereby it can be chosen whether this comparison of EXAFS shall be performed in the $k$ space
or in the $R$ space. The latter case has the advantage of a direct comparison of
single coordination radii where others can be ignored easily, but it requires
the complex $\tilde{\chi}$ to compare and not just the absolute value.

\section{Discussion of Results and Method}\label{sec:discussion}
The main result, that sample $243$ has a diffusion length
of $\delta_\mathrm{243} = \SI{73.69\pm71.25}{\nano\m}$
and $\delta_\mathrm{247} = \SI{44.56\pm9.07}{\nano\m}$ for sample $247$,
tells that the mid-T treated sample $243$ indeed shows a higher diffusivity
than the EuXFEL reference $247$. The obvious explanation hereby is,
that the low-T range of $247$ is not able to dissolve the
natural oxide layer of the niobium surface, thus 
a lower amount of oxygen atoms diffuse into the material,
whereas the mid-T sample $243$ fulfills this property, leading to a higher amount of diffusion.
Sample $245$ on the other hand shows no sign of a diffusion process,
which could be explained by a too narrow range of depths chosen for measuring.
If the range was chosen wider, an actual profile may be visible in the data,
but this would mean that the corresponding diffusion length is way greater than for
the other two samples, contradicting the precalculations in sec. \ref{sec:treatments}.
Then either the assumption of an \textrm{erfc}-related concentration profile (Fickian diffusion)
is wrong, so a deviation of the conventional model is indicated,
or the overall interpretation of the peak migration data in fig. \ref{dia:concentration-fits}
is misleading and instead due to strong noise in the measurements.\\

It has to be concluded, that the measurement uncertainties are way to high to
deduce resilient statements about the lattice, as it gets clear by the fact that
only one of $18$ diagrams of peak migration in app. \ref{app:peak-migration}
was usable for at least some statement.\\ 

Performing an EXAFS simulation in fluoresence as considered in sec. \ref{sec:deductive},
would already be possible with the collected data for this thesis.
The outcome is not expected to be enlightening, but more like an attempt to
get the maximum of what is possible with the available expertise out of the analysis.
Since the possiblities of the simulation software programs turned out to not be designed
for verifying an EXAFS analysis but rather investigate another academic question.\\

Compared to the analysis by Luth \cite{luth-2024}, which only allowed
a sight on the white lines\footnote{The white line of an XAFS spectrum is defined as the first major peak above the absorption edge and thus part of NEXAFS. Because it depends on the inner-atomic distances, interstitial oxygen with a higher electronegativity attract the bound electrons of niobium. Therefore the presence of oxygen interstitials is encoded roughly in the amplitude of the white line.} of the spectra, to get a rough and indirect estimation for
the oxygen concentration, the present thesis allowed more insight.
This is due to the application of the fourier transformation
and the analysis of the amplitude behaviour. 

\subsection{Error Discussion}\label{sec:error-discussion}
The main source of the measurement uncertainties definitely lies in the low brilliance
of the DELTA storage ring and it's wiggler SCW (see sec. \ref{sec:outlook}).
With this boundary condition and a beamtime of five days, integration time and energy step were chosen accordingly,
but $28$ of $148$ total exafsscans (see app. \ref{app:raw-measurements})
were useless whatsoever and thus due to other reasons.
An obvious reason is that construction workers next to the
DELTA facility caused vibrations with heavy machinery, strong enough to disturb
the experiment during the measurement aquisition, noticable through an oscilloscope.
After complaint, they shared their working hours (scheduled aspecially in the morning) with us and we adapted to that,
but exceptions occured and made their way as noise into the data. 
Furthermore, the double crystal monochromator had to be recalibrated ongoingly
via the piezo element, to ensure a brilliant beam with a constant intensity
and the correct photon energy.
Even though it is not plausible to check it strictly,
some displacements could have occured because of inaccurate or irregular
calibration, which then resulted into systematic errors.
Also the measurements during night were vulnerable to disruptions and malfunctions,
due to the non-living surveillance. Those exafsscans were checked later by
verifing the absorption curves and graph of the electron beam current inside the storage ring.
In addition, the synchrotron beam switched multiple times because of reinjections of the
electron beam. In the hypothetical case some systematic errors inside the storage ring occured, those
could have been in effect only in one electron beam and not the other. But aspecially because almost
every exafsscan during the day got repeated during the night and in most cases those measurements
were performed with another beam, errors like systematic glitches mostly would have been averaged out.
Beyond that, a varying amount of exafsscans was taken redundantly for averaging.
This implies a varying fluctuation for averaged spectra
and thus influences error estimation, strictly speaking.\\

Moreover, it can be assumed that many glitches appear in the spectra. One distinct glitch
occured in every exafsscan, was located at $+$\SI[round-precision=1]{226.1}{\electronvolt} above the $K$-edge
(or \SI[round-precision=1]{19212.6}{\electronvolt} as absolute value) and already
identified during data aquisition. Possible explanations for this and other glitches in general are
defects inside the monochromator crystals (perturbation of Bragg reflection), distortions in other optical instruments
(like the kapton plastic foils), incorrect gas composition in the ionization chamber $I_0$
(which could disturb the intensity measurement $\rightarrow$ lower
measured intensity results into a higher calculated absorption coefficient and
potentially produces a glitch) or problems inside the synchrotron/ wiggler (some absorbing
component manipulating the beam before entering the experimental hut). 
But apart from the discussed glitch, only a few additional
glitches were removed. Aspecially further back in the spectra the data quality
was noisy (see fig. \ref{fig:measured-spectra}) and with more experience and courage,
the quality could have been improved
manually even more.
The distinct feature at about $+\SI[round-precision=0]{345}{\electronvolt}$
above the $K$-edge ($\SI[round-precision=1]{19330.5}{\electronvolt}$ as absolute value),
which occurs exactly only in the mid-T treated samples, is suspicious and probably due to
the mid-T treatment and not some external disturbance.
Regarding roughness considerations,
the manufacturing of each samples caused parallel machining marks on the surface
during cutting. Trivial considerations beforehand suggested that an alignment of
those marks parallel to the beamline reduces effects of roughness in the
measurements, but due to forgetfulness this wasn't realized. Instead
the approximate orientation of the machining marks on the first mounted sample in the setup
was at least kept the same for the other two samples. An alternative influence
of roughness are fingerprints and dust which would cause a minor effect in the data.
As for the filling gas in the ionization chambers, a proportion of krypton in the argon gas
would be imaginable to further increase the intensity measurement and improve the poisson statistics,
but due to the about $18\,\%$ absorption in the chambers already, this is a minor consideration.\\

Looking at the processing procedure afterwards, it would have been possible to
set the so called \texttt{rbkg} parameter manually for each spectrum, to further optimize
background subtraction and thus
the extraction of the fine structure, but an uniform
value of $1$ was chosen instead (see sec. \ref{sec:extracting-the-fine-structure}).
Same applies for the \texttt{spline clamps}. Regarding the interpretation of
the measurements, the surjective character\footnote{By that I mean that multiple
crystal structures can look the same in a measurement of the fine structure $\chi$.} of 
EXAFS spectroscopy allows multiple conclusions about a feature in the data.
Common practice is to take previous analysis of others into account to
get to the most plausible interpretation, as it has been done in this thesis as well.
Therefore it is worth emphasizing again that this field of spectroscopy relies on
this approach. Taking the third and fourth coordination radii for example
(see fig. \ref{fig:measured-fourier-pendepth}), it would be plausible to
interpret their peaks in the $R$ space as a Ramsauer-Townsend phenomenon.
Only the reference to a previous analysis of \nb lattices, where these peaks were investigated
in more detail, makes it clear that they indeed represent independent coordination radii.
Same applies for the peak at about \SI[round-precision=0]{2}{\angstrom}, which looks appealingly
like an independent peak, but is instead a side lobe of the peak next to it.\\

Furthermore, it is debatable whether the performed error analysis in sec. \ref{sec:error-analysis}
is appropriate for the performed experiment and it's discussion about SRF cavities in the MSL group.
The estimation of $\delta k$ is already rather detailed compared with the
simple approach of defining the energy step as $\delta E$ and
then calculate $\delta k$, while $\delta R$ is simply calculated by
the upper fourier range $k_\mathrm{max}$ based on the fourier relation and thus
expandable. 
The approach for the $\chi$ fluctuation $\varepsilon_k$ on the other hand is a bit difficult to rate.
On the one side, taking the RMS of the back values of $|\tilde{\chi}|$ in $R$ space
turned out empirically to result into the highest estimations for $\varepsilon_k$,
compared with the approaches of Poisson counting statistics or 
taking the direct RMS of the absorption data in the pre-edge
or post-edge regimes (following \cite{newville-1999}). And because only values in the
range between \SI[round-precision=0]{8}{\angstrom} and \SI[round-precision=0]{10}{\angstrom}
could be considered, the resulting uncertainty is yet again bigger, which suggests an upper limit.
On the other side, assuming a constant\footnote{With \enquote{constant} is meant that $\varepsilon_k$ doesn't depend directly on $k$. It is indeed depending indirectly on $k$ due to $\delta k(k)$ in eq. \ref{equ:epsilon-k}.}
$\varepsilon_k$ does not acknowledge major glitches or systematic errors (like bad background removal)
where a variable uncertainty would be necessary. In these cases the chosen approach cannot
represent an upper limit. This can be concluded for $\varepsilon_R$ as well due to it's
connection to $\varepsilon_k$ through the fourier transform, which is the
primary uncertainty of interest for this thesis
regarding the useful measurements (peak amplitudes, see fig. \ref{dia:concentration-fits}).
Without changing the method, $\varepsilon_R$ can be improved by measuring a larger range in $R$ space.
All in all, the calculated uncertainties proof, that with the available data,
a great noise background can be assumed.
The $\chi^2$ values in the fittings of fig. \ref{dia:concentration-fits} are quite low,
which often implies an overestimation of uncertainties in the data.
In this case on the other hand, it is more plausible that the data suggests a
way too low diffusion length due to fluctuations in a rather small depth range.
Improving $\varepsilon_R$ or enlarging the depth range would help to improve the fits.

\subsection{Methodological Comparison}\label{sec:methodological-comparison}
Besides studying the oxygen concentration profile via EXAFS spectrocopy, other
analyzation methods are common and they reveal more about how to look at
EXAFS for SRF research and the results of this thesis.
LE-$\mu$SR (Low energy muon spin rotation spectroscopy) utilizes
muons as local magnetic probes, in order to gain information about the
magnetic field inside a sample through the muon decay and it's spin disturbance by
that local magnetic field.
In an investigation
of the three samples discussed in this thesis, by Ghanbari et al.\footnote{Yet to be published.},
their magnetic field distribution was measured and the corresponding
oxygen concentration distribution deduced. The preliminary result is that
$\delta_\mathrm{243} = \SI{150\pm3.4}{\nano\m}$,
$\delta_\mathrm{247} = \SI{30\pm14}{\nano\m}$ and no plausible value
for $\delta_\mathrm{245}$ can be determined. This is consistent with
the determined values in fig. \ref{dia:concentration-fits},
where $\delta_\mathrm{243}$ is clearly bigger than $\delta_\mathrm{247}$
and sample $245$ shows no significant \textrm{erfc} behaviour.
But it seems to be clear as well that the EXAFS analysis did not do well
in terms of quantitative accuracy, already due to the large uncertainty of the fit
($\delta_\mathrm{243} = \SI{73.69\pm71.25}{\nano\m}$
and $\delta_\mathrm{247} = \SI{44.56\pm9.07}{\nano\m}$).
Moreover because the results of Ghanbari et al. can be understood
as a lower bound to the true values due to a sensitivity of that method only
until about \SI[round-precision=0]{100}{\nano\meter} depth.
Another approach is the SIMS analysis (secondary ion mass spectrometry) where an ion beam
is sputtered onto the sample surface in order to knock out the lattice atoms as
\textit{secondary ions}, whose detection is utilized for a statistical analysis of
the removed atom counts as a function of depth. Because the natural oxide layers
are negligibly thin compared with the diffusion length, the oxygen counts directly
give a measure for the interstitial concentration distribution.
Significant for this sputtering method is that unlike EXAFS or LE-$\mu$SR spectroscopy
it is destructive, thus a sample cannot be analyzed twice.
A study using \textit{time-of-flight} SIMS by Bate et al. in
2025\cite{bate-2024-sims}
found that out of six different mid-T treated samples at least three
samples showed strong consistency between the theoretical Fickian diffusion length
(see eq. \ref{equ:penetration-depth-Tt}) and the SIMS-measured one, while
the other three show clearly higher diffusion lengths.
A plausible assumption the paper made, are grain boundaries in which the diffusivity
is higher than in the lattice. This applies for this thesis too,
as already touched in sec. \ref{sec:treatments}.
Beyond that, a number of other methods utilizing X-ray diffraction techniques or scanning electron microscopy
can also analyze suboxides in cavities, but only poorly oxygen interstitial profiles
(see e.g. \cite{2019-xrd-o-concentration}).

\subsection{Outlook}\label{sec:outlook}
Further oxygen profile investigations at DELTA are unlikely to yield much better results,
it seems the optimum with this machine got reached. The PETRA III facility at DESY on the other hand,
provides an enormously higher brilliance, compared to DELTA. While this analysis
utilized a brilliance in the order of
$10^{15}\,\frac{1}{\mathrm{s\,mrad^2\,mm^2}\,\num[round-precision=1]{0.1}\,\%\,\mathrm{BW}}$ at
DELTA\footnote{This order of magnitude is valid for the superconducting asymmetric wiggler (SAW) at DELTA, the predecessor of the SCW.}, 
a possible beamline experiment at PETRA III would have access to an order
of $10^{24}\,\frac{1}{\mathrm{s\,mrad^2\,mm^2}\,\num[round-precision=1]{0.1}\,\%\,\mathrm{BW}}$!\cite{delta-brilliance}\cite{petra-brilliance}
This would open the gates for a far more comprehensive study, in terms
of the amount of analyzable samples, integration time, energy step and
angle of incidence settings. The single exafsscan with the parameters in this study within
a time of \SI[round-precision=0]{23}{\minute}, would shrink
to ~$\SI[round-precision=0]{1}{\micro\s}$. This is a purely mathematical value, because
the high brilliance applied to the present setup would smear the photon energies and thus
prohibit data aquisition. So an intentional delay of the measurements would be necessary or
another setup, like for energy-dispersive EXAFS, which is far beyond the scope of this thesis, could be
utilized.\\

These benefits seem very convincing, but the crucial point here for future projects is, whether EXAFS can
actually deliver the desired quantitative results by a raised brilliance
or systematic problems in this method rather suggest to focus the ressources on a more reliable
method. Due to a missing simulation approach for EXAFS data, it cannot be judged if
the measurements follow theoretical considerations in principle and a reliable oxygen concentration
profile could be obtained.
But previous EXAFS measurements \cite{klaes-2020} at PETRA III showed that very distinct
fine structures can be resolved and thus confidence in case of oxygen interstitials is justified.\\

Furthermore the selection of samples can be of interest for future analyses.
The intentional focus on only three samples was due to the bad experience with
EXAFS measurements at DELTA before. If a higher brilliance is accessible,
it will become attractive to analyze more samples again, like aluminium coated
samples which serve the purpose of preventing the suboxides to disappear into the vacuum
while heating. These resulting oxygen profiles are of great interest as well.

\newpage
\section{Conclusion}
This thesis provided an analysis of the oxygen diffusion profile of three
differently heat treated samples,
two in form of experimental mid-T baking and the remaining one as a reference for the standard
low-T baseline baking, utilized in currently operating cavities, installed e.g. in the European XFEL in Hamburg.
The measurement of EXAFS at the DELTA facility suffered under high noise levels,
thus the quantitative results are not trustworthy and even qualitatively barely
statements can be provided. Given that the trend in measurement
is not due to random fluctuations, no deviation from the basic Fickian diffusion
behaviour could be found and as expected, the regular mid-T treatment
results into a higher diffusion length than the EuXFEL recipe.
Irritating is the missing outcome for the the second, long duration
mid-T treatment, which can most likely be explained by an overshadowed trend due to measurement noise.
All in all, we have at maximum a confirmation of the known and beyond that,
noise that invites to speculation. This goes hand in hand with the
conclusion that more or less the maximum of possibilities at the DELTA facility got provoked.
Further interpretation was thus up to the identification of errors.
Various noise sources and glitches have been pointed out, from which future EXAFS analyses will benefit.
The provided error analysis confirms a wide range of possible results
for the concentration profile due to large calculated uncertainties.
Hereby the extent of uncertainty estimation allows several
different approaches, whereby the overall handling of error determination is generally underdeveloped
in the XAFS community. The hope of simulating EXAFS spectra in order to provide a comparison
with DELTA measurements, turned out to be inappropriate for the given experiment,
upcoming fluoresence analyses with less noisy data could at maximum utilize simulations
to focus on the investigation of similarities in the optical behaviour of suboxide layers and
oxygen interstitials. In case of a future analysis in transmission mode,
EXAFS simulations can actually be used to verify the success of a measurement.
The classification of EXAFS in alternative analysis methods, showed that this thesis 
did not contradict previous analyses, but potential for future investigations utilizing EXAFS exists.
The favoured repetition of this investigation at the PETRA III facility at DESY,
would enable a far more detailed analysis in terms of sample number, 
material depth resolution and accuracy.

\newpage
\printbibliography
\addcontentsline{toc}{section}{References}
\newpage
\appendix

\section{Raw Measurements}\label{app:raw-measurements} 

\begin{table}[htbp]
    \centering
    \caption{Overview of collected exafsscans.}

\begin{tabular}{cllllllllllllll}
    \toprule
    \rotatebox{90}{Sample} & \rotatebox{90}{angle $\theta$ / \unit{\degree}} && \rotatebox{90}{\# total} & \rotatebox{90}{\# total@{\color{yellow!70!black}day}} & \rotatebox{90}{\# total@{\color{blue}night}} && \rotatebox{90}{\# used {\footnotesize(\& averaged)}} & \rotatebox{90}{\# used@{\color{yellow!70!black}day}} & \rotatebox{90}{\# used@{\color{blue}night}} && \rotatebox{90}{\# dismissed} & \rotatebox{90}{\# dismissed@{\color{yellow!70!black}day}} & \rotatebox{90}{\# dismissed@{\color{blue}night}} \\
    \midrule
    \multirow{8}*{\rotatebox{90}{243}} & \num[round-precision=1]{0.1}  && \textbf{7} &{\color{yellow!70!black}\textbf{4}}&{\color{blue}\textbf{3}}&& 4 &{\color{yellow!70!black}3}&{\color{blue}1}&& 3 &{\color{yellow!70!black}1}&{\color{blue}2}\\
    & \num{0.15}                                                       && \textbf{8} &{\color{yellow!70!black}\textbf{5}}&{\color{blue}\textbf{3}}&& 3 &{\color{yellow!70!black}2}&{\color{blue}1}&& 5 &{\color{yellow!70!black}3}&{\color{blue}2}\\
     & \num{0.17}                                                      && \textbf{7} &{\color{yellow!70!black}\textbf{4}}&{\color{blue}\textbf{3}}&& 4 &{\color{yellow!70!black}4}&{\color{blue}0}&& 3 &{\color{yellow!70!black}0}&{\color{blue}3}\\
     & \num[round-precision=1]{0.2}                                    && \textbf{8} &{\color{yellow!70!black}\textbf{3}}&{\color{blue}\textbf{5}}&& 3 &{\color{yellow!70!black}3}&{\color{blue}0}&& 5 &{\color{yellow!70!black}0}&{\color{blue}5}\\
     & \num[round-precision=1]{0.3}                                    && \textbf{5} &{\color{yellow!70!black}\textbf{3}}&{\color{blue}\textbf{2}}&& 4 &{\color{yellow!70!black}3}&{\color{blue}1}&& 1 &{\color{yellow!70!black}0}&{\color{blue}1}\\
     & \num[round-precision=1]{0.5}                                    && \textbf{4} &{\color{yellow!70!black}\textbf{3}}&{\color{blue}\textbf{1}}&& 4 &{\color{yellow!70!black}3}&{\color{blue}1}&& 0 &{\color{yellow!70!black}0}&{\color{blue}0}\\
     & \num[round-precision=1]{0.7}                                    && \textbf{4} &{\color{yellow!70!black}\textbf{3}}&{\color{blue}\textbf{1}}&& 4 &{\color{yellow!70!black}3}&{\color{blue}1}&& 0 &{\color{yellow!70!black}0}&{\color{blue}0}\\
     & \num[round-precision=0]{1}                                      && \textbf{4} &{\color{yellow!70!black}\textbf{0}}&{\color{blue}\textbf{4}}&& 4 &{\color{yellow!70!black}0}&{\color{blue}4}&& 0 &{\color{yellow!70!black}0}&{\color{blue}0}\\
    \midrule
    \multirow{8}*{\rotatebox{90}{245}} & \num[round-precision=1]{0.1}  && \textbf{8} &{\color{yellow!70!black}\textbf{6}}&{\color{blue}\textbf{2}}&& 8 &{\color{yellow!70!black}6}&{\color{blue}2}&& 0 &{\color{yellow!70!black}0}&{\color{blue}0}\\
     & \num{0.15}                                                      && \textbf{8} &{\color{yellow!70!black}\textbf{3}}&{\color{blue}\textbf{5}}&& 7 &{\color{yellow!70!black}2}&{\color{blue}5}&& 1 &{\color{yellow!70!black}1}&{\color{blue}0}\\
     & \num{0.17}                                                      && \textbf{8} &{\color{yellow!70!black}\textbf{8}}&{\color{blue}\textbf{0}}&& 6 &{\color{yellow!70!black}6}&{\color{blue}0}&& 2 &{\color{yellow!70!black}2}&{\color{blue}0}\\
     & \num[round-precision=1]{0.2}                                    && \textbf{5} &{\color{yellow!70!black}\textbf{2}}&{\color{blue}\textbf{3}}&& 5 &{\color{yellow!70!black}2}&{\color{blue}3}&& 0 &{\color{yellow!70!black}0}&{\color{blue}0}\\
     & \num[round-precision=1]{0.3}                                    && \textbf{6} &{\color{yellow!70!black}\textbf{3}}&{\color{blue}\textbf{3}}&& 4 &{\color{yellow!70!black}3}&{\color{blue}1}&& 2 &{\color{yellow!70!black}0}&{\color{blue}2}\\
     & \num[round-precision=1]{0.5}                                    && \textbf{6} &{\color{yellow!70!black}\textbf{3}}&{\color{blue}\textbf{3}}&& 5 &{\color{yellow!70!black}3}&{\color{blue}2}&& 1 &{\color{yellow!70!black}0}&{\color{blue}1}\\
     & \num[round-precision=1]{0.7}                                    && \textbf{6} &{\color{yellow!70!black}\textbf{3}}&{\color{blue}\textbf{3}}&& 5 &{\color{yellow!70!black}3}&{\color{blue}2}&& 1 &{\color{yellow!70!black}0}&{\color{blue}1}\\
     & \num[round-precision=0]{1}                                      && \textbf{4} &{\color{yellow!70!black}\textbf{1}}&{\color{blue}\textbf{3}}&& 4 &{\color{yellow!70!black}1}&{\color{blue}3}&& 0 &{\color{yellow!70!black}0}&{\color{blue}0}\\
    \midrule
    \multirow{8}*{\rotatebox{90}{247}} & \num[round-precision=1]{0.1}  && \textbf{6} &{\color{yellow!70!black}\textbf{3}}&{\color{blue}\textbf{3}}&& 6 &{\color{yellow!70!black}3}&{\color{blue}3}&& 0 &{\color{yellow!70!black}0}&{\color{blue}0}\\
     & \num{0.15}                                                      && \textbf{7} &{\color{yellow!70!black}\textbf{3}}&{\color{blue}\textbf{4}}&& 6 &{\color{yellow!70!black}2}&{\color{blue}4}&& 1 &{\color{yellow!70!black}1}&{\color{blue}0}\\
     & \num{0.17}                                                      && \textbf{7} &{\color{yellow!70!black}\textbf{3}}&{\color{blue}\textbf{4}}&& 7 &{\color{yellow!70!black}3}&{\color{blue}4}&& 0 &{\color{yellow!70!black}0}&{\color{blue}0}\\
     & \num[round-precision=1]{0.2}                                    && \textbf{6} &{\color{yellow!70!black}\textbf{3}}&{\color{blue}\textbf{3}}&& 6 &{\color{yellow!70!black}3}&{\color{blue}3}&& 0 &{\color{yellow!70!black}0}&{\color{blue}0}\\
     & \num[round-precision=1]{0.3}                                    && \textbf{6} &{\color{yellow!70!black}\textbf{4}}&{\color{blue}\textbf{2}}&& 6 &{\color{yellow!70!black}4}&{\color{blue}2}&& 0 &{\color{yellow!70!black}0}&{\color{blue}0}\\
     & \num[round-precision=1]{0.5}                                    && \textbf{6} &{\color{yellow!70!black}\textbf{4}}&{\color{blue}\textbf{2}}&& 6 &{\color{yellow!70!black}4}&{\color{blue}2}&& 0 &{\color{yellow!70!black}0}&{\color{blue}0}\\
     & \num[round-precision=1]{0.7}                                    && \textbf{6} &{\color{yellow!70!black}\textbf{4}}&{\color{blue}\textbf{2}}&& 5 &{\color{yellow!70!black}3}&{\color{blue}2}&& 1 &{\color{yellow!70!black}1}&{\color{blue}0}\\
     & \num[round-precision=0]{1}                                      && \textbf{6} &{\color{yellow!70!black}\textbf{1}}&{\color{blue}\textbf{5}}&& 4 &{\color{yellow!70!black}0}&{\color{blue}4}&& 2 &{\color{yellow!70!black}1}&{\color{blue}1}\\
     \bottomrule
\end{tabular}

    \label{tab:measurements-sorted-out}
\end{table}

\newpage
\section{Peak Migration}\label{app:peak-migration} 

\begin{figure}[h!]
    \centering
    
    \inputsubfigureimg{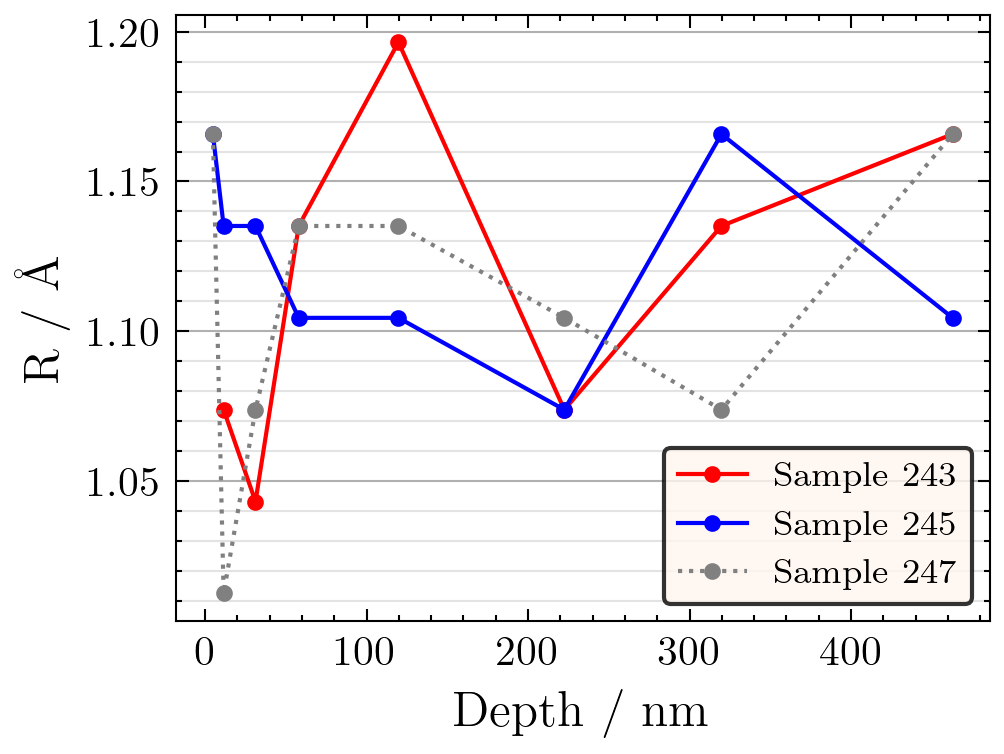}{width=0.9\textwidth}{0.485\textwidth}{}{1. $\mathrm{Nb-O}$ distance}
    \inputsubfigureimg{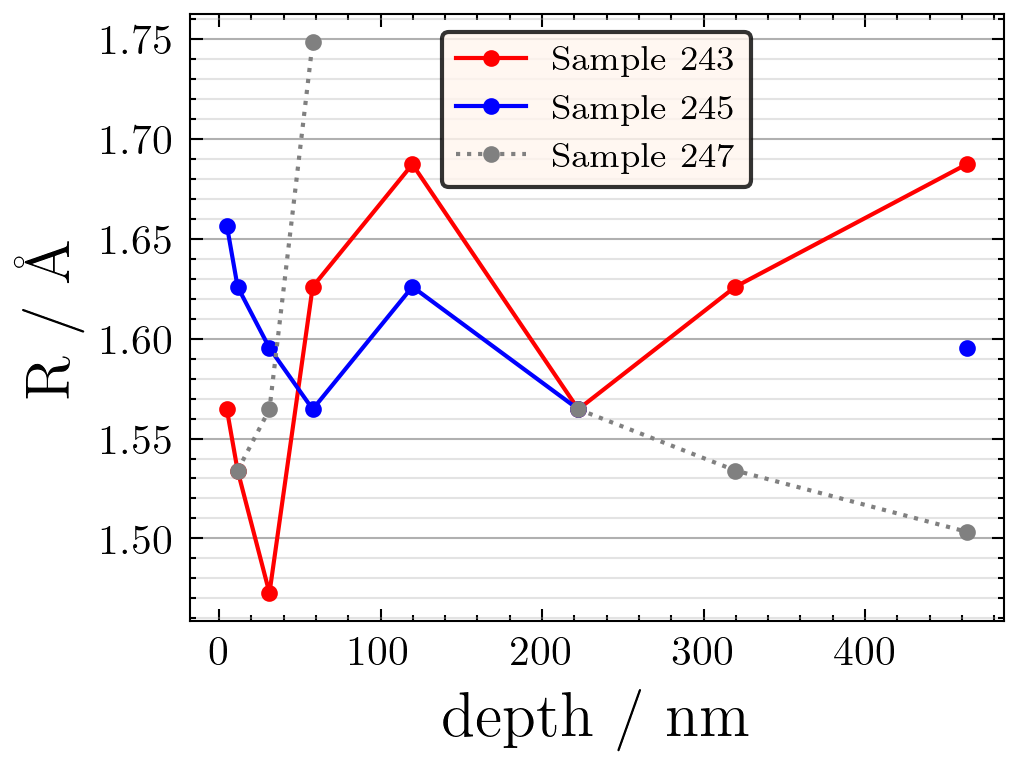}{width=0.9\textwidth}{0.485\textwidth}{}{2. $\mathrm{Nb-O}$ distance}
    \inputsubfigureimg{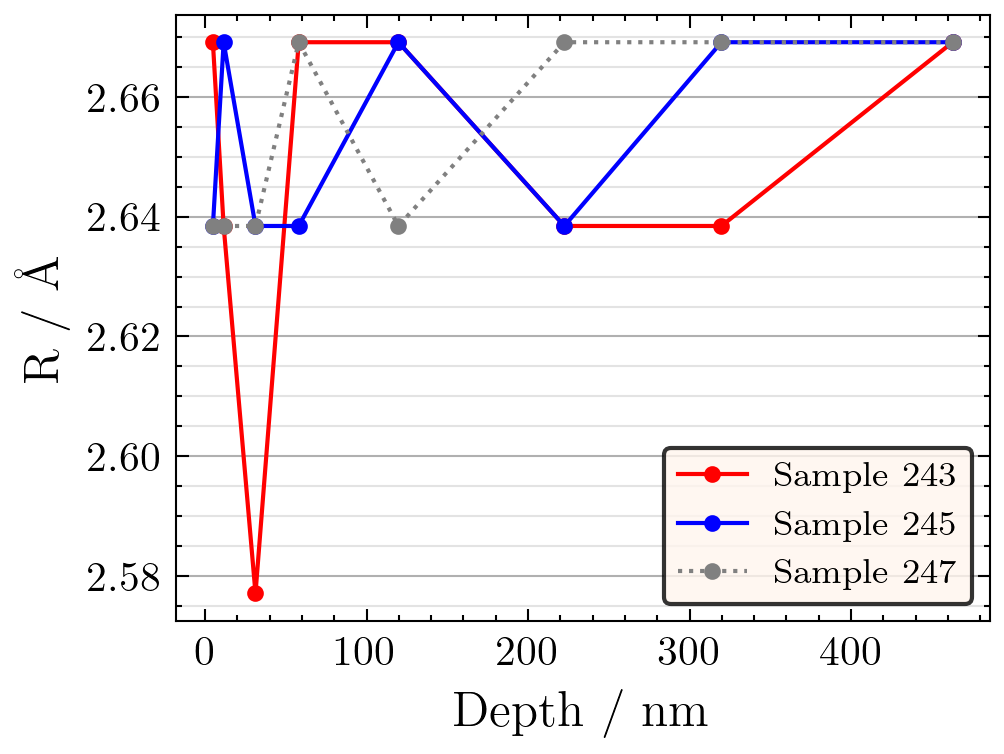}{width=0.9\textwidth}{0.485\textwidth}{}{1. coordination radius ($\mathrm{Nb-Nb}$)}
    \inputsubfigureimg{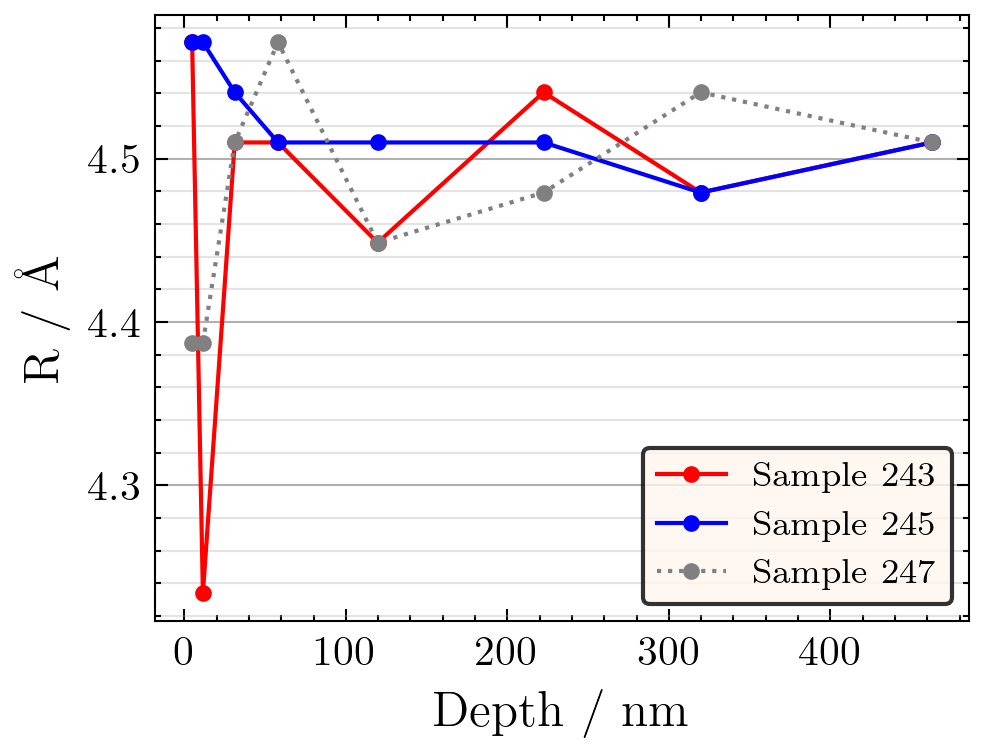}{width=0.9\textwidth}{0.485\textwidth}{}{3. coordination radius ($\mathrm{Nb-Nb}$)}
    \inputsubfigureimg{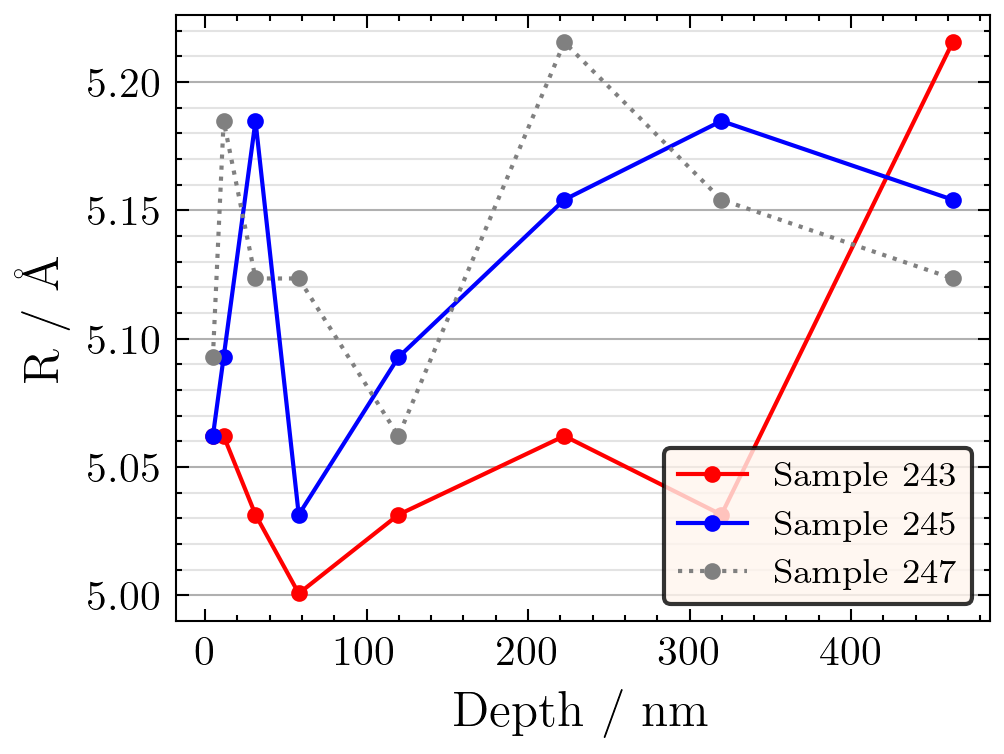}{width=0.9\textwidth}{0.485\textwidth}{}{4. coordination radius ($\mathrm{Nb-Nb}$)}
    \inputsubfigureimg{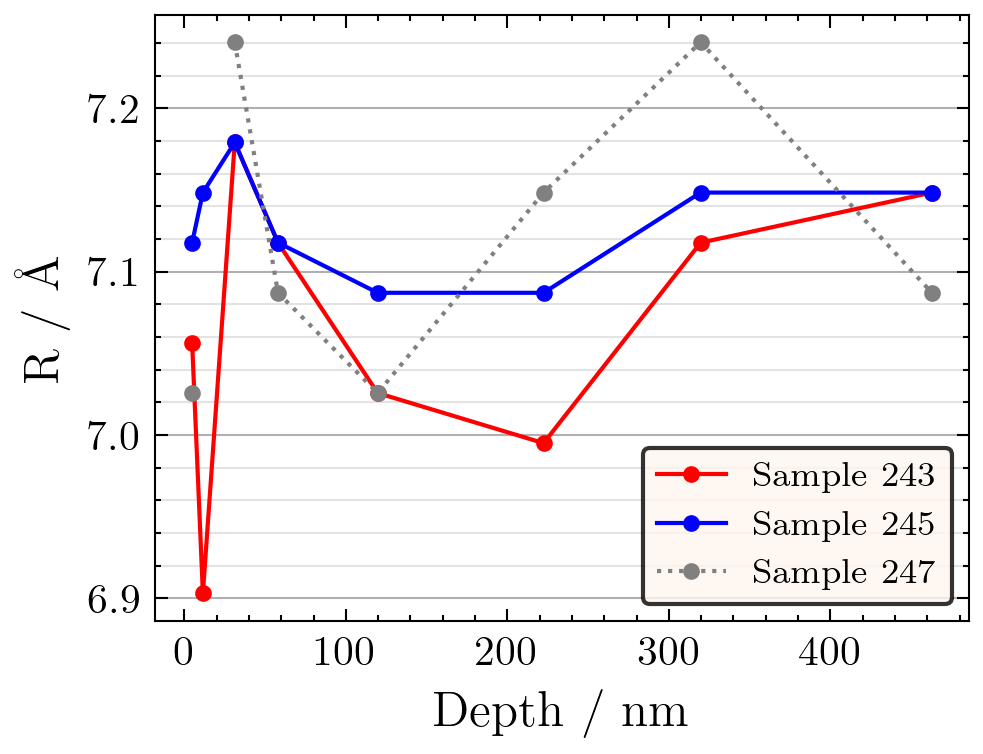}{width=0.9\textwidth}{0.485\textwidth}{}{5. coordination radius ($\mathrm{Nb-Nb}$)}

    \caption{Peak position (and thus, to a certain extent, lattice distances) as a function of the penetration depth. The lines are for illustrative purposes only and do not represent data, same applies for fig. \ref{fig:peak-migrations-amplitude} and \ref{fig:peak-migrations-width}.}
    \label{fig:peak-migrations-positions}
\end{figure}

\begin{figure}
    \centering
    
    \inputsubfigureimg{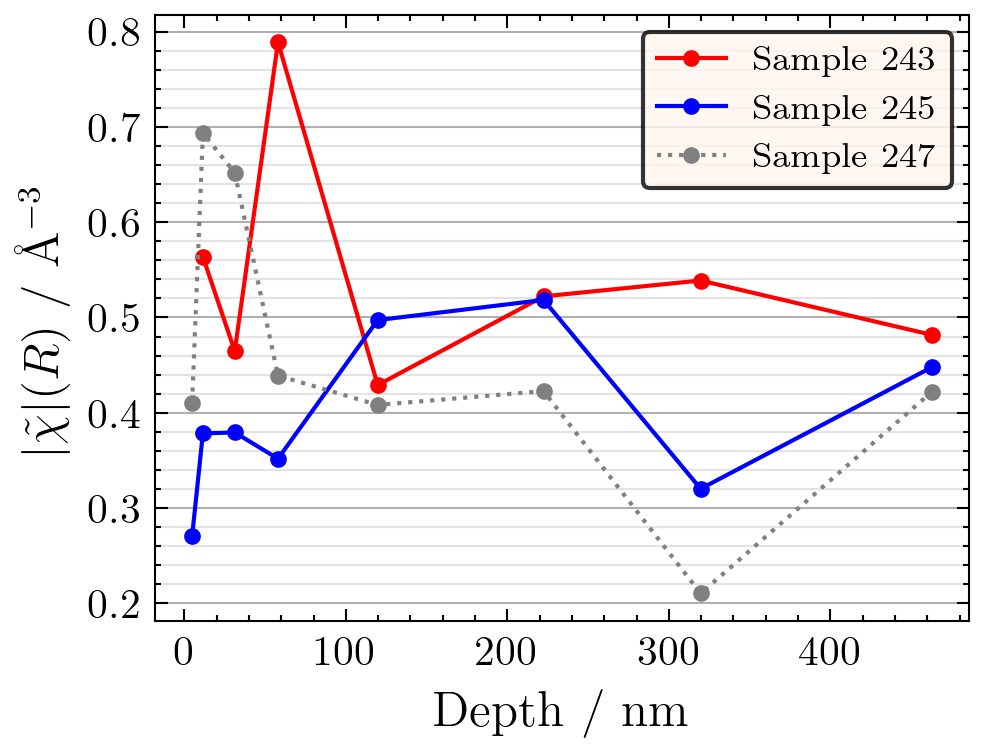}{width=\textwidth}{0.485\textwidth}{}{1. $\mathrm{Nb-O}$ distance}
    \inputsubfigureimg{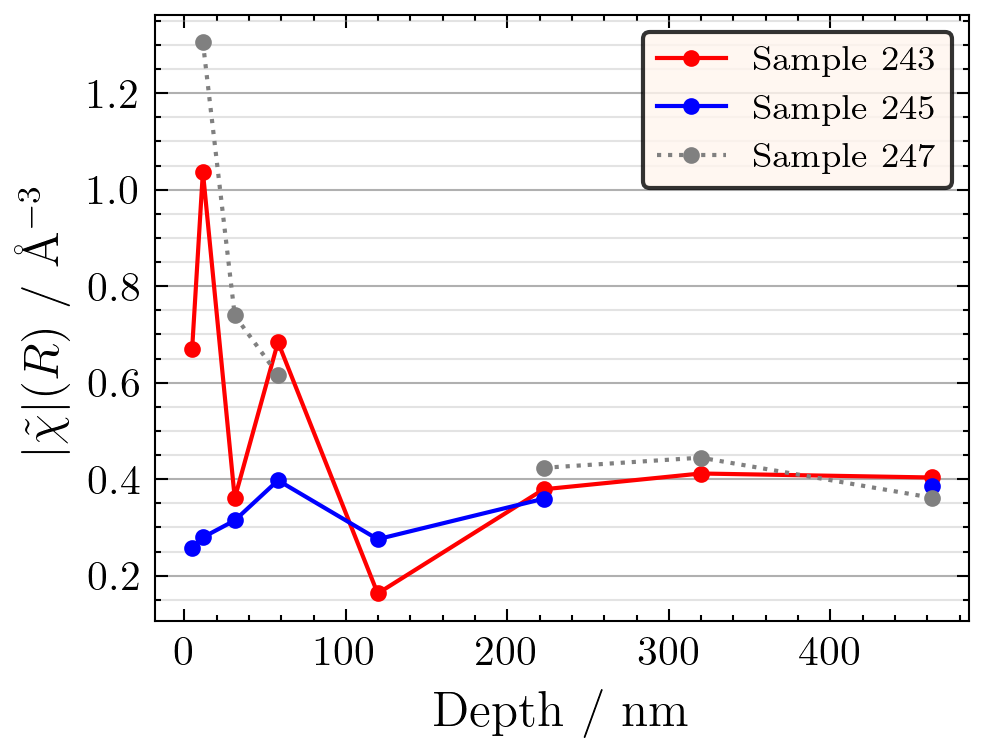}{width=\textwidth}{0.485\textwidth}{afterprepeak}{2. $\mathrm{Nb-O}$ distance}
    \inputsubfigureimg{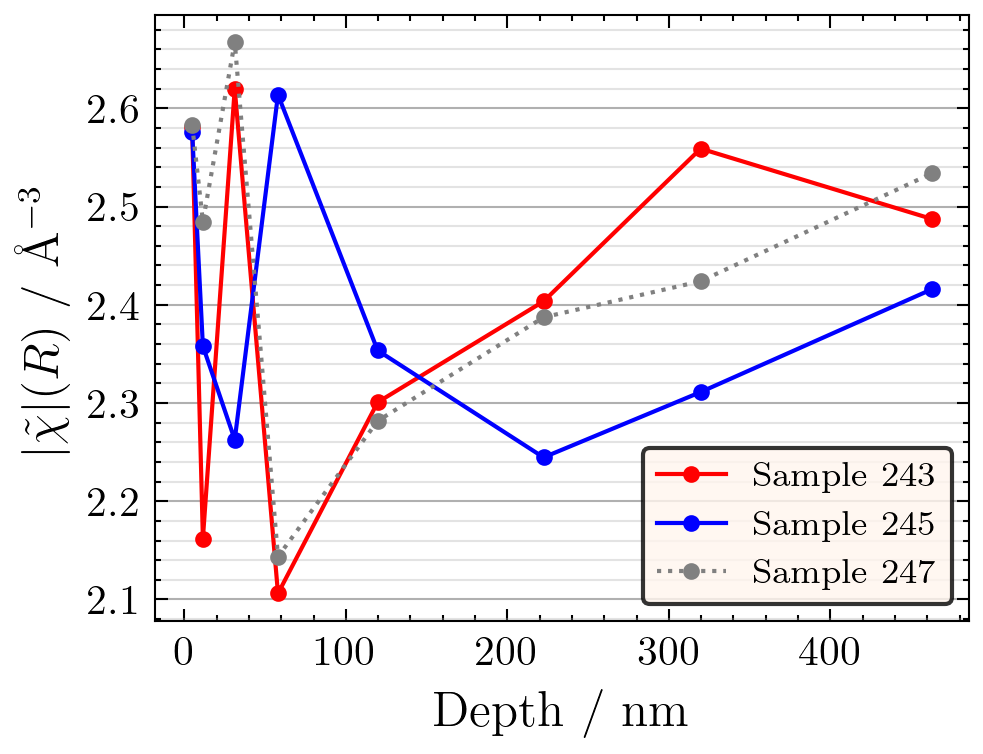}{width=\textwidth}{0.485\textwidth}{}{1. coordination radius ($\mathrm{Nb-Nb}$)}
    \inputsubfigureimg{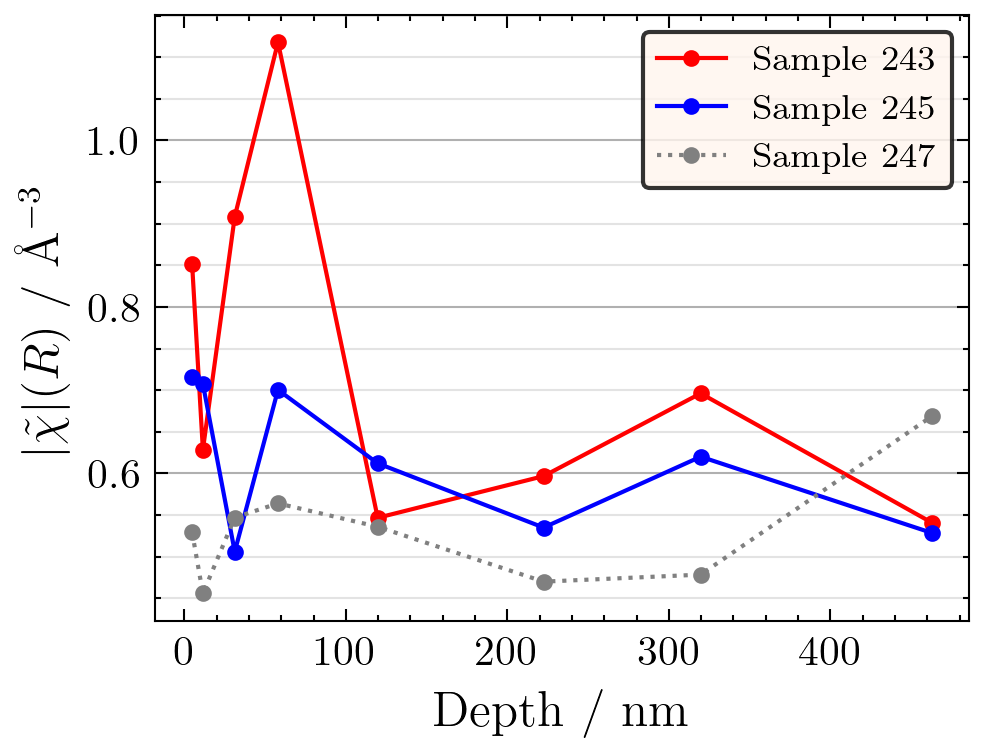}{width=\textwidth}{0.485\textwidth}{}{3. coordination radius ($\mathrm{Nb-Nb}$)}
    \inputsubfigureimg{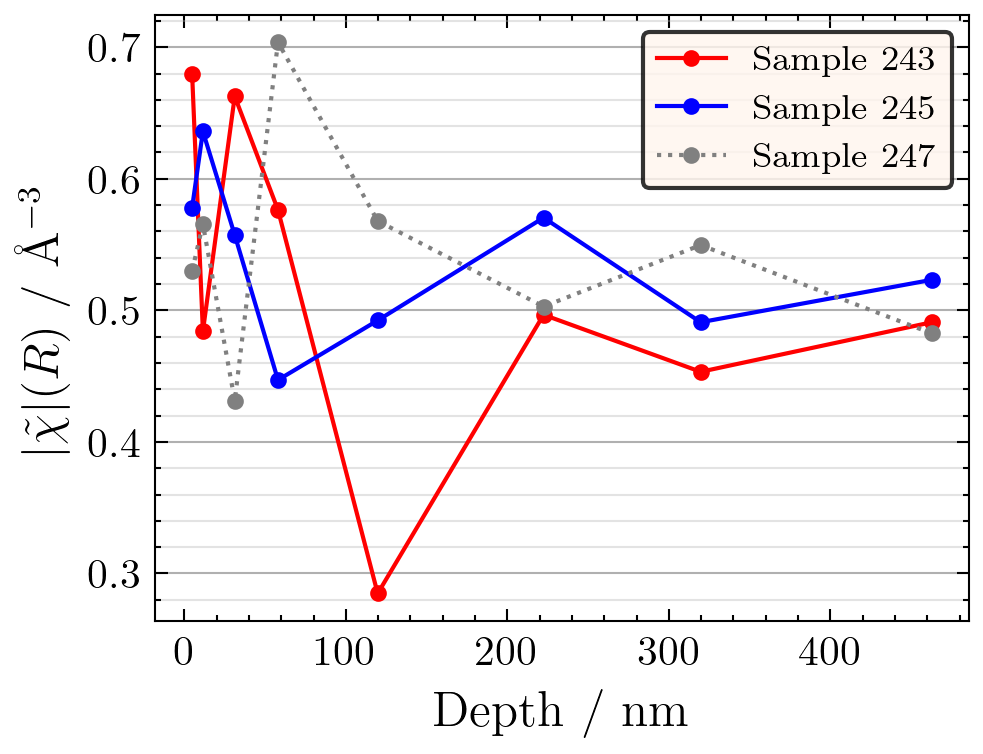}{width=\textwidth}{0.485\textwidth}{}{4. coordination radius ($\mathrm{Nb-Nb}$)}
    \inputsubfigureimg{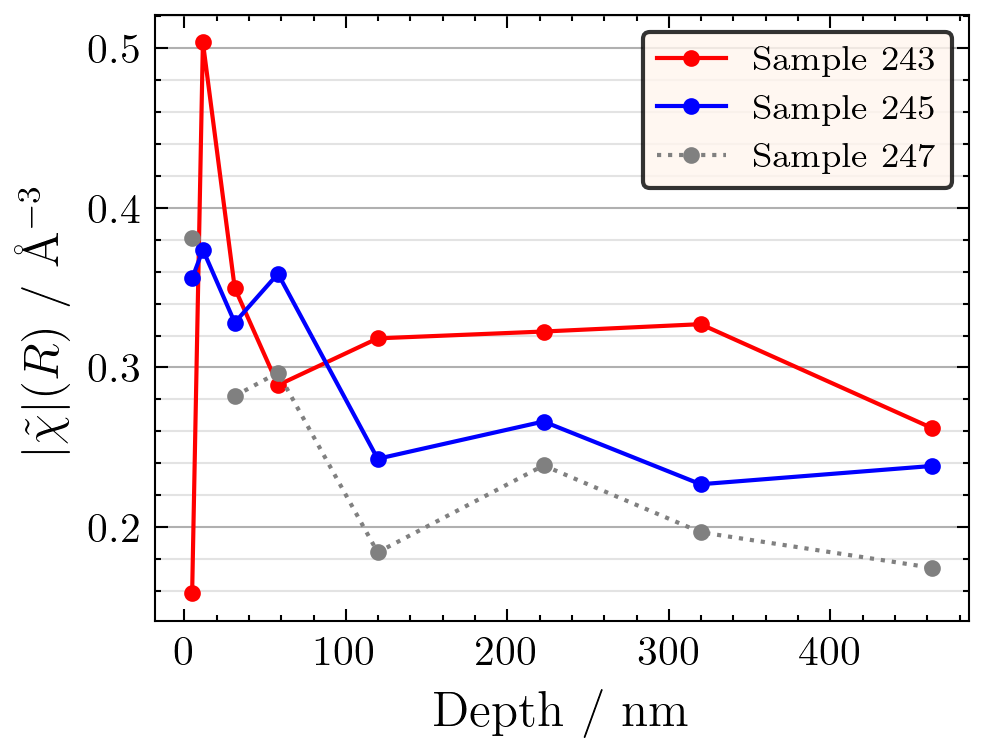}{width=\textwidth}{0.485\textwidth}{}{5. coordination radius ($\mathrm{Nb-Nb}$)}

    \caption{Peak amplitude as a function of the penetration depth.}
    \label{fig:peak-migrations-amplitude}
\end{figure}

\begin{figure}
    \centering
    
    \inputsubfigureimg{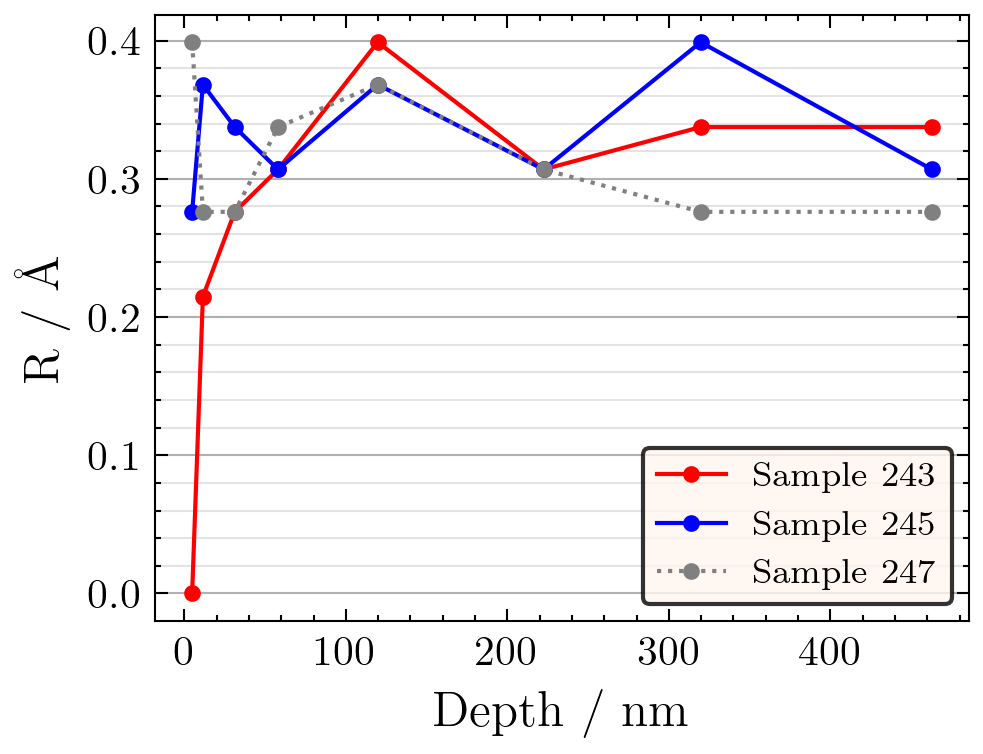}{width=\textwidth}{0.485\textwidth}{}{1. $\mathrm{Nb-O}$ distance}
    \inputsubfigureimg{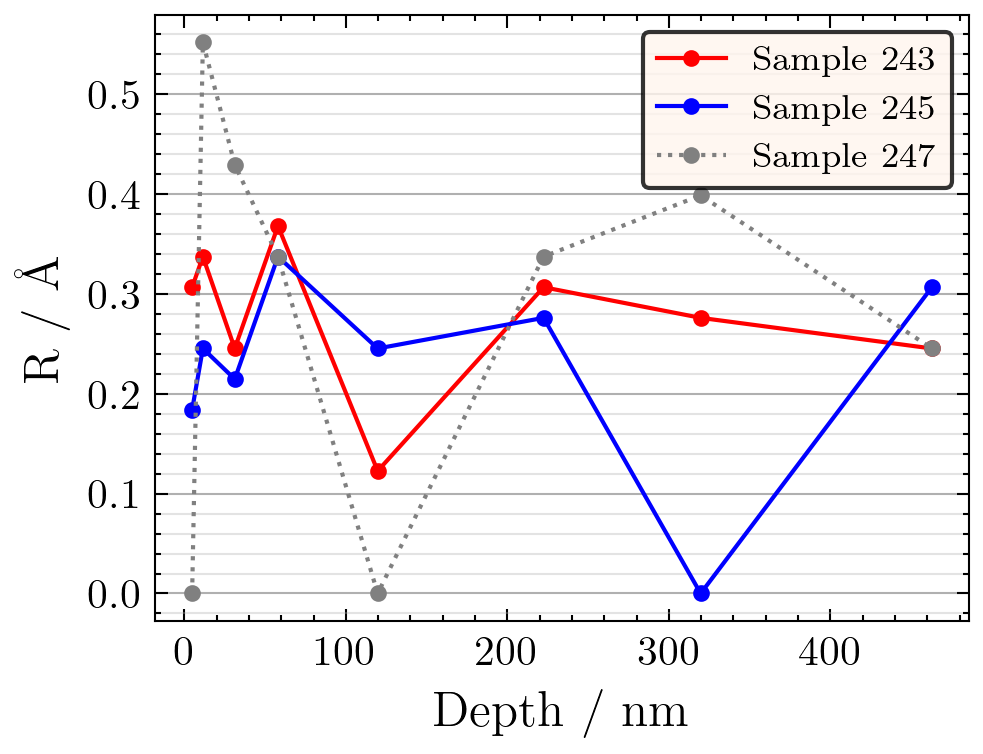}{width=\textwidth}{0.485\textwidth}{}{2. $\mathrm{Nb-O}$ distance}
    \inputsubfigureimg{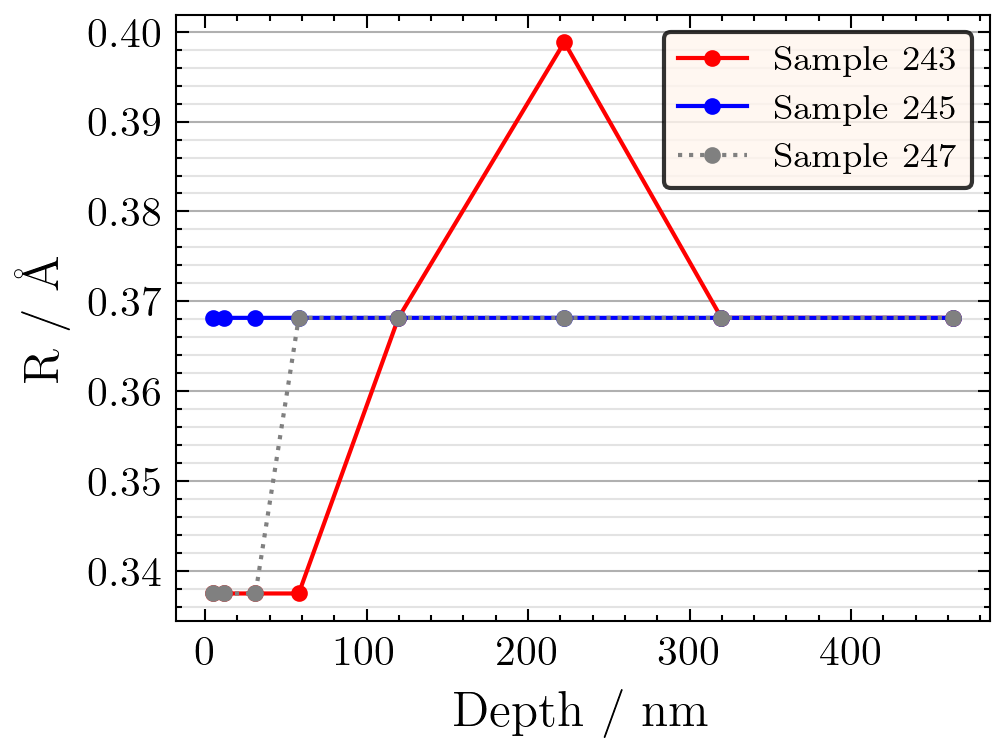}{width=\textwidth}{0.485\textwidth}{}{1. coordination radius ($\mathrm{Nb-Nb}$)}
    \inputsubfigureimg{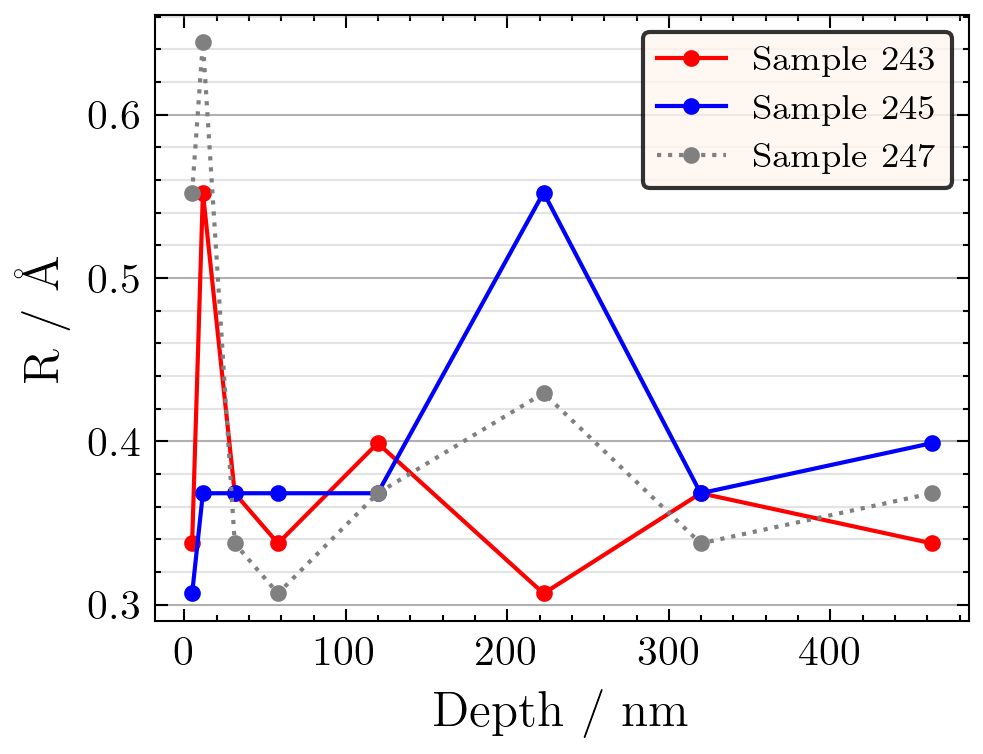}{width=\textwidth}{0.485\textwidth}{}{3. coordination radius ($\mathrm{Nb-Nb}$)}
    \inputsubfigureimg{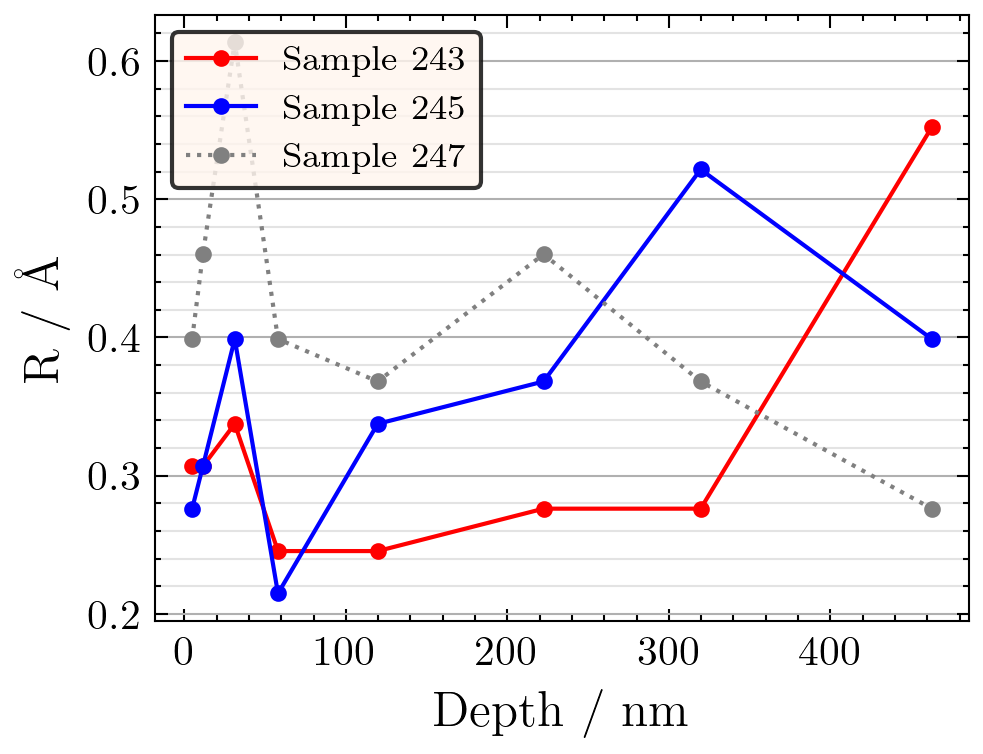}{width=\textwidth}{0.485\textwidth}{}{4. coordination radius ($\mathrm{Nb-Nb}$)}
    \inputsubfigureimg{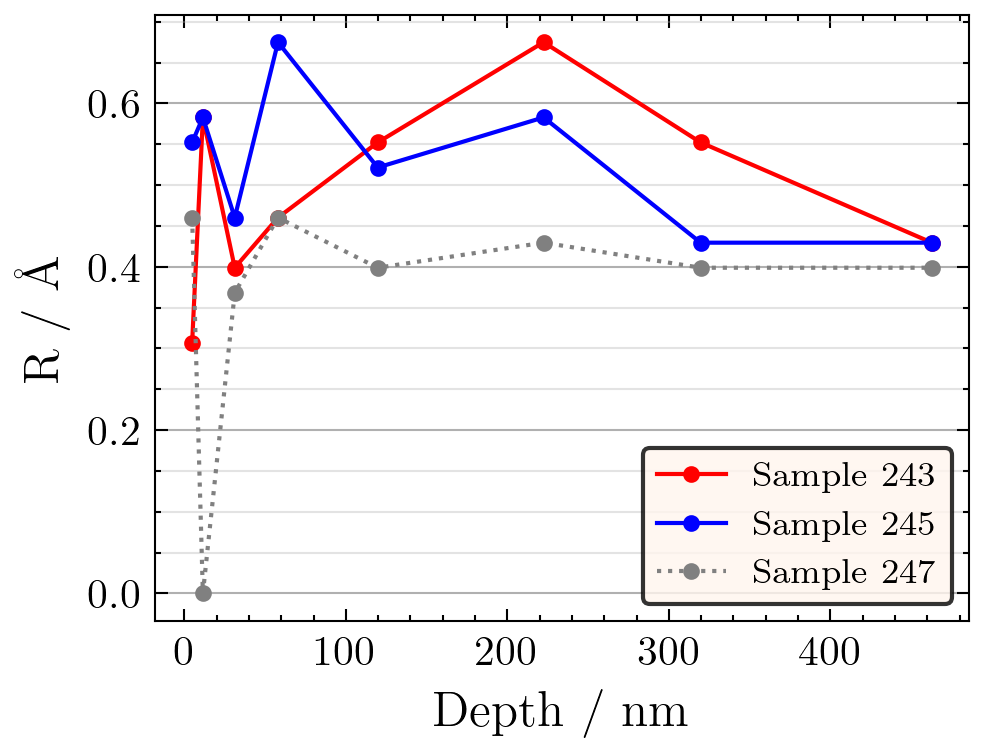}{width=\textwidth}{0.485\textwidth}{}{5. coordination radius ($\mathrm{Nb-Nb}$)}

    \caption{Peak width as a function of the penetration depth.}
    \label{fig:peak-migrations-width}
\end{figure}

\newpage
\section{Error Plots}\label{app:error-plots} 

\inputimgh{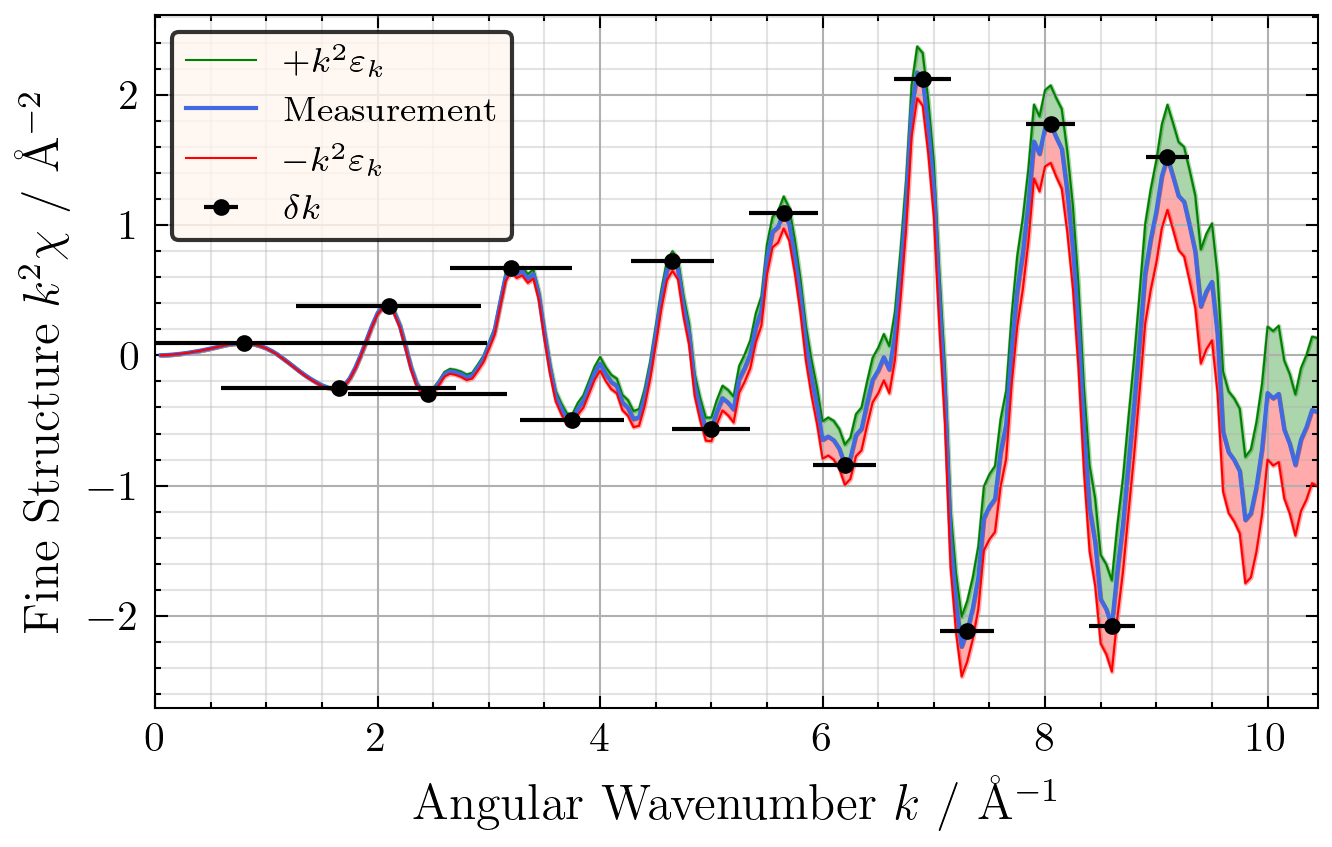}{width=0.85\textwidth}{fs-error}{Fluctuation $\varepsilon_k$ ($k^2$-weighted) and resolution $\delta k$ exemplarily of sample $245$ with an angle of incidence $\theta = \SI[round-precision=1]{0.5}{\degree}$, according to error analysis in sec. \ref{sec:error-analysis}.}
\inputimgh{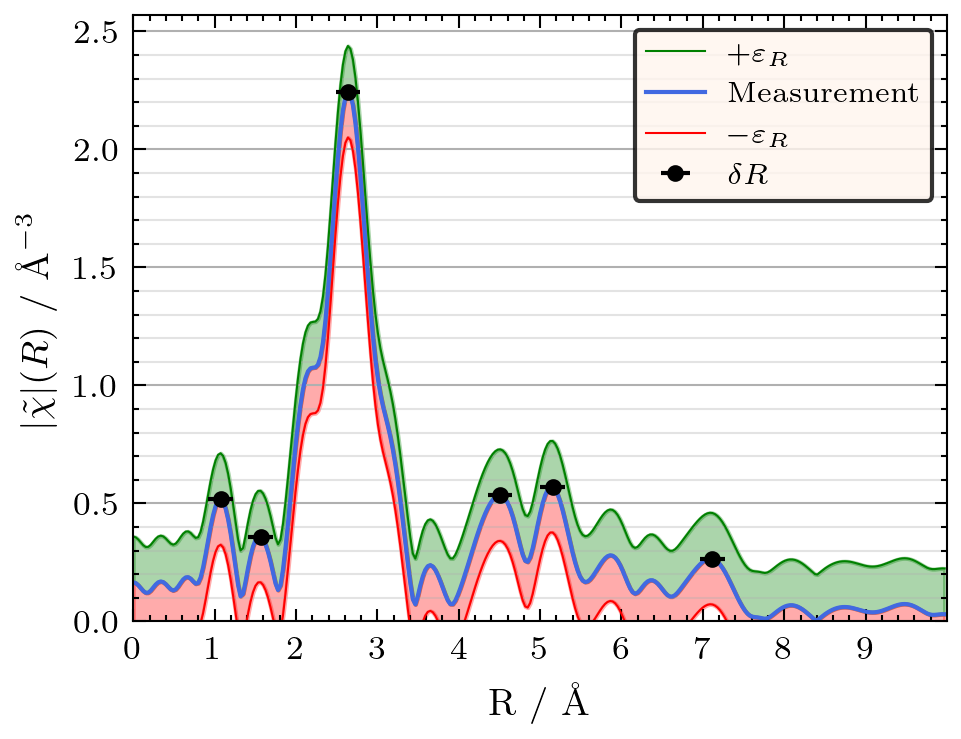}{width=0.85\textwidth}{fourier-error}{Fluctuation $\varepsilon_R$ and resolution $\delta R$ exemplarily of sample $245$ with an angle of incidence $\theta = \SI[round-precision=1]{0.5}{\degree}$, according to error analysis in sec. \ref{sec:error-analysis}.}

\section{\color{violet}Basic Terms of Self-Referencing Systems}\label{app:self-referencialism}

Practicing physics, both in the theoretical approach of discussing and inventing models and
the experimental approach of discussing and performing analyses, requires
an understanding of the research field in order to know which research will be productive and 
which meaningless\footnote{Hereby, research without a meaningful result isn't necessarily meaningless, because it can help to achieve progress. I take this thesis as a prime example.}. 
This understanding exists either in the form of knowing what the task is
(performing a predefined experiment, calculating a family of integrals for a more general problem, 
comparing the predictions of several known models following literature, etc.)
or knowing how to define an appropriate task (which experiments are promising for a successful
analysis of my sample, which models are able to explain a given many-body problem, etc.).
Though both forms act together, the latter and only the latter tends to ask the question
of \enquote{What should I do to achieve the good?} and
is due to the little word \enquote{good}, philosophical.
Furthermore it's describing a situation where the yet to be given answer
(the appropriate task) depends on the situation as a whole (with all it's values, influences
and interests) and thus
especially on the questioner. In other words, it describes a self-referencial
situation.\\

To return to the point and conclude, practicing physics requires to identificate and deal with
self-referencial situations.
The more independent a physicist has to work -- already increasingly during the process of
his/her education --, the more necessary this ability becomes,
as best exemplified by the reorientation of research following paradigm shifts,
which can occur without warning. In fields where methods, techniques and experience
are missing, knowledge about handling self-referencial systems helps moving forward.
This does not concern specific fields only, but strictly speaking every single field in science.\\

For this purpose, terms suitable for analyzing self-referencial systems will be proposed below.
The general picture in which we operate is that the physicist is
acting as him-/herself, observes natural phenomena inductively,
some of which he/she doesn't understand and others trying to
describe deductively, in a way that every other physicist could
observe them as well, due to that description.\\

Beginning by defining the term \textbf{seref} as representative for self-referencial,
the circumstance of knowing what the task is \textbf{calcism} and the circumstance
of finding the proper task \textbf{launism}. The idea is,
that these terms represent a more abstract meaning.
When a scientist observes reality (nature), he/she is first of all overwhelmed by
\textit{launistic} impressions (launism of nature itself), phenomena that he/she does not question but just accepts.
This way of experiencing a \textbf{seref system} shall be called \textbf{serlau}
and is supplemented by the term \textbf{sercal} for reducing the phenomena of
a \textit{seref system} to \textit{calculistic} (e.g. mathematically describable)
relations. Because every proposition about a \textit{seref system}
(\textbf{seref proposition}) can be denied (e.g. the Schr\"odinger equation of a
charged quantum particle in a potential $V(r)$ can be denied as not fundamental
by argumenting that a potential cannot be given without taking the solution of
the equation of motion into account, which influences the potential $V(r)$ again
via the coulomb interaction), it is an intrinsic property of a
\textit{seref proposition} that it can be doubted. Such doubts should therefore be called
\textbf{seref doubts}. And the
motivation of the physicist to further make propositions although they can be doubted
by \textit{seref doubts} is given by the awareness that some proposition will
turn out to be \textit{serefically true}\footnote{Meaning that such a proposition is of
the property that every other physicist could verify it by him-/herself.} for sure.
This motivation can thus be called \textbf{seref pleasure}.\\

Now \textit{seref propositions} can be categorized in the case of the proposer
in either a proposition motivated through \textit{seref pleasure} due to pure \textit{calculistic}
considerations of the proposer (let those be called \textbf{sercal pleasure} induced)
and motivated through knowingly \textit{launistic} considerations (accordingly called \textbf{serlau pleasure} induced).
And in case of the doubter of a given \textit{seref proposition},
this can lead to a doubt, that is only some obvious \textit{seref doubt}
and thus we obtain a \textbf{sercal doubted} \textit{seref proposition}
or a doubt, that references to some other \textit{seref proposition}
which contradicts the initial one (leading to a \textbf{serlau doubted}
\textit{seref proposition}).\\

Reflecting again, \textit{seref systems} do not have a clear rule to determine which proposition about
it is true and which is false.
Ultimately, however, this can be determined by examining what the case actually was
(e.g. the motion of a double pendulum cannot be predicted theoretically due to deterministic chaos,
meaning that propositions about it beforehand cannot be called true or false.
After observing it's development over time however, such propositions can be
verified afterwards). So some \textit{seref propositions} will turn out to be true,
even though they could never be proven logically. To describe this
dynamic entity of an ongoing (flowing) verification of \textit{seref propositions},
not due to logic but the verdict of reality, it shall be called \textbf{flarameter},
which in most cases of physics is given by the continuous parameter of time.\\

To allow a beneficial use of analyzing \textit{seref systems}
for the work of the physicist, it shall rather be discussed by the physicist,
which of the propositions he/she made and encountered are
induced through \textbf{sercal} or \textbf{serlau pleasure}
and \textbf{sercally} or \textbf{serlauily doubted}.
The \textbf{flarameter} will further show, which propositions
find confirmation and which not, and thus inspire the researching scientist
in his/her more resilient method of critical rationalism. It's
discussion will help further research to understand the context back then,
to gain imagination for the following. \\

Finally, a productive research in terms of \enquote{knowing how to define an appropriate task}
is accomplished through the following three-part program:
First identify those propositions that can be expressed as \textit{seref propositions}.
Secondly making sure that all expressed \textit{seref propositions}
are induced by \textit{serlau pleasure} and all challenged
\textit{seref propositions} (of cited literature) are \textit{serlau doubted}. 
And lastly providing scenarios of outcomes through the \textit{flarameter}
regarding the discussed \textit{seref propositions}.


\newpage
\begin{center}
     \textbf{\Large \scshape \underline{\underline{Acknowled}g\underline{{ements}}}}
\end{center}
\label{sec:acknowledgements}

\addcontentsline{toc}{section}{Acknowledgements}%
\vspace{1em}

This thesis shall be shared with several other people who contributed to it's creation.
I want to thank Lennard Luth for introducing me into the field of EXAFS
by sharing his knowledge through his own thesis. For performing the beamline experiment,
we have learned a lot from the previous attempt which he analyzed.
My supervisor Marc Wenskat was regularly available for discussions
and helped carry out the experiment in Dortmund. His character trait of
developing a deep fascination for his field of research and enthusiastically
expressing this to others helped me greatly to find motivation for this work.
I can't say that his ideas and visions in general and for this thesis specifically
always stand up to the reality of implementation. This lead to
repeated failures of considerations, frustration and forced me to make
spontaneous changes to our plans.
But I think I can say that this has taught me even more about the scientific way of operating
and I admire his optimism and willingness to try things out on principle and
to link areas that would otherwise probably rarely come together.
It matches my ideal understanding of experimental research
and I hope he can maintain this ability for himself.
Rezvan Ghanbari was very supportive, helped to perform the beamline experiment as well, %
gave advice for it's success and offered her expertise to gain deeper understanding of the matter.
I am very thankful for this, especially since she is very busy with her doctoral thesis.
As the representing professor, Wolfgang Hillert expressed critical opinions which I %
understood as a counterbalance to Marc. I always had his scepticism in mind,
which made me confident enough to not content myself with half-hearted answers. 
Dirk L\"utzenkirchen-Hecht gave main expertise at the DELTA facility during beamtime %
and in the theory of EXAFS afterwards, though the communication primarily had to take place virtually
between Hamburg and Wuppertal. It was interesting to see a synchrotron experiment
performed with all the experience needed in the regular case of unforseen problems.
Also I want to thank my friend Ivo Ziesche for TikZ'ing the experimental
setup, whose experience with this package is greater than mine and
has relieved me of the stress of having to worry about that as well in recent weeks.\\

Furthermore I'd like to thank Jessica Pfeffer, whose organizational skills helped me
transition from my student assistant status to that of a bachelor's student and print
this thesis, among other things.
Lea Steder familiarized me as supervisor with the
general experimental practice of SRF research in the AMTF hall during my student assistant time before.
This experience was surely helpful while writing this thesis,
to be aware of what the actual purpose of the MSL group is.
During the period of water damage in our group's HF lab,
I would like to thank everyone who helped with the renovation, especially 
Carsten Müller, Thorsten Büttner and Detlef Reschke.
For me, it was an opportunity to move into a different office that I found more comfortable,
but as a student, you benefit greatly from the social space that the HF lab provides.
Moreover, I want to thank the student night watch at DELTA during the beam time,
who had even longer shifts than I did. Although they locked me in the facility several times,
they always unlocked it for me again.\\

Ultimately, I thank my brother Till, who allowed me to turn the apartment into utter
chaos during the demanding last few weeks surrounding our grandmother's illness and
the completion of this thesis.

\newpage
\includepdf{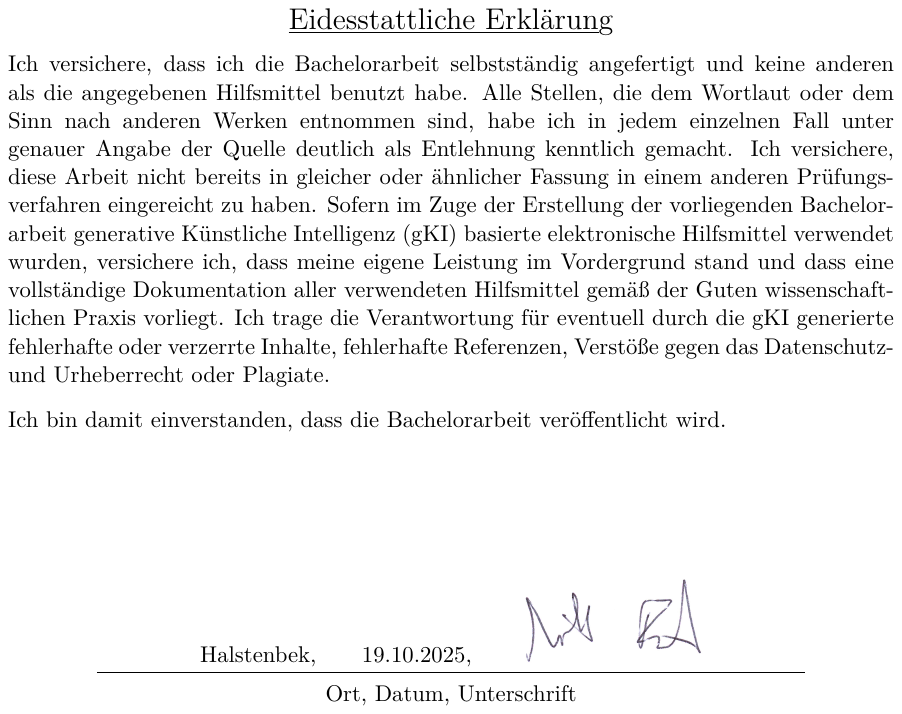}
\end{document}